\documentclass[iop, apj, iopart]{emulateapj}
\usepackage{color}
\usepackage[encapsulated]{CJK}
\usepackage{ucs}
\usepackage[utf8x]{inputenc}
\usepackage{gensymb}
\usepackage{amsmath}
\usepackage{amssymb}
\usepackage[flushleft]{threeparttable}
\usepackage{array}
\newcolumntype{x}[1]{>{\centering\arraybackslash\hspace{0pt}}p{#1}}

\newenvironment{Contfigure*}{%
\renewcommand{\thefigure}{\arabic{figure} (Cont.)}%
\addtocounter{figure}{-1}%
\begin{figure*}}{%
\renewcommand{\thefigure}{\arabic{figure}}%
\end{figure*}}

\newcommand{\kms}{$\mathrm{km\,s^{-1}}$}

\begin{document}
\begin{CJK}{UTF8}{gbsn}

\title{The redshift evolution of rest-UV spectroscopic properties in Lyman Break Galaxies at $z\sim2-4$}

\author{Xinnan Du (杜辛楠)\altaffilmark{1}, Alice E. Shapley\altaffilmark{1}, Naveen A. Reddy\altaffilmark{2}, Tucker Jones\altaffilmark{3}, Daniel P. Stark\altaffilmark{4}, 
Charles C. Steidel\altaffilmark{5}, Allison L. Strom\altaffilmark{6}, Gwen C. Rudie \altaffilmark{6}, Dawn K. Erb\altaffilmark{7}, Richard S. Ellis\altaffilmark{8}, $\&$ Max Pettini\altaffilmark{9}}

\altaffiltext{1}{University of California, Los Angeles, CA 90095}

\altaffiltext{2}{University of California, Riverside, CA 92521}

\altaffiltext{3}{University of California, Davis, CA 95616}

\altaffiltext{4}{University of Arizona, Tucson, AZ 85719}

\altaffiltext{5}{California Institute of Technology, Pasadena, CA 91125}

\altaffiltext{6}{Carnegie Observatories, Pasadena, CA 91101}

\altaffiltext{7}{University of Wisconsin Milwaukee, Milwaukee, WI 53211}

\altaffiltext{8}{University College London, London, WC1E 6BT}

\altaffiltext{9}{Institute of Astronomy, Madingley Road, Cambridge CB3 OHA, UK}

\slugcomment{Draft Version \today}

\shorttitle{LBG evolution}
\shortauthors{Du}

\begin{abstract}
We present the first comprehensive evolutionary analysis of the rest-frame UV spectroscopic properties of star-forming galaxies at $z\sim2-4$. We match samples at different redshifts in UV luminosity and stellar mass, 
and perform systematic measurements of spectral features and stellar population modeling. By creating composite spectra grouped according to $\mbox{Ly}\alpha$ equivalent width (EW), and various galaxy properties, we study the 
evolutionary trends among $\mbox{Ly}\alpha$, low- and high-ionization interstellar (LIS and HIS) absorption features, and integrated galaxy properties. We also examine the redshift evolution of $\mbox{Ly}\alpha$ and LIS
absorption kinematics, and fine-structure emission EWs. The connections among the strengths of 
$\mbox{Ly}\alpha$, LIS lines, and dust extinction are redshift-independent, as is the decoupling of Ly$\alpha$ and HIS line strengths, and the bulk outflow kinematics as traced by LIS lines. Stronger $\mbox{Ly}\alpha$ emission is observed at higher 
redshift at fixed UV luminosity, stellar mass, SFR, and age. Much of this variation in average Ly$\alpha$ strength with redshift, and the variation in Ly$\alpha$ strength at fixed redshift, can be explained in terms of variations in neutral gas covering fraction and/or dust content in the ISM and CGM. However, based on the connection between Ly$\alpha$ and \textrm{C}~\textsc{iii}] emission strengths, we additionally find evidence for variations in the {\it intrinsic} 
production rate of Ly$\alpha$ photons at the highest Ly$\alpha$ EWs. The challenge now is to understand the observed evolution in neutral gas covering fraction and dust extinction within a coherent model for galaxy formation, and make robust predictions for the escape of ionizing radiation at $z>6$.
\end{abstract}


\section{INTRODUCTION}
\label{sec:Intro}

The rest-frame ultraviolet (UV) spectra of star-forming galaxies provide rich insights into the physical properties of not only their massive stars, but also their gas. This gas includes the multi-phase interstellar medium (ISM) and circumgalactic medium (CGM), which extends to the virial radius \citep{Tumlinson2017}. Both low- and high-ionization interstellar (LIS and HIS, respectively) absorption lines are probes of the ISM and CGM, which typically reflect the kinematic signatures of galaxy-wide outflows in star-forming galaxies at these redshifts \citep[e.g.,][]{Shapley2003,Steidel2010,Jones2012}. LIS features primarily trace the neutral phase of outflows while HIS lines mainly trace the ionized phase. Fine-structure emission lines (e.g., \textrm{Si}~\textsc{ii}*, \textrm{Fe}~\textsc{ii}*) arise from the emission of a photon corresponding to a drop down to the excited ground state following resonant absorption. These fine structure features are typically coupled with corresponding resonant LIS absorption, and are thus effective tracers of the structure and the spatial extent of the CGM \citep[e.g.,][]{Erb2012,Jones2012,Kornei2013}. The \textrm{H}~\textsc{i} $\mbox{Ly}\alpha$ feature, on the other hand, has a more complex nature. $\mbox{Ly}\alpha$ photons are produced by recombination in \textrm{H}~\textsc{ii} regions, and then propagate through the ISM, interacting with both neutral hydrogen and dust grains. As such, $\mbox{Ly}\alpha$ photons potentially offer insights into the properties of both their HII regions of origin, and the more extended ISM and CGM through which they propagate. Nebular emission lines, such as \textrm{C}~\textsc{iii}]$\lambda\lambda$1907,1909 and [\textrm{O}~\textsc{iii}]$\lambda\lambda$1661, 1666, are produced in \textrm{H}~\textsc{ii} regions, and serve as useful probes of the ionized ISM and radiation field produced by massive stellar populations \citep[e.g.,][]{Erb2010,Berg2016,Stark2016,Du2017,Senchyna2017}.
When measured over a wide range of redshifts, the properties of rest-UV absorption and emission features provide a window into the evolution of ISM/CGM properties. Properties of particular interest include the evolving gas covering fraction, which modulates the emergent $\mbox{Ly}\alpha$ properties and escape of ionizing radiation; and the characteristics of galaxy outflows, which play a crucial role in the formation and evolution of galaxies.

As $\mbox{Ly}\alpha$ photons are produced in \textrm{H}~\textsc{ii} regions and scattered and absorbed in the dusty ISM, $\mbox{Ly}\alpha$ is a key probe of the physical conditions of interstellar gas (e.g., metallicity, ionization parameter, \textrm{H}~\textsc{i} column density and covering fraction, and dust distribution) given its unique detectability especially at high redshift. Numerous studies have been carried out to understand the factors modulating the strength of $\mbox{Ly}\alpha$ emission at $z\sim2-6$. Observationally, larger $\mbox{Ly}\alpha$ equivalent width (EW) is typically found in galaxies with bluer UV color, lower metalliticity, lower stellar mass, lower UV luminosity, and lower SFR \citep[e.g.,][]{Shapley2003, Reddy2006,Kornei2010,Pentericci2010,Stark2010,Berry2012,Jones2012,Trainor2016,Erb2016,Hathi2016}. Furthermore, understanding the mechanisms underlying the existence of strong $\mbox{Ly}\alpha$ emitters (LAEs; rest-frame $\mbox{Ly}\alpha$ EW $>20\mbox{\AA}$) is critical for our interpretation of the galaxy populations during the reionization epoch, where, thus far, the majority of data are only photometric.  

The connections between $\mbox{Ly}\alpha$ and other spectral features, including LIS and HIS absorption lines and nebular emission lines, have been extensively studied in star-forming galaxies. Stronger $\mbox{Ly}\alpha$ appears to be associated with weaker LIS absorption lines \citep{Shapley2003,Jones2012,Berry2012} and stronger nebular emission lines \citep{Shapley2003,Stark2014,Stark2015,Rigby2015,Jaskot2016,Erb2016,Trainor2016}. In contrast, $\mbox{Ly}\alpha$ EW does not show significant correlation with the strength of HIS absorption lines \citep{Shapley2003,Berry2012}. The shape of the $\mbox{Ly}\alpha$ profile (e.g., peak location, number of peaks, asymmetry) can also be used as a probe of the kinematics and density distribution of outflowing neutral gas \citep[e.g.,][]{Verhamme2006,Steidel2010,Kulas2012}.

While much of the work on $\mbox{Ly}\alpha$ emission and other rest-UV spectroscopic properties of star-forming galaxies at high redshift has been focused on characterizing samples at individual redshifts \citep[e.g.,][]{Shapley2003,Pentericci2007,Pentericci2009,Kornei2010,Steidel2010}, there have been limited studies on the evolution of these spectroscopic properties at $z\gtrsim2$. LAEs are found to be more prevalent at higher redshifts up to $z\sim6$ \citep{Reddy2008,Stark2010,Schenker2014}, and stronger $\mbox{Ly}\alpha$ and weaker LIS absorption are observed with increasing redshift at fixed UV luminosity \citep{Stark2010,Jones2012}. Specifically, by comparing a sample of $z\sim4$ LBG spectra to the $z\sim3$ LBG composites from \citet{Shapley2003}, \citet{Jones2012} found no evidence for evolution of the correlation between $\mbox{Ly}\alpha$ and LIS absorption EWs. However, these authors discovered smaller characteristic radii of fine-structure emission and overall weaker LIS absorption at higher redshifts. While \citet{Jones2012} created a subsample of $z\sim4$ galaxies to match the $z\sim3$ galaxies in UV absolute magnitude for a fair comparison, the study was limited in that the SED-derived galaxy properties (e.g., stellar mass) were unavailable at that time for the majority of the galaxies in the $z\sim3$ comparison sample from \citet{Shapley2003}. Furthermore, the $z\sim3$ spectroscopic measurements (e.g., $\mbox{Ly}\alpha$ and LIS absorption EWs) were taken from \citet{Shapley2003} and not necessarily performed in a manner consistent with that of \citet{Jones2012}. When conducting evolutionary studies on the connection between rest-UV spectroscopic and galaxy properties, both the spectroscopic measurements and the modeling of galaxy photometry must be performed in a uniform, controlled fashion.

The redshift range $z\sim2-4$ spans nearly 2 billion years of cosmic time and covers the peak epoch of star formation and beyond \citep{Madau2014}. The active star formation and mass assembly over this period must have been shaped by various processes (e.g., feedback, mergers, gas accretion), that will also be reflected in the physical characteristics of the ISM/CGM. Accordingly, it is of key interest to trace the evolving ISM/CGM with galaxy spectra.

Here we present the first comprehensive study of the evolution of the ISM and CGM as probed by rest-UV spectra in star-forming galaxies at $z\sim2-4$. In order to perform meaningful comparisons between galaxy samples at different redshifts, we carefully construct these samples with galaxies spanning the same range in UV absolute magnitude and stellar mass. Furthermore, we model the stellar populations and measure the spectral features in all redshift samples in a uniform manner to avoid potential systematic biases. By comparing the rest-UV spectroscopic properties in the controlled samples at different redshifts, we aim to investigate the evolution of covering fractions of gas and dust, structure and kinematics of the multi-phase ISM/CGM, and the intrinsic $\mbox{Ly}\alpha$ production, which, in combination, provide rich insights into galaxy evolution. 

We provide a brief overview of the observations and data in Section~\ref{sec:data}, along with a description of galaxy SED modeling and the construction of controlled samples in UV luminosity and stellar mass. In Section~\ref{sec:meas}, we describe methods for creating composite spectra, and measuring $\mbox{Ly}\alpha$ emission and interstellar absorption profiles. We present multiple $z\sim2-4$ evolutionary trends in Section \ref{sec:res1}, including the connections among key spectroscopic features ($\mbox{Ly}\alpha$, and LIS and HIS absorption features), between $\mbox{Ly}\alpha$ and galaxy stellar populations, and fine-structure emission properties. We also investigate the correlation between $\mbox{Ly}\alpha$ and \textrm{C}~\textsc{iii}] emission. In Section \ref{sec:res2}, we examine the evolution of outflow kinematics as probed by both $\mbox{Ly}\alpha$ and LIS absorption features over the same redshift range. In Section \ref{sec:dis}, we connect our results to the evolution of outflow kinematics over $z\sim0-2$, and present a physical picture for the evolving distribution of $\mbox{Ly}\alpha$ emission and interstellar absorption measurements at $z\sim2-4$. Finally, we summarize our results in Section \ref{sec:sum}. In Appendix \ref{sec:AGN}, we consider the rest-UV spectroscopic properties of galaxies with mid-IR SED excesses, another topic of interest that can be explored with our data.

Throughout this paper, we adopt a standard $\Lambda$CDM cosmology with $\Omega_{m}=0.3$, $\Omega_{\Lambda}=0.7$ and $H_{0}=$70 km $\mbox{s}^{-1}$. All wavelengths are measured in the vacuum frame. Magnitudes are on the AB system.

\section{Observations, Data Reduction and Samples}
\label{sec:data}

In this section, we describe the $z\sim2-4$ LBG data used in this study, along with the derivation of stellar population parameters from spectral energy distributions (SEDs), and the sample properties. We refer readers to the original papers in which these data were presented for a more in-depth discussion of sample selection, data reduction, and photometric measurements of the samples presented here. 

\subsection{Samples}
\label{sec:sample}

\subsubsection{LRIS Sample}
\label{sec:z2z3}

The $z\sim2-3$ sample was drawn from the UV-selected galaxy surveys described in \citet{Steidel2003,Steidel2004} and \citet{Reddy2008}. These galaxies were preselected photometrically according to the $U_{n}GR$ color cuts to an apparent magnitude limit of $R_{AB}=25.5$, and spanned 15 fields covering a total area of $\sim1900$ $\mbox{arcmin}^{2}$. The candidates were then followed up spectroscopically with the Low Resolution Imager and Spectrometer \citep[LRIS;][]{Oke1995,Steidel2004} on the Keck I telescope. We note that galaxies selected in this manner at $z\sim2$ are not technically LBGs, as their actual Lyman break falls bluewards of the $U_{n}$ band. However, as described in \citet{Steidel2004}, the $z\sim2$ $U_{n}GR$-selected galaxies have very similar properties to those of the $z\sim3$ LBGs. Therefore, we also refer to the $z\sim2$ $U_{n}GR$-selected galaxies as LBGs hereafter for simplicity. 

The data were collected during multiple observing runs from 1997 to 2009 using $1."2$ slits for the multislit masks. The majority of the objects in the LRIS sample were observed with the 400-line $\mbox{mm}^{-1}$ ($\sim57\%$) and 600-line $\mbox{mm}^{-1}$ ($\sim33\%$) grisms. A small fraction of the data was obtained using the 300-line $\mbox{mm}^{-1}$ grism ($\sim3\%$), and the 300-line $\mbox{mm}^{-1}$ grating ($\sim4\%$), the latter taken before the LRIS-B upgrade in 2000 \citep{Steidel2004}. Twenty-one spectra ($\sim2\%$) were combined from multiple observations with a mixture of the 400- and 600-line $\mbox{mm}^{-1}$ grisms. The effective spectral resolutions of the 400-, 600-, 300-line $\mbox{mm}^{-1}$ grisms, and the 300-line $\mbox{mm}^{-1}$ grating are $R\sim800$, 1330, 530, and 670, respectively. The typical integration time for each slit mask is 1.5 hours, and is significantly longer ($5-8$ hours) for a small subset of the slitmasks observed as part of specialized investigations \citep[e.g., searches for Lyman continuum radiation][Steidel in prep.]{Reddy2016}. All two-dimensional spectra were flat fielded, cleaned of cosmic rays, background subtracted, extracted, wavelength and flux calibrated.

Ideally, the systemic redshift can be robustly measured from strong rest-frame optical nebular emission lines (e.g., $[\textrm{O}~\textsc{ii}]\lambda3727$, $\mbox{H}\beta\lambda4861$, $[\textrm{O}~\textsc{iii}]\lambda\lambda4959,5007$, and $\mbox{H}\alpha\lambda6563$). However, nebular lines were not measured for the vast majority of the $z\sim2-3$ objects in our sample, so the systemic redshift of these objects was estimated instead based on the redshift of $\mbox{Ly}\alpha$ emission and/or low-ionization interstellar (LIS) absorption lines. Due to the presence of outflowing neutral gas in the galaxies, $\mbox{Ly}\alpha$ emission and the LIS absorption features are typically observed to be redshifted and blueshifted, respectively, relative to each other, and therefore do not trace the galaxy systemic velocity. Instead, the systemic redshift was determined following the procedure described in\citet{Rudie2012} assuming that $\mbox{Ly}\alpha$ is redshifted by +300 \kms from the systemic velocity, and the LIS absorption lines are blueshifted by $-160$ \kms. The $\mbox{Ly}\alpha$ velocity correction was applied to all spectra with measurable $\mbox{Ly}\alpha$, and the LIS correction was applied to those with LIS redshift measurements only. In this manner, we obtained the systemic redshift for all the $z\sim2-3$ LBGs observed with LRIS, and the typical uncertainty on the systemic redshift is $\sim125$ \kms. The systemic redshift, as estimated above, was used to transform each spectrum to the rest frame.

The resulting LRIS sample includes 1297 LBGs spanning the redshift range $1.7\lesssim z \lesssim3.65$. These galaxies have secure redshift measurements, spectral coverage of $\mbox{Ly}\alpha$, and detections in the $K$- band and/or IRAC channel 1 or 2, enabling reliable stellar population modeling.

\subsubsection{DEIMOS/FORS2 Sample}
\label{sec:z4}

To probe $z\sim4$ LBGs, we primarily use the spectra presented in \citet{Jones2012}. This sample includes 70 objects (81 spectra accounting for duplicate observations) with $3.5 < z < 4.5$ and apparent magnitude of $z^{'}_{AB} < 26.0$. These galaxies were selected as $B$-band dropouts in the two GOODS fields \citep{Gia2004} and have a spectroscopic completeness of $\sim100\%$ at $z^{'}_{AB} < 24.75$ (Stark et al. in prep.). Detailed descriptions of sample selection and photometric measurements can be found in \citet{Stark2009,Stark2010} and \citet{Jones2012}. In the $z\sim4$ sample, 42 out of 70 objects (53 out of 81 spectra) were observed with the Deep Imaging Multi-Object Spectrograph \citep[DEIMOS;][]{Faber2003} on the Keck II telescope. The DEIMOS spectra have a constant resolution of $\simeq3.5\mbox{\AA}$ across the spectral range of $4100-9300\mbox{\AA}$. The typical integration time for the DEIMOS spectra is $5-7$ hours per mask. The other 28 objects (28 spectra) were obtained from the archival spectroscopic database of the FOcal Reducer and low dispersion Spectrograph 2 \citep[FORS2;][]{Vanzella2005,Vanzella2006,Vanzella2008,Vanzella2009}. The FORS2 data have a spectral resolution of $R\sim660$ over $6000-10000\mbox{\AA}$, corresponding to $\sim9.0$ $\mbox{\AA}$ near the observed wavelength of $\mbox{Ly}\alpha$. The FORS2 spectra typically have $4-6$ hr integrations for each mask. Similar to the $z\sim2-3$ LRIS data, standard data reduction procedures (flat fielding, cosmic ray rejection, background subtraction, extraction, wavelength and flux calibration, and transformation to the vacuum wavelength frame) were performed on the DEIMOS and FORS2 data to obtain calibrated one-dimensional spectra.

The systemic redshift of the $z\sim4$ galaxies was also estimated based on $\mbox{Ly}\alpha$ emission and LIS absorption features, using a procedure analogous to the one described in Section \ref{sec:z2z3}. To correct $\mbox{Ly}\alpha$ and LIS absorption redshifts to the systemic value, \citet{Jones2012} assumed a velocity offset of +330 \kms for $\mbox{Ly}\alpha$ and $-190$ \kms for the LIS features. The systemic redshift of individual galaxies was determined from the $\mbox{Ly}\alpha$ redshift when available, given its robustness compared to that of the LIS features, and from the LIS absorption redshift when the $\mbox{Ly}\alpha$ redshift was not measurable.

\subsubsection{Redshift Samples}
\label{sec:zsam}

\begin{figure}
\includegraphics[width=1.0\linewidth]{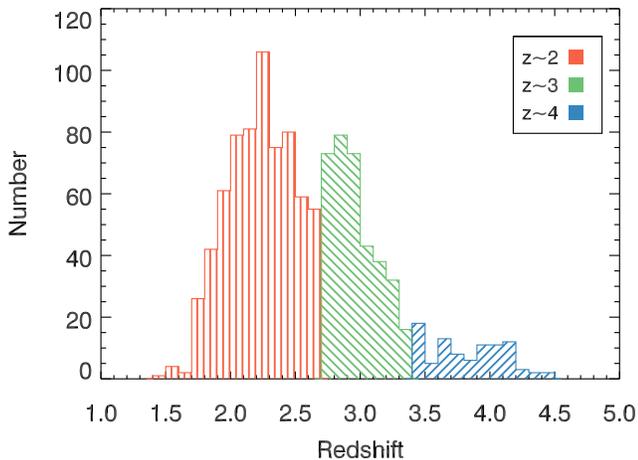}
\caption{Redshift distribution of our sample, divided according to boundaries at $z=2.7$ and $z=3.4$. The $z\sim2$ sample (671 objects) is indicated in red, the $z\sim3$ sample (352 objects) in green, and the $z\sim4$ sample (80 objects, 91 spectra) in blue. UV absolute magnitude and mass cut offs have not been applied here.}
\label{fig:zhist}
\end{figure}

We applied fixed redshift cuts to the LRIS, DEIMOS, and FORS2 datasets to define the $z\sim2, 3$, and $4$ samples. We considered galaxies with $z < 2.7$ to be in the $z\sim2$ sample, galaxies with $2.7 \leqslant z < 3.4$ to be in the $z\sim3$ sample, and those with $z \geqslant 3.4$ to be in the $z\sim4$ sample. The application of the redshift cuts results in 671, 352, and 80 objects (91 spectra) in the $z\sim2$, $z\sim3$, and $z\sim4$ samples, respectively. Figure \ref{fig:zhist} shows the redshift histograms of the $z\sim2, 3$ and $4$ samples defined by fixed redshift cuts, with median redshifts of 2.25, 2.93, and 3.86, respectively. In practice, the $z\sim2$ and $z\sim3$ samples are all probed with LRIS data, while the $z\sim4$ sample is mainly covered by DEIMOS and FORS2 spectra, with a small addition of objects observed with LRIS.

\subsection{SED Fitting}
\label{sec:sed}

\begin{table*}
\centering
\begin{threeparttable}
  \caption{Photometric Bands Used in SED Modeling\tnote{*}}
  \label{tab:phot}
%  \small
{\renewcommand{\arraystretch}{1.3}
  \begin{tabular}{c x{10cm}}
  \hline
  \hline
    Fields & Photometric bands \\
     & \\
  \hline
& $U$, $F435W$, $B$, $G$, $V$, $F606W$, $R$, ${R}_{s}$, $i$, $F775W$, \\
 GOODS-N\tnote{1} & $z$, $F850LP$, $F125W$,  $F140W$, $J$, $H$, $F160W$, \\
 & $Ks$, $IRAC1$, $IRAC2$, $IRAC3$, $IRAC4$ \\ 
 \hline
 & ${U}_{38}$, $U$, $F435W$, $B$, $V$, ${F606W}_{candels}$, \\
 & $F606W$, $R$, $Rc$, $F775W$, $i$, ${F814W}_{candels}$, \\
 GOODS-S\tnote{1} & $F850LP$, ${F850LP}_{candels}$, $F125W$, \\
 & $J$, ${J}_{tenis}$, $F140W$, $H$, $F160W$, ${K}_{tenis}$, $Ks$, \\ 
 & $IRAC1$, $IRAC2$, $IRAC3$, $IRAC4$ \\ 
\hline
 Q0100, Q0142, Q0449, Q1009, & ${U}_{n}$\tnote{2}, $G$\tnote{2}, ${R}_{s}$\tnote{2}, ${J1}$\tnote{3}, ${J2}$\tnote{3}, ${J3}$\tnote{3}, ${J}$\tnote{4}, $F140W$, ${H}_{s}$\tnote{3}, ${H}$\tnote{3}, \\
Q1217, Q1549, Q2343 & $F160W$, ${H}_{l}$\tnote{3}, ${K}$\tnote{4}, $IRAC1$, $IRAC2$, $IRAC3$, $IRAC4$ \\
\hline
Q0933, Q1307, Q1422, Q1623, & ${U}_{n}$\tnote{2}, $G$\tnote{2}, ${R}_{s}$\tnote{2}, $i$\tnote{5}, $J$\tnote{6}, $H$\tnote{7}, $K$\tnote{6}, $F160W$, \\
Q1700, Q2206, Q2346 & $IRAC1$, $IRAC2$, $IRAC3$, $IRAC4$ \\
\hline
 \end{tabular}}
\begin{tablenotes}
 \item[*] While some objects in our samples miss photometric data from a subset of the bands listed here, we ensured that every object in our final, controlled samples has detections in the $K$- band and/or one of the IRAC channels.
 \item[1] Photometric bands as described in the 3D-HST catalog \citep{Skelton2014, Brammer2012}. 
 \item[2] Observed with LRIS.
 \item[3] Observed with the Four Star IR camera on the Magellan Baade Telescope.
 \item[4] Observed with Multi-Object Spectrometer For Infra-Red Exploration (MOSFIRE) at the Keck I telescope.
 \item[5] Observed with the Kitt Peak 4-m Mayall telescope.
 \item[6] Observed with the Palomar 5.08-m telescope.
 \item[7] Observed with the Wide-field InfraRed Camera (WIRCam) at the Palomar 200-in Hale telescope.
 \end{tablenotes}
\end{threeparttable}
\end{table*}

\begin{figure}
\includegraphics[width=1.0\linewidth]{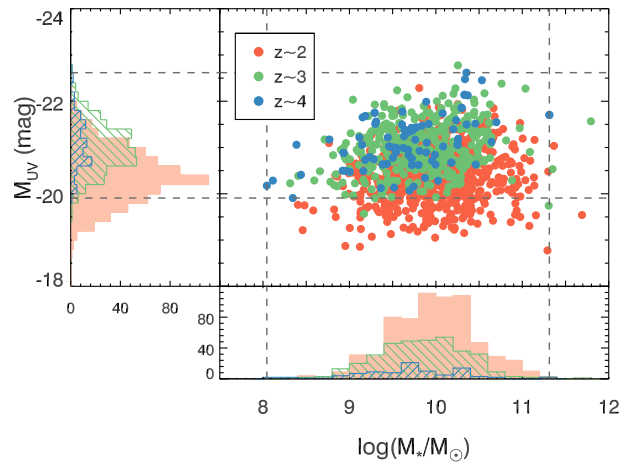}
\caption{UV absolute magnitude vs. stellar mass for the $z\sim2$ (red), $z\sim3$ (green), and $z\sim4$ (blue) samples. Histograms in M$_{*}$ and M$_{UV}$ are indicated, respectively, along the x- and y-axes. The vertical and horizontal gray dashed lines indicate the cuts in stellar mass and UV absolute magnitude, respectively, imposed onto the $z\sim2$ and $z\sim3$ samples, determined by the dynamic range of these properties in the $z\sim4$ sample.}
\label{fig:cuts}
\end{figure}

In order to make fair comparisons among LBGs at different redshifts, one important aspect of our study is the measurement of the spectral and galaxy properties in a systematic, uniform manner. In this subsection, we describe the SED fitting procedure adopted for fitting all the galaxies in our samples from $z\sim2-4$.

We collected the most up-to-date photometry for all objects in the redshift samples, and the photometric bands available for SED modeling are listed in Table \ref{tab:phot}. For objects observed in the GOODS fields (145 objects in the $z\sim2$ sample, 59 objects in the $z\sim3$ sample, and 69 objects in the $z\sim4$ sample), we acquired the publicly available photometric catalog from the 3D-HST survey \citep{Skelton2014} and matched the objects in our samples with the targets observed in the GOODS-N and GOODS-S fields. The objects were matched according to their RA, Dec, and redshift. We excluded objects from the samples if there was no unique, unambiguous match in the 3D-HST $F606W$ science image. In this way, we removed 7 objects from the $z\sim2$ sample, 1 object from the $z\sim3$ sample, and 3 objects from the $z\sim4$ sample because we were unable to identify them in the 3D-HST photometric catalog. For galaxies in the $z\sim2-3$ samples in fields other than GOODS-N and GOODS-S, we utilized the updated photometric measurements from \citet[][private communication]{Steidel2003,Steidel2004}, \citet{Reddy2012}, and \citet{Strom2017}.

Based on the photometric measurements of the objects in our samples, we aimed to derive key galaxy properties (e.g., stellar mass, dust extinction, age, and SFR) by fitting the SEDs of individual galaxies. In order to ensure the robustness of the stellar population modeling, we required at least one photometric measurement redward of the Balmer break. 12 objects in the $z\sim2$ sample and 28 objects in the $z\sim3$ sample were removed due to a lack of near- and mid-IR photometry. 

We fit the galaxy SEDs with stellar population templates from \citet {BC03} assuming a \citet{Chabrier2003} initial mass function (IMF). We note that \citet{BC03} models do not account for emission lines or nebular continuum emission. Strong emission lines (e.g., $\mbox{H}\alpha$, $[\textrm{O}~\textsc{iii}]\lambda\lambda4959,5007$) could bias the shape of the SED redward of the Balmer break, resulting in an older best-fit galaxy age than the value derived from the stellar continuum alone. In the case of our $z\sim2-4$ galaxies, $[\textrm{O}~\textsc{iii}]$ and $\mbox{H}\alpha$ fall in the $H$ and $K$ bands, respectively, for the $z\sim2$ sample, $[\textrm{O}~\textsc{iii}]$ falls in the $K$ band for the $z\sim3$ sample, and $\mbox{H}\alpha$ falls within IRAC channel 1 for galaxies with $z\geqslant3.8$ (38 out of 70 objects) in the $z\sim4$ sample. As discussed in Section \ref{sec:lya_galprop}, the contamination from $[\textrm{O}~\textsc{iii}]$ has a significant impact on the age estimate for the $z\sim3$ galaxies, but nebular emission does not appear to significantly bias results from the $z\sim2$ and $z\sim4$ samples (bottom panel of Figure \ref{fig:galprop}). Hence, we excluded $K$-band photometry from the SED fits for the $z\sim3$ sample \footnote{Excluding $K$-band photometry resulted in another 41 objects being removed from the $z\sim3$ sample, as $K$-band was the reddest photometric bands available for those objects.}, but did not reject any photometric data in the $H$ and $K$ bands (IRAC channel 1) for the $z\sim2$ ($z\sim4$) sample.

Following \citet{Reddy2017}, we adopted two combinations of metallicity and extinction curves for the SED modeling. These include 1.4 solar metallicity ($Z_{\sun}=0.014$) with the \citet{Calzetti2000} attenuation curve (hereafter ``1.4 $Z_{\sun}$+Calzetti"), and 0.28 $Z_{\sun}$ with the SMC extinction curve (hereafter ``0.28 $Z_{\sun}$+SMC"). The grid of each model includes different star-formation histories (exponentially declining, constant and rising), age ranging from 10 Myr to 5 Gyr, and $E(B-V)$ ranging from $E(B-V) = 0.00$ to $0.60$. 

Although our SED-fitting grid enables numerous combinations of stellar parameters, we applied several constraints on the best-fit stellar population model based on our current knowledge of the galaxies in our samples. For all the redshift samples, we considered the constant SFR model as a satisfactory description of the star-formation history for the typical star-forming galaxies at $z\sim2-4$ \citep{Reddy2012,Steidel2014,Strom2017}. Furthermore, we required the age of the $z\sim2$ and $z\sim3$ galaxies to be no younger than 50 Myr based on the typical dynamical timescales of these galaxies \citep{Reddy2012}. With these constraints, we calculated a set of best-fit stellar parameters for individual galaxies in our samples for both the 1.4 $Z_{\sun}$+Calzetti and the 0.28 $Z_{\sun}$+SMC models. While the former model has been traditionally used to describe the SEDs of the $L^{*}$ galaxies at $z\gtrsim2$, recent work suggests that sub-solar metallicity models with an SMC curve provide a better description of the IRX-$\beta$ relation for these objects \citep[e.g.,][]{Reddy2017}, especially at higher redshifts \citep[$z\gtrsim4$;][]{Oesch2013,Bouwens2016}. Therefore, we adopted the best-fit stellar parameters of the 0.28 $Z_{\sun}$+SMC model as the final best-fit parameters for the $z\sim4$ galaxies, given that these objects are in general younger and likely less enriched on average, compared with the samples at lower redshift. For the $z\sim2$ and $z\sim3$ objects, we found that the 1.4 $Z_{\sun}$+Calzetti model gave a systemically better fit to the observed SEDs than the 0.28 $Z_{\sun}$+SMC model above a given stellar mass threshold, which is consistent with the presence of the mass-metallicity relation \citep{Sanders2015,Steidel2014,Onodera2016}. As a result, we adopted the best-fit stellar parameters yielded by the 1.4 $Z_{\sun}$+Calzetti model above $\log(\mbox{M}_{*, 0.28 Z_{\sun}+SMC}/\mbox{M}_{\sun})=10.45$ (10.65) for the $z\sim2$ ($z\sim3$) sample, and used those from the 0.28 $Z_{\sun}$+SMC model for galaxies below the corresponding stellar mass thresholds.

\subsubsection{Controlled Samples}
\label{sec:control}

In order to conduct a well-controlled comparison, we need to ensure that we are comparing galaxies at different redshifts with similar galaxy properties. Two key properties that can be used to constrain the galaxy populations are luminosity and stellar mass. To determine the UV absolute magnitude for individual objects in our samples, we selected the corresponding photometric bands that cover the rest-frame wavelength $\lambda_{rest}=1500 \mbox{\AA}$. $\lambda_{rest}=1500 \mbox{\AA}$ falls in the $U$-, $G$-, $V$- bands and $F435W$ for the $z\sim2$ sample depending on the exact galaxy redshift, in the $R_{s}$ band for the $z\sim3$ sample, and in $F775W$ for the $z\sim4$ sample. The observed magnitude in the corresponding band was then converted into UV absolute magnitude by accounting for the monochromatic luminosity distance (since AB magnitudes correspond to flux density). The UV absolute magnitude estimated from photometry agrees well with that calculated from the best-fit galaxy SED at $\lambda_{rest}=1500 \mbox{\AA}$, and using SED-based $\mbox{M}_{UV}$ instead of $\mbox{M}_{UV}$ estimated from the photometry has negligible effect on the sample selection as well as the Ly$\alpha$ trends discussed in Section \ref{sec:lya_galprop}.

The $z\sim4$ sample spans a narrower range in UV absolute magnitude and stellar mass than the $z\sim2$ and $z\sim3$ samples. Therefore, we selected a subsample of galaxies from each of the $z\sim2$ and $z\sim3$ samples that span the same range in UV absolute magnitude ($-22.62 < \mbox{M}_{UV} < -19.91$) and stellar mass ($8.04 < \log(\mbox{M}_{*}/\mbox{M}_{\sun}) <11.31$) as the full $z\sim4$ sample. In Figure \ref{fig:cuts}, we plot the $\mbox{M}_{UV}$ vs. $\mbox{M}_{*}$ diagram for the $z\sim2$, 3, and 4 samples defined according to the fixed redshift boundaries (the same sets of objects as in Figure \ref{fig:zhist}), along with the cuts in UV absolute magnitude and stellar mass we applied to construct the final, controlled samples.
In the following sections, we present results based on the analyses of the controlled samples restricted in UV absolute magnitude and stellar mass. The resulting sample sizes for the $z\sim2, 3$, and $4$ samples after being matched in UV absolute magnitude and stellar mass are 539, 309, and 91, respectively.

\begin{deluxetable*}{cccccccc}
\tablewidth{0pt}
  \tablecaption{Median Galaxy Properties in Controlled $z\sim2-4$ Samples}
  \tablehead{
    \colhead{Sample} &
    \colhead{Redshift} &
    \colhead{$\mbox{M}_{UV}$} &
    \colhead{$\log(\mbox{M}_{*}/\mbox{M}_{\sun})$} &
    \colhead{$E(B-V)$} &
    \colhead{Age} &
    \colhead{SFR} &
    \colhead{Sample Size} \\
    \colhead{} &
    \colhead{} &
    \colhead{(Magnitude)} &
    \colhead{} &
    \colhead{} &
    \colhead{(Myr)} &
    \colhead{($\mbox{M}_{\sun} \mbox{yr}^{-1}$)} &
    \colhead{} \\
    }
  \startdata
$z\sim2$ & 2.267 &-20.51 & 10.00 & 0.09 & 806 & 13 & 539\\ 
$z\sim3$ & 2.925 & -21.00 & 9.87 & 0.08 & 404 & 16 & 309\\ 
$z\sim4$ & 3.856 &-21.06 & 9.72 & 0.04 & 321 & 13 & 91\\ 
 \enddata
\label{tab:galprop}
\end{deluxetable*}

We list the median galaxy properties of the controlled $z\sim2-4$ samples in Table \ref{tab:galprop}. Although the final redshift samples are constrained within a range in UV luminosity and stellar mass, the median values of these properties and other stellar parameters differ slightly among the samples. The $z\sim2$ sample median is $\sim0.5$ magnitude fainter than in the $z\sim3$ and $z\sim4$ samples, and the median stellar masses agree to within a factor of two, with $z\sim2$ ($z\sim4$) having the highest (lowest) mass. We note that objects falling onto the faint tail of UV luminosity in the $z\sim2$ sample will not affect the evolutionary trends we explore here. Specifically, we have verified that the results presented in Sections \ref{sec:res1} and \ref{sec:res2} do not change significantly if the $z\sim2$ sample is further restricted to have a similar median UV luminosity and stellar mass to those of the $z\sim3$ and $z\sim4$ samples (i.e., by removing the high-mass, faint wedge in the $z\sim2$ sample that doesn't overlap with the $z\sim3$ and $z\sim4$ samples in Figure \ref{fig:cuts}). As shown in the top right panel of Figure \ref{fig:galprop}, the strength of $\mbox{Ly}\alpha$ barely changes with $\mbox{M}_{UV}$, indicating that having a slightly larger portion of fainter objects will not affect the collective neutral ISM/CGM properties noticeably at $z\sim2$. Similarly, given that $\mbox{Ly}\alpha$ EW has little variation with stellar mass on the higher-mass end (middle left panel of Figure \ref{fig:galprop}), we do not consider the $z\sim2$ sample having a relatively larger median stellar mass as a potential bias against the results presented in Sections \ref {sec:res1} and \ref {sec:res2}. As for other galaxy properties, the lower redshift samples on average have a higher level of dust extinction and older age. The median SFRs of the samples are comparable, with that of the $z\sim3$ sample being slightly higher than the SFRs of $z\sim2$ and $z\sim4$ samples.

\begin{figure*}
\includegraphics[width=1.0\linewidth]{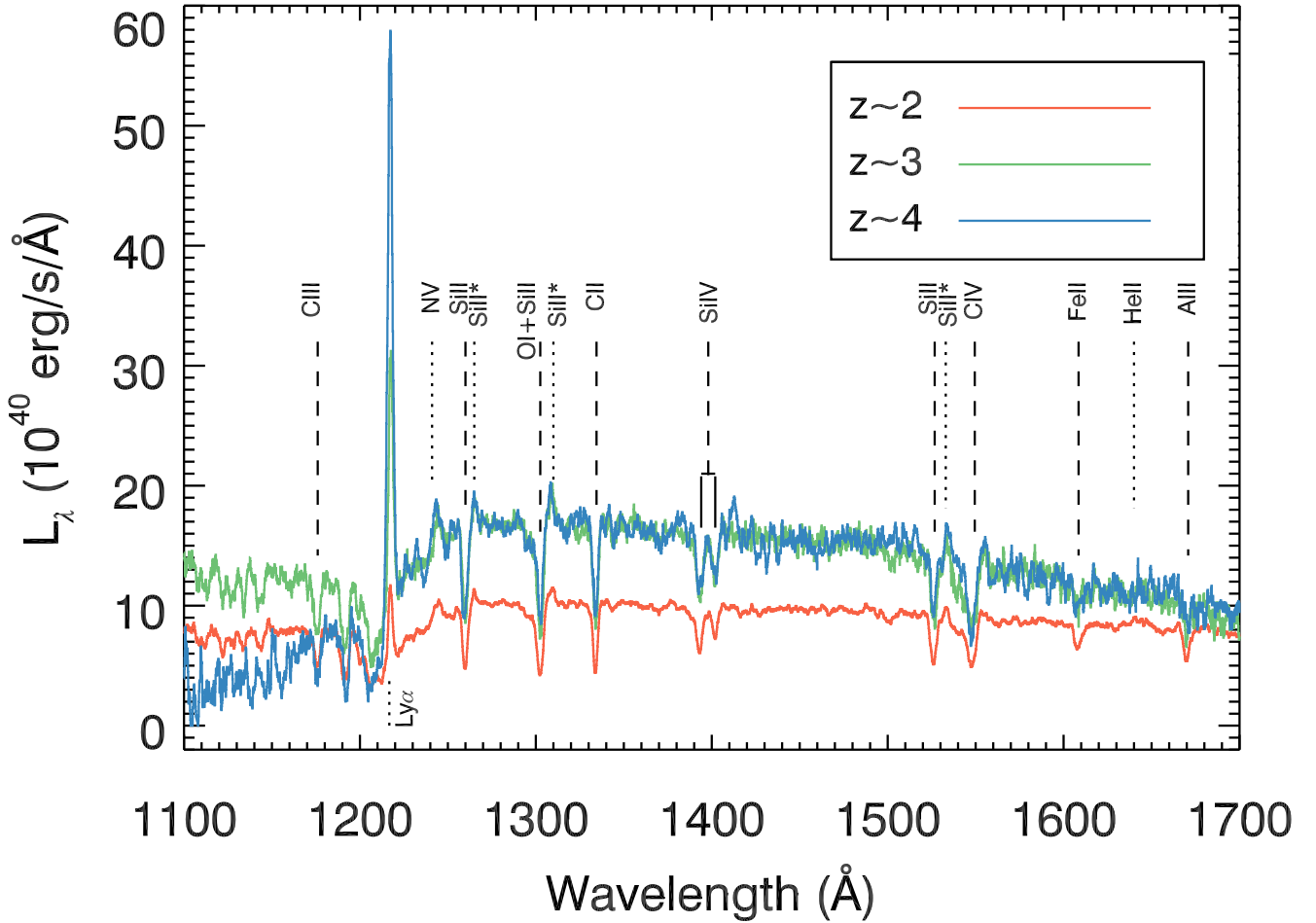}
\caption{Composite UV spectra of the LBGs in the $z\sim2$ (red), $z\sim3$ (green), and $z\sim4$ (blue) samples with UV luminosity and stellar mass constrained. Emission (dotted lines) and absorption (dashed lines) lines analyzed in this study are identified. At the resolution of the LRIS spectra ($z\sim2$ and $3$ objects) and FORS2 spectra ($z\sim4$ objects), \textrm{O}~\textsc{i}$\lambda1302$ and \textrm{Si}~\textsc{ii}$\lambda1304$ are blended, as is the \textrm{C}~\textsc{iv}$\lambda\lambda1548, 1550$ doublet.}
\label{fig:spec}
\end{figure*}

\section{Measurements}
\label{sec:meas}

The rest-UV spectral range covered by the $z\sim2-3$ LRIS spectra and the $z\sim4$ DEIMOS and FORS2 spectra includes multiple interstellar absorption lines and fine-structure emission features in addition to $\mbox{Ly}\alpha$. These include the low-ionization absorption lines \textrm{Si}~\textsc{ii}$\lambda$1260, \textrm{O}~\textsc{i}$\lambda$1302+\textrm{Si}~\textsc{ii}$\lambda$1304, \textrm{C}~\textsc{ii}$\lambda$1334, and \textrm{Si}~\textsc{ii}$\lambda$1527, the high-ionization features \textrm{Si}~\textsc{iv}$\lambda\lambda1393,1402$, and \textrm{C}~\textsc{iv}$\lambda\lambda1548,1550$, and the \textrm{Si}~\textsc{ii}*$\lambda$1265,1309,1533 fine-structure emission lines. While $\mbox{Ly}\alpha$ and the strong absorption features are detected individually in many but not all cases, composite spectra enable us to study the universal correlations between the spectral and galaxy properties by utilizing all objects in the samples. Moreover, composite spectra provide significantly higher continuum $S/N$ than individual detections, which potentially allows the measurement of weak emission and absorption lines that are not typically detected on an individual basis. In light of these advantages, we investigate the correlations among spectral and galaxy properties by studying the spectral features in the composite spectra at different redshifts. In this section, we describe our methods of creating composite spectra, and measuring the rest-frame EW of both the $\mbox{Ly}\alpha$ feature and interstellar absorption lines.

\subsection{Composite Spectra}
\label{sec:comp}

To examine how $\mbox{Ly}\alpha$ correlates with other spectral lines and galaxy properties, we divided each redshift sample into 4 bins in $\mbox{Ly}\alpha$ EW, UV luminosity, stellar mass, $E(B-V)$, age, and SFR, with each bin containing nearly the same number of galaxies. Given the fact that galaxies in the same bin have different rest-frame spectral coverage, resulting from the difference in their redshifts and observed-frame coverage, we further required that, in every composite spectrum, the same set of objects contribute to all the wavelengths. For the measurements of $\mbox{Ly}\alpha$, \textrm{Si}~\textsc{ii}$\lambda$1260, \textrm{Si}~\textsc{ii}*$\lambda$1265, \textrm{O}~\textsc{i}$\lambda$1302+\textrm{Si}~\textsc{ii}$\lambda$1304, and \textrm{C}~\textsc{ii}$\lambda$1334 in the composite spectra, we required individual contributing objects to have spectral coverage from $\mbox{Ly}\alpha$ to at least $1340\mbox{\AA}$. Similarly, we set the minimum reddest wavelength of individual contributing spectra to be $1410\mbox{\AA}$ for measuring \textrm{Si}~\textsc{iv}$\lambda\lambda$1393, 1402, and $1560\mbox{\AA}$ for measuring \textrm{Si}~\textsc{ii}$\lambda$1527, \textrm{Si}~\textsc{ii}*$\lambda$1533 and \textrm{C}~\textsc{iv}$\lambda\lambda$1448, 1550 from the composites. In general, the extra requirement on spectral coverage has an insignificant impact on the sample size for the measurement near $\mbox{Ly}\alpha$ (up to \textrm{C}~\textsc{ii}$\lambda$1334; $0\%$, $6\%$, and $8\%$ reduction for the $z\sim2$, 3, and 4 samples, respectively). Objects included in the composites made for the measurement of \textrm{Si}~\textsc{iv} are of $99\%$, $60\%$, and $96\%$ of the full $z\sim2$, 3, and 4 samples, respectively, and $97\%$, $53\%$, and $81\%$ for the spectral line measurements near \textrm{C}~\textsc{iv}.

We constructed the composite spectra according to the following steps. First, we converted the individual spectra from $F_{\nu}$ space to $L_{\lambda}$ space, in order to capture the contribution of their $intrinsic$ luminosity to the overall composite. The 600-line spectra were smoothed to the resolution of the 400-line spectra for the $z\sim2$ and $z\sim3$ samples,\footnote{Given the negligible fraction of the 300-line spectra we have in our samples, and the similarity of the spectral resolution of the 400- and the 300-line spectra, we did not further smooth the spectra down to the resolution of the 300-line spectra.} and the DEIMOS and LRIS spectra were smoothed to match the resolution of the FORS2 spectra in the $z\sim4$ sample. Next, all the individual spectra were interpolated onto a grid with $0.15\mbox{\AA}$ increments in wavelength, and scaled to have the same median value over the wavelength range $1270-1290\mbox{\AA}$, which itself represents the median of the median values estimated over  $1270-1290\mbox{\AA}$ of all individual contributing spectra. After the scaling, the median of all input spectra was estimated at each wavelength increment to create a composite spectrum. Figure \ref{fig:spec} shows the overall composite spectra of the $z\sim2$, 3, and 4 samples created from the luminosity- and stellar mass-constrained galaxies, with strong emission and absorption features marked. We describe our method of estimating individual and composite error spectra in Appendix \ref{sec:comperr}. 

We note that adopting a different stacking method (e.g., sigma-clipped mean) has a negligible effect ($\lesssim5\%$ difference) on the EW and centroid measurements of the interstellar absorption and fine-structure and nebular emission features, as the measurements were performed on the continuum-normalized spectra. As for the Ly$\alpha$ measurements, median stacking yields a systemically lower rest-frame Ly$\alpha$ EW, and a more redshifted Ly$\alpha$ centroid. However, given that our main results about Ly$\alpha$ are presented in a relative sense among the redshift samples, and that the spectra were analyzed in a consistent manner, they are robust regardless of the manner in which the composite spectra are produced. 

Although the establishment of systemic redshift was not verified for individual galaxy spectra in our samples, the rest-frame composite spectra at all three redshifts indicate a close alignment with the systemic velocity. The centroid velocities of \textrm{C}~\textsc{iii}$\lambda$1176, a stellar absorption feature that typically indicates the galaxy systemic velocity (i.e., $v=0$), are measured to be $-22\pm20$ \kms, $-24\pm27$ \kms, and $18\pm123$ \kms, respectively, at $z\sim2$, $z\sim3$, and $z\sim4$ in the overall composites controlled in UV luminosity and stellar mass (Figure \ref{fig:spec}).

\subsection{$\mbox{Ly}\alpha$ Measurement} 
\label{sec:lya}

\begin{figure}
\includegraphics[width=1.0\linewidth]{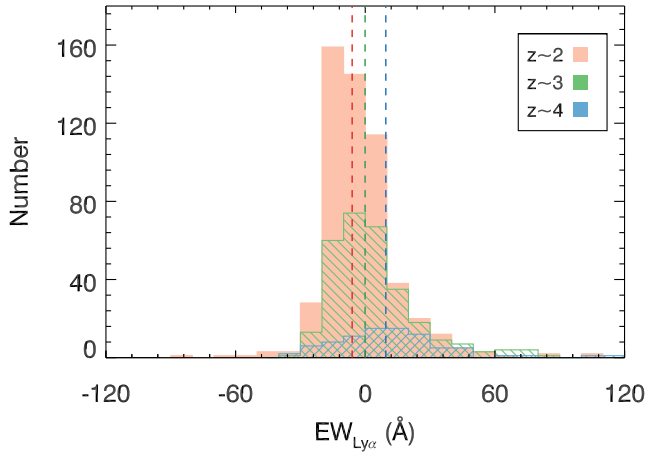}
\caption{$\mbox{Ly}\alpha$ rest-frame EW distribution for the $z\sim2$ (red), $z\sim3$ (green), and $z\sim4$ (blue) samples constrained in UV luminosity and stellar mass. The color dashed lines mark the median $\mbox{Ly}\alpha$ EW in corresponding samples: $-6.07\mbox{\AA}$, $-0.08\mbox{\AA}$, and $9.51\mbox{\AA}$ at $z\sim2$, $z\sim3$, and $z\sim4$, respectively.}
\label{fig:lya_hist}
\end{figure}

To measure the $\mbox{Ly}\alpha$ rest-frame EW in both the individual and composite spectra (the former being a property for binning), we adopted the procedure described in \citet{Kornei2010}. In brief, the morphology of $\mbox{Ly}\alpha$ in individual galaxy spectra is classified into 4 categories through visual inspection: ``emission," ``absorption," ``combination," and ``noise." The blue and red side continuum levels were measured over the wavelength range of $1225-1255\mbox{\AA}$ and $1120-1180\mbox{\AA}$, respectively. The integrated $\mbox{Ly}\alpha$ flux was calculated between the blue and red wavelength ``boundaries," $\lambda_{blue}$ and $\lambda_{red}$, which are defined as where the flux density level first meets the blue and red side continuum, respectively, on either side of the $\mbox{Ly}\alpha$ feature. However, without additional constraints imposed on $\lambda_{blue}$, the measured $\mbox{Ly}\alpha$ EWs appear to have a bi-modal distribution, with an apparent deficit near $\mbox{EW}_{Ly\alpha} \simeq 0$. A solution to this issue is fixing $\lambda_{blue}$ at $1208\mbox{\AA}$ for the ``emission" objects, and requiring $\lambda_{blue}$ to be no bluer than $1208\mbox{\AA}$ for the ``combination" objects. The $\mbox{Ly}\alpha$ rest-frame EW was then estimated by dividing the enclosed $\mbox{Ly}\alpha$ flux by the red side continuum level. 

While the method described above worked well for measuring the $\mbox{Ly}\alpha$ EW in individual galaxy spectra at $z\sim2$ and $z\sim3$, it failed to handle the $z\sim4$ spectra due to the limited spectral coverage blueward of $\mbox{Ly}\alpha$. Five out of 91 spectra do not have $\mbox{Ly}\alpha$ coverage, and an additional 19 spectra do not have a full spectral coverage of $1120-1180\mbox{\AA}$, the default window used to determine the blue side continuum in \citet{Kornei2010}. Out of these 19 spectra, 16 have their bluest wavelength greater than $1160\mbox{\AA}$, such that the spectral region available for estimating the blue-side continuum level is less than $20\mbox{\AA}$, and therefore insufficient for obtaining a robust value. In order to estimate the $\mbox{Ly}\alpha$ EW for these objects, we measured the relative level of the blue and red side continua from objects with sufficient spectral coverage on both sides within the $z\sim4$ sample. The median blue-to-red continuum ratio is 0.434, the value we adopted to obtain a rough estimate of the blue side continuum for the objects without sufficient blue-side spectral coverage. The same method as described above was then applied to calculate the $\mbox{Ly}\alpha$ EW for these 16 objects. Figure \ref{fig:lya_hist} shows the rest-frame EW distribution of $\mbox{Ly}\alpha$ in the controlled redshift samples. While spanning the same range in UV absolute magnitude and stellar mass, objects at higher redshift in general have a higher $\mbox{Ly}\alpha$ EW than the lower-redshift counterparts. The median $\mbox{Ly}\alpha$ EWs in the three redshift samples are consistent with the $\mbox{Ly}\alpha$ EWs measured from the overall composites (Figure \ref{fig:spec}), which are $-4.27\pm0.09\mbox{\AA}$, $-0.12\pm0.12\mbox{\AA}$, and $8.02\pm1.15\mbox{\AA}$ for the $z\sim2$, $z\sim3$, and $z\sim4$ samples, respectively. 

The $\mbox{Ly}\alpha$ EW of composite spectra was measured in the same manner as in the individual spectra at $z\sim2-3$, except that the composite spectra have much higher $S/N$ and accordingly never fall in the ``noise'' category. In order to obtain the uncertainty associated with the $\mbox{Ly}\alpha$ measurement in the composites, we perturbed in the composite science spectra 100 times with the corresponding composite error spectra. The $\mbox{Ly}\alpha$ profiles in the 100 fake composite spectra were measured, and we adopted the average and standard deviation of the measurements as the composite $\mbox{Ly}\alpha$ EW and the $1\sigma$ error bar, respectively.

\subsection{Absorption Line Measurement} 
\label{sec:lis}

Many interstellar absorption lines are covered within the $z\sim2-4$ rest-frame UV spectra described here. These include the low-ionization features of \textrm{Si}~\textsc{ii}$\lambda$1260, \textrm{O}~\textsc{i}$\lambda$1302+\textrm{Si}~\textsc{ii}$\lambda$1304, \textrm{C}~\textsc{ii}$\lambda$1334, and \textrm{Si}~\textsc{ii}$\lambda$1527. As for the high-ionization lines, \textrm{Si}~\textsc{iv}$\lambda\lambda1393,1402$ and \textrm{C}~\textsc{iv}$\lambda\lambda1548,1550$ are covered in a majority of the spectra. Interstellar absorption lines were only measured in composite spectra, which have a fairly low spectral resolution. Hence, we adopted single-component Gaussian fits as the simplest possible functional form to describe the interstellar absorption lines in the composites. The key best-fit parameters from such fits are the line centroid and EW.

We continuum normalized the rest-frame composite spectra using spectral regions (`windows') that are clean of spectral features defined by \citet{Rix2004}. Based on these windows, we modeled the continuum for all composite spectra with the IRAF $continuum$ routine, using a $spline3$ function of order $=16$ in order to provide a reasonable fit near $\mbox{Ly}\alpha$, where the continuum is fairly curved. In cases where the fitted continuum level did not provide a proper description of the observed spectrum due to the limited coverage of windows from \citet{Rix2004}, additional windows customized for each object were added to provide reasonable constraints on the fit.

The absorption line profile fitting was performed on the continuum-normalized composite spectra. In general, we used the IDL program MPFIT \citep{Mark2009} with the initial values 
of continuum flux level, line centroid, EW and Gaussian FWHM estimated from the program $splot$ in IRAF. The best-fit was then determined where the $\chi^{2}$ of the fit reached a minimum. We iterated the fitting over a narrower wavelength range for all the interstellar absorption lines: centroid$-4\sigma < \lambda < $centroid$+4\sigma$, where the centroid and $\sigma$ are, respectively, the returned central wavelength and standard deviation of the best-fit Gaussian profile from the initial MPFIT fit to respective lines over $\lambda_{rest}-10\mbox{\AA}$ to $\lambda_{rest}-10\mbox{\AA}$. 

Some extra care was required for the high-ionization lines, \textrm{Si}~\textsc{iv}$\lambda\lambda1393,1402$ and the blended \textrm{C}~\textsc{iv}$\lambda\lambda1548,1550$ feature. As \textrm{Si}~\textsc{iv} is a spectrally resolved doublet, each doublet member was individually fit with a single-component Gaussian profile, and the centroids of the doublet members were fixed at the rest-wavelength ratio in both the initial and iterated fits. As for \textrm{C}~\textsc{iv}, the overall absorption profile includes a P-Cygni profile originating from the stellar wind common in the most massive stars. Without proper treatment, the presence of stellar absorption may bias the measurement of interstellar \textrm{C}~\textsc{iv}. Accordingly, we removed the stellar component following the method presented in \citet{Du2016}, and the remaining interstellar absorption trough was then fit with a single-Gaussian profile for the EW measurement.\footnote{We did not try to model the individual \textrm{C}~\textsc{iv} doublet members as done in \citet{Du2016}, as in the present study we did not infer any kinematic information from the interstellar \textrm{C}~\textsc{iv} profile. The single-component Gaussian profile is sufficient in characterizing the line profile for the purpose of EW measurement.}

\section{Line Strength}
\label{sec:res1}

A key goal of our study is to examine the redshift evolution of the neutral and ionized ISM/CGM using luminosity- and mass-controlled samples. In particular, in this section we investigate the relations in line strength among multiple rest-far-UV spectral features in an evolutionary context. These include the relations between $\mbox{Ly}\alpha$ and (1) both LIS and HIS absorption features (Section \ref{sec:lya_spec}); (2) multiple galaxy properties (Section \ref{sec:lya_galprop}); and (3) nebular \textrm{C}~\textsc{iii}] emission (Section \ref{sec:ciii}). We also investigate the mutual relationships among $\mbox{Ly}\alpha$, LIS absorption lines and dust extinction (Section \ref{sec:relat}) and fine-structure emission properties at $z\sim2-4$ (Section \ref{sec:finestruct}). 

\subsection {$\mbox{Ly}\alpha$ vs. Interstellar Absorption Lines}
\label{sec:lya_spec}

\begin{figure}
\includegraphics[width=1.0\linewidth]{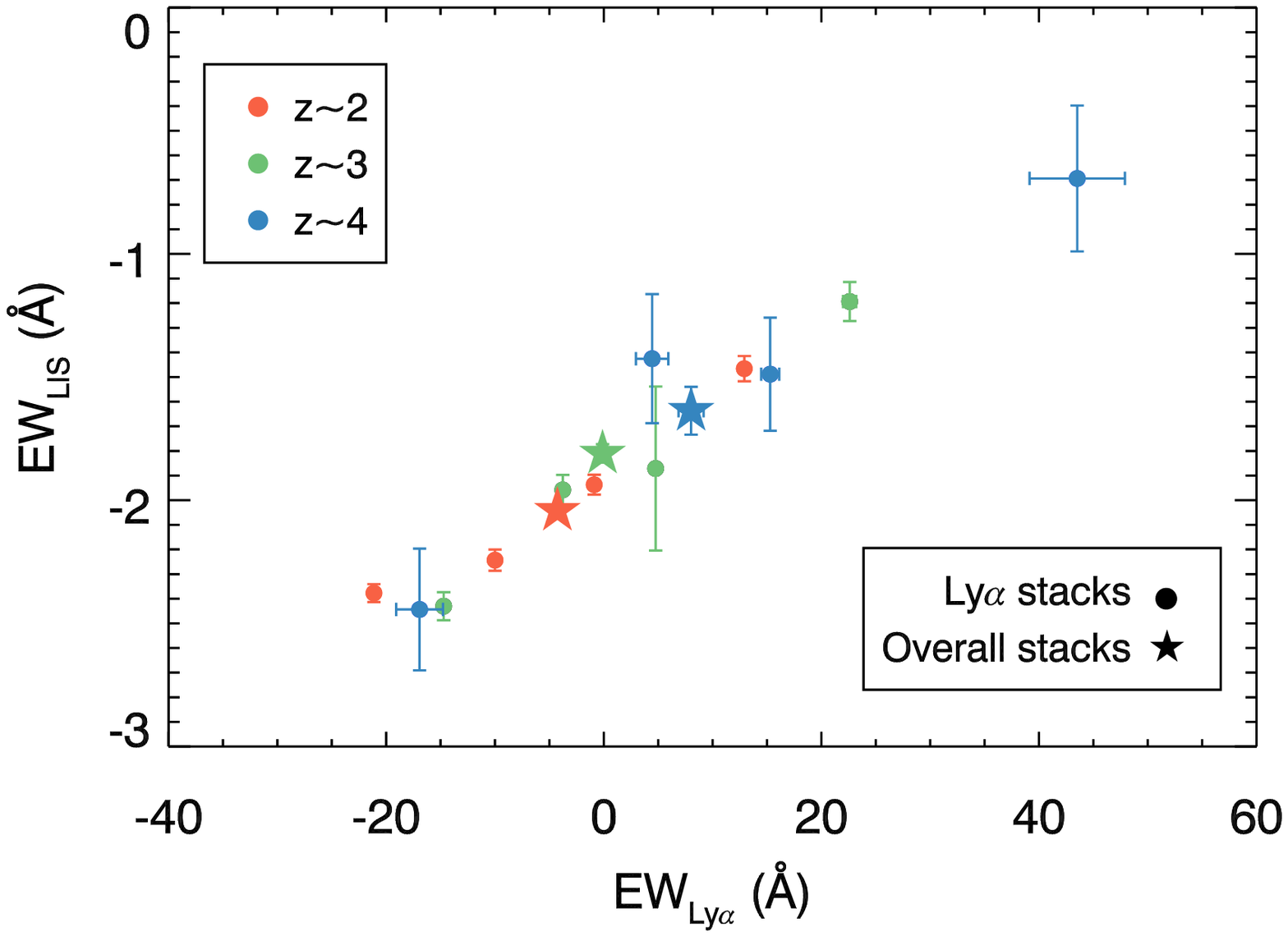}
\includegraphics[width=1.0\linewidth]{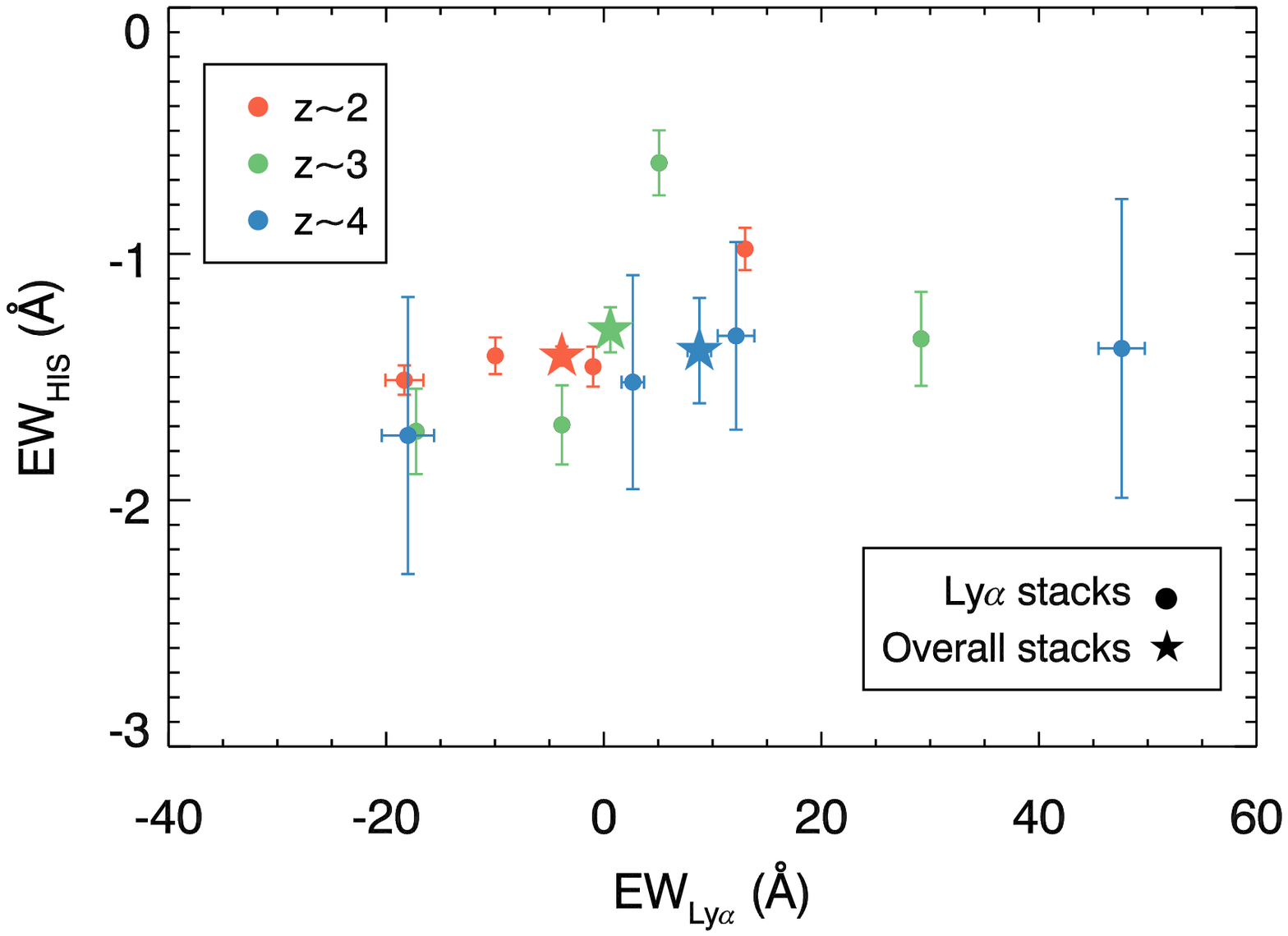}
\caption{\textbf{Top:} $\mbox{EW}_{LIS}$ vs. $\mbox{EW}_{Ly\alpha}$ in the composite spectra binned according to $\mbox{EW}_{Ly\alpha}$ (circles) and in the overall composite spectra (stars, as measured from the spectra shown in Figure \ref{fig:spec}). The plotted EW for LIS absorption features was calculated by averaging the EW of \textrm{Si}~\textsc{ii}$\lambda$1260, \textrm{O}~\textsc{i}$\lambda$1302+\textrm{Si}~\textsc{ii}$\lambda$1304, \textrm{C}~\textsc{ii}$\lambda$1334, and \textrm{Si}~\textsc{ii}$\lambda$1527. \textbf{Bottom:} $\mbox{EW}_{HIS}$ vs. $\mbox{EW}_{Ly\alpha}$ in the same set of composites. Color coding of the symbols is the same as in the top panel. The combined EW of \textrm{Si}~\textsc{iv}$\lambda\lambda1393,1402$ and \textrm{C}~\textsc{iv}$\lambda\lambda1548,1550$ is shown.}
\label{fig:lya_spec}
\end{figure}

Previous studies have shown that the strength of $\mbox{Ly}\alpha$ is tightly correlated with that of LIS absorption lines in LBGs at $z\sim3-4$, such that stronger $\mbox{Ly}\alpha$ emission corresponds to weaker LIS absorption \citep[e.g., ][]{Shapley2003,Quider2009,Vanzella2009,Jones2012}. In contrast, no significant correlation has been found between $\mbox{Ly}\alpha$ and HIS absorption lines \citep[e.g., ][]{Shapley2003}. These observations can be explained by a physical model in which clumpy neutral clouds are embedded in a more smoothly distributed halo of ionized gas \citep{Shapley2003}. Both the EWs of LIS absorption features, which are typically saturated in the galaxy spectra in our samples, and $\mbox{Ly}\alpha$ emission are modulated by the covering fraction of neutral gas: the higher covering fraction of the neutral ISM/CGM, the higher the fraction of $\mbox{Ly}\alpha$ photons that are resonantly scattered out of the line of sight, resulting in a weaker $\mbox{Ly}\alpha$ emission profile \citep[e.g.,][]{Reddy2016}. At the same time, an increasing covering fraction of neutral gas leads to stronger LIS absorption lines as these neutral and singly-ionized metal species primarily arise in neutral gas. The lack of direct link between HIS absorption and $\mbox{Ly}\alpha$ emission strengths is consistent with a picture in which the HIS absorption primarily arises in different gas from the phase responsible for the radiative transfer of $\mbox{Ly}\alpha$. Furthermore, recent work suggests that while the LIS absorption occurs spatially closer to the galaxy, HIS gas is seen with high covering fraction out past the virial radius (Rudie in prep.). Thus, the different spatial distribution of the LIS and HIS absorbers is likely to be another factor responsible for the different behaviors of LIS and HIS with $\mbox{Ly}\alpha$ in EW.

Motivated by this physical picture, we investigated whether the correlations between $\mbox{Ly}\alpha$ and the absorption features from different ionization states evolve with time from $z\sim4$ down to $z\sim2$, using carefully controlled comparison samples and uniformly measured spectra. Figure \ref{fig:lya_spec} shows the relation between rest-frame $\mbox{Ly}\alpha$ EW ($\mbox{EW}_{Ly\alpha}$) and the rest-frame LIS (top) and HIS (bottom) absorption line EWs ($\mbox{EW}_{LIS}$ and $\mbox{EW}_{HIS}$, respectively) at $z\sim2-4$ from the composite spectra binned according to $\mbox{EW}_{Ly\alpha}$. We also show the stacks of the full samples at each redshift. $\mbox{EW}_{LIS}$ was calculated by averaging the EW of \textrm{Si}~\textsc{ii}$\lambda$1260, \textrm{O}~\textsc{i}$\lambda$1302+\textrm{Si}~\textsc{ii}$\lambda$1304, \textrm{C}~\textsc{ii}$\lambda$1334, and \textrm{Si}~\textsc{ii}$\lambda$1527, and $\mbox{EW}_{HIS}$ was estimated by averaging the EW of \textrm{Si}~\textsc{iv}$\lambda\lambda$1393, 1402 and interstellar \textrm{C}~\textsc{iv}$\lambda\lambda$1548, 1550. 

From the $\mbox{EW}_{LIS}$ vs. $\mbox{EW}_{Ly\alpha}$ relation, we find that the $z\sim4$ sample spans a wider dynamic range in $\mbox{EW}_{Ly\alpha}$ extending towards larger emission EW than the $z\sim2$ and $z\sim3$ samples. In addition, the $z\sim4$ galaxies on average show stronger $\mbox{Ly}\alpha$ emission and weaker LIS absorption in the overall stacks than the lower-redshift samples, as suggested by the measurements from the overall stacks. This result is in agreement with what \citet{Jones2012} found when comparing the strengths of $\mbox{Ly}\alpha$ and the LIS lines in the $z\sim4$ and $z\sim3$ LBG composites. In terms of redshift evolution, it is important to notice that there is no apparent evolutionary trend associated with the $\mbox{EW}_{LIS}$ vs. $\mbox{EW}_{Ly\alpha}$ correlation (Figure \ref{fig:lya_spec}). The absence of evolution in the trend itself suggests that $\mbox{Ly}\alpha$ and the LIS absorption features are fundamentally related, indicating that the gas giving rise to LIS absorption modulates the radiative transfer of $\mbox{Ly}\alpha$ photons. While not fixing the galaxies in UV luminosity and stellar mass would introduce a slightly larger scatter into the $\mbox{EW}_{LIS}$ vs. $\mbox{EW}_{Ly\alpha}$ relation, this trend remains qualitatively the same in terms of the absence of redshift evolution, offering additional support to the picture of this relationship being fundamental.

As for the HIS absorption lines, we notice that they appear to be decoupled from $\mbox{Ly}\alpha$, in that little variation in the HIS absorption strength is observed with increasing $\mbox{EW}_{Ly\alpha}$. This result is consistent with previous findings \citep[e.g.,][]{Shapley2003} and provides further evidence that the HIS absorption features are produced in a different phase of gas from that controlling the radiative transfer of $\mbox{Ly}\alpha$. In addition, we discover that the ``flat" trend of HIS absorption with $\mbox{EW}_{Ly\alpha}$ does not seem to evolve with redshift either. Instead, the measured $\mbox{EW}_{HIS}$ at all redshifts seems to scatter around a constant value of $\sim-1.5\mbox{\AA}$. We note that in contrast to the LIS absorption lines, \textrm{Si}~\textsc{iv} appears to be at least partially optically thin, based on the observed $\mbox{EW}_{\textrm{Si}~\textsc{iv}\lambda1393}$/$\mbox{EW}_{\textrm{Si}~\textsc{iv}\lambda1402}$ doublet ratios that are greater than unity at all redshifts we study here. 
In the optically-thin regime, the rest-frame EW of \textrm{Si}~\textsc{iv}, $\mbox{EW}_{\textrm{Si}~\textsc{iv}}$, does not depend on only the ionized gas covering fraction and velocity range, but is also determined by the column density. In the overall composite spectra, the \textrm{Si}~\textsc{iv} doublet ratio decreases from $2.13\pm0.13\mbox{\AA}$ at $z\sim2$, to $1.67\pm0.18\mbox{\AA}$ and $1.70\pm0.55\mbox{\AA}$ at $z\sim3$ and $z\sim4$, respectively. The decreasing doublet ratio suggests that a larger fraction of gas is optically thick at $z\sim3-4$ than at $z\sim2$, and that the column density of \textrm{Si}~\textsc{iv} increases with increasing redshifts. In the meantime, the total $\mbox{EW}_{\textrm{Si}~\textsc{iv}}$ and the FWHM of individual doublet members remain almost constant, suggesting that the covering fraction of the \textrm{Si}~\textsc{iv} gas may decrease at higher redshifts. We will expand the discussion of the covering fraction of the ionized gas and relate it to a physical picture for the evolving ISM/CGM in Section \ref{sec:pic}.

\subsection{$\mbox{Ly}\alpha$ vs. Galaxy Properties}
\label{sec:lya_galprop}

\begin{figure*}
\includegraphics[width=0.5\linewidth]{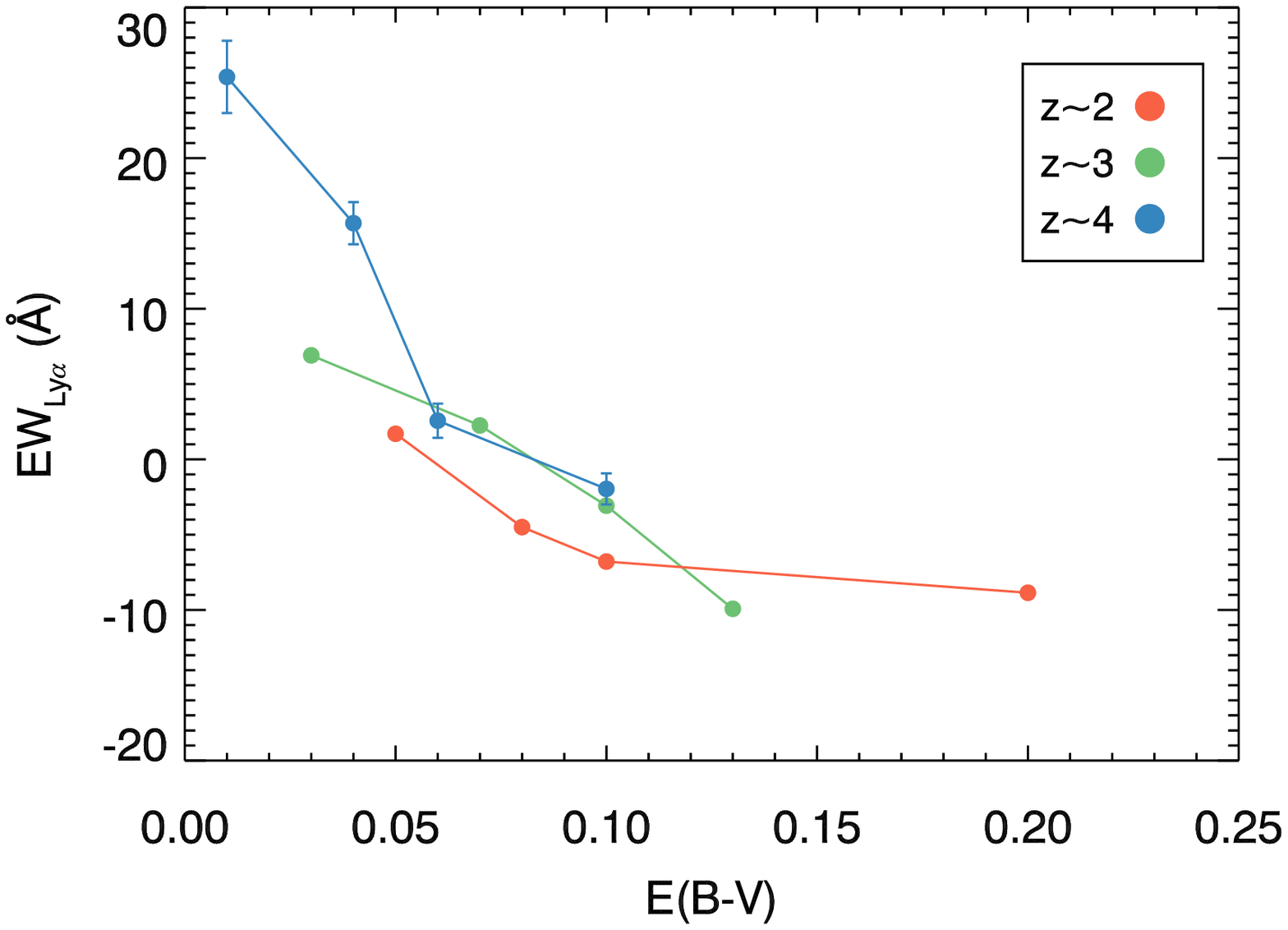}
\includegraphics[width=0.5\linewidth]{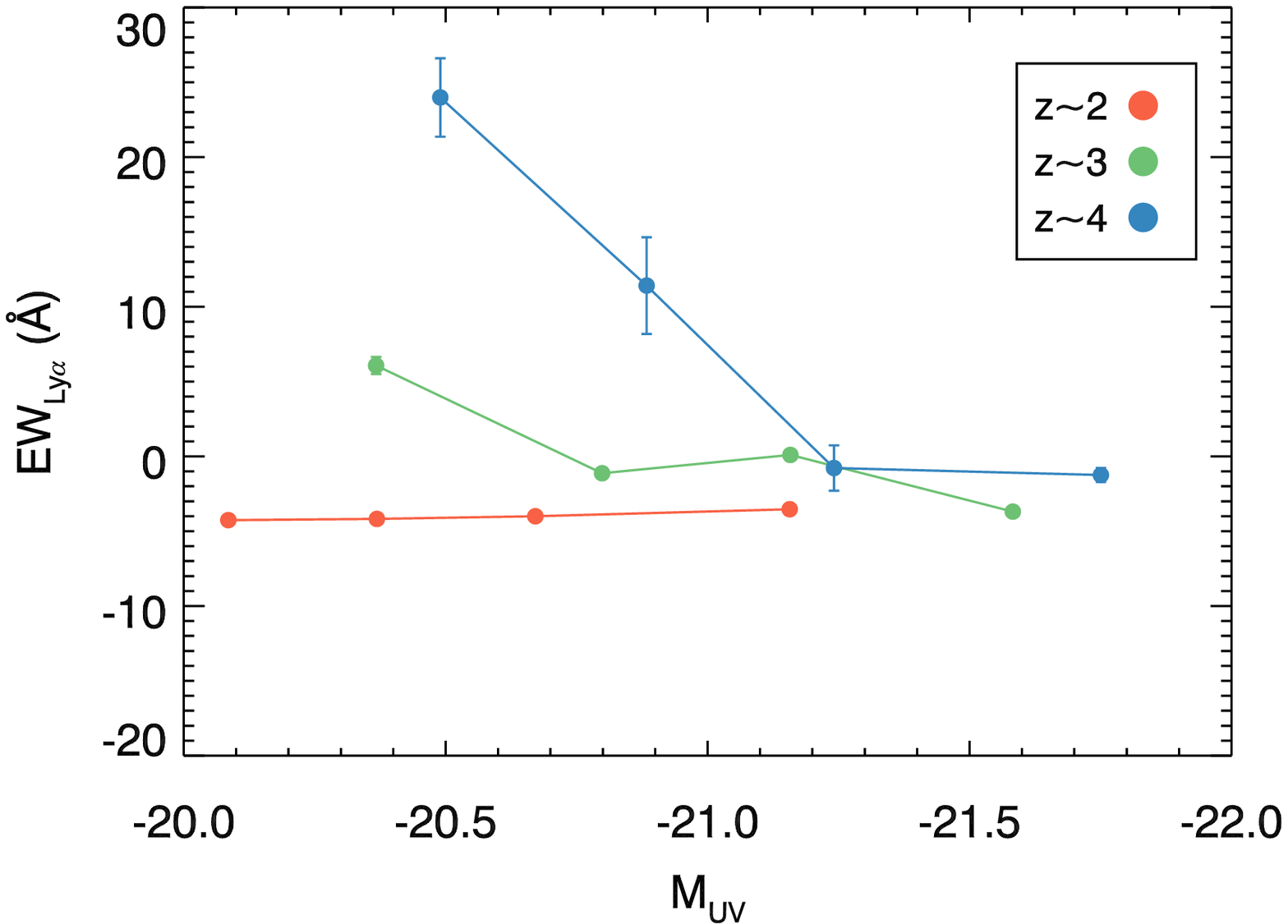}
\includegraphics[width=0.5\linewidth]{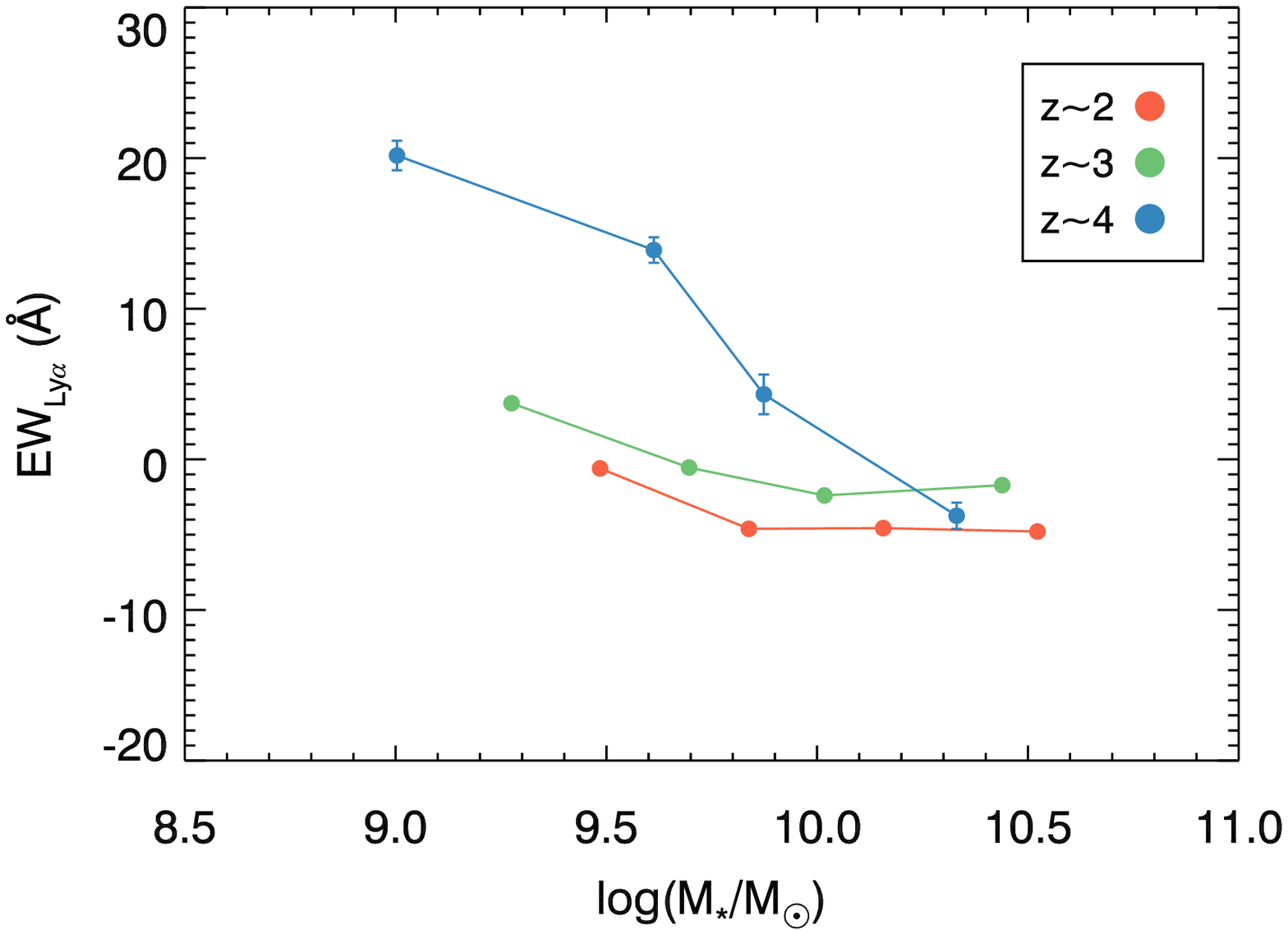}
\includegraphics[width=0.5\linewidth]{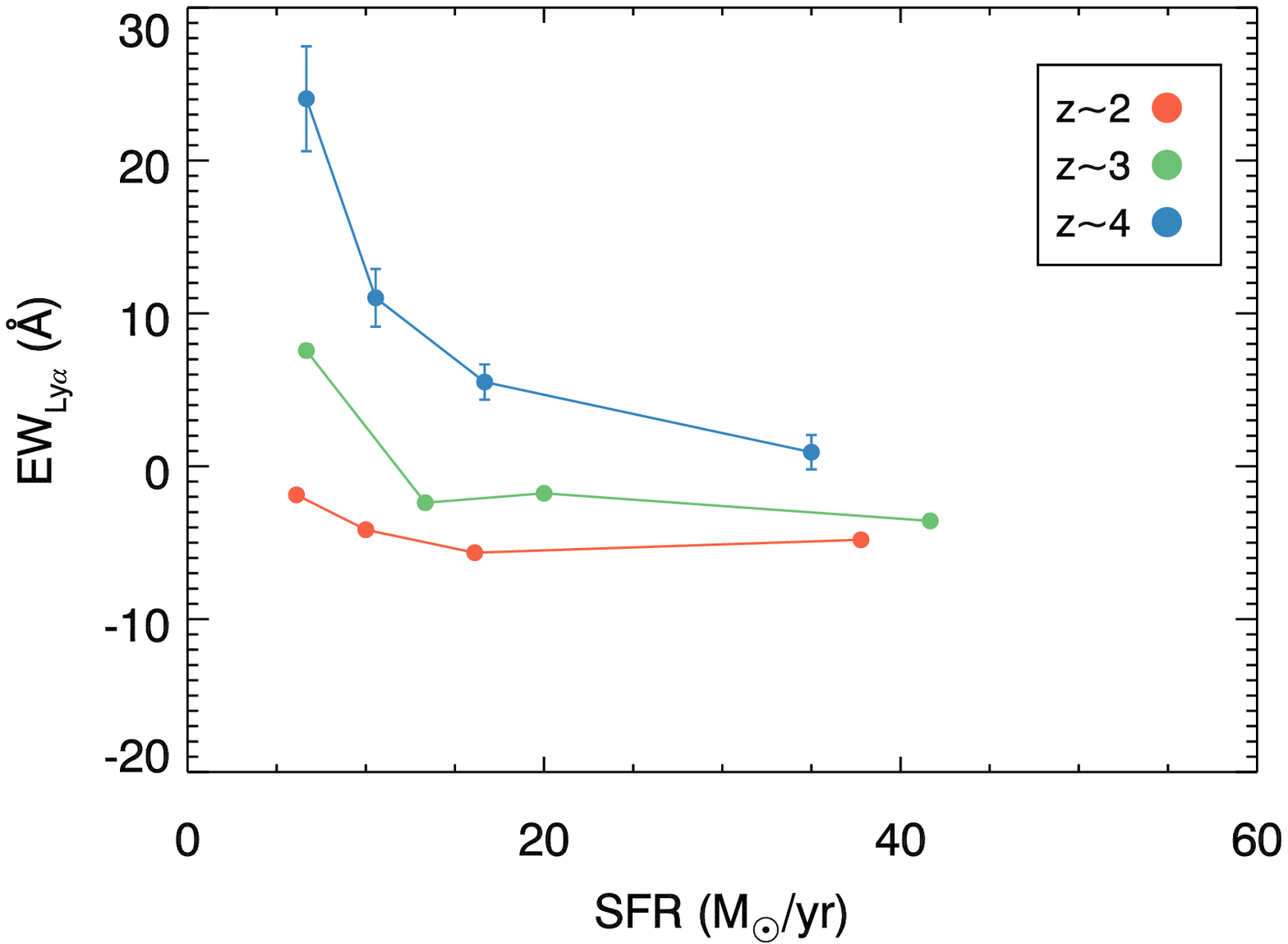}
\begin{center}
\includegraphics[width=0.5\linewidth]{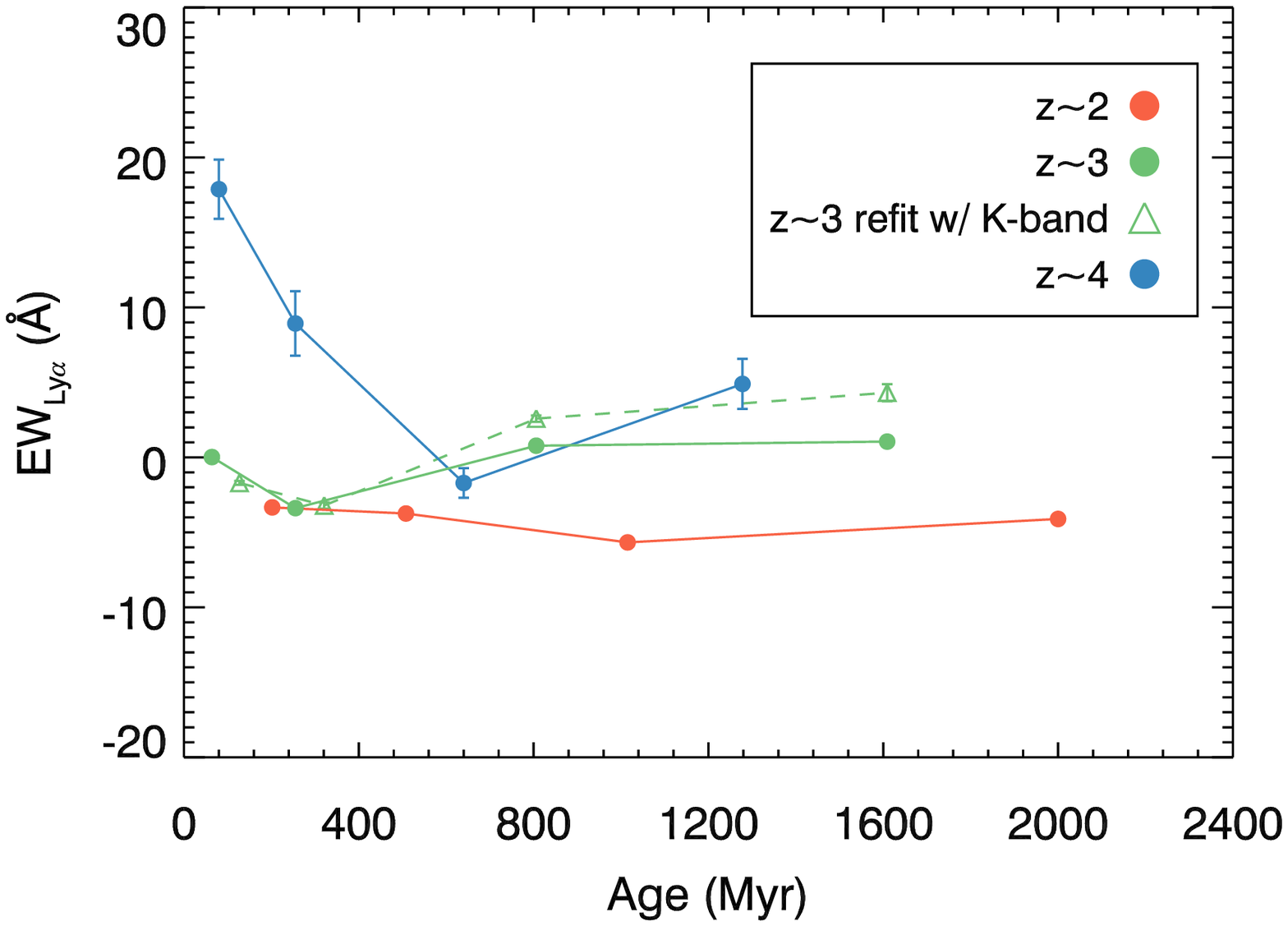}
\caption{\textbf{Top to bottom, left to right:} $\mbox{EW}_{Ly\alpha}$ vs. UV absolute magnitude, stellar mass, dust extinction, star-formation rate, and galaxy age in the composite spectra. The triangles and dashed lines in the bottom panel represent the measurements from fitting the $z\sim3$ galaxy SED with the $K$-band photometry included.} 
\label{fig:galprop}
\end{center}
\end{figure*}

Extensive studies have been carried out to examine the connection between $\mbox{Ly}\alpha$ emission and galaxy properties, in an effort to understand the factors modulating the escape fraction of $\mbox{Ly}\alpha$ photons. Among all the relations, the one for which there is the most consensus is between $\mbox{Ly}\alpha$ and rest-frame UV color (alternatively dust reddening, i.e., $E(B-V)$), according to which galaxies with stronger $\mbox{Ly}\alpha$ emission have bluer UV continuua \citep[e.g.,][]{Shapley2003, Vanzella2009,Kornei2010, Pentericci2010,Berry2012, Jones2012, Erb2016}. In contrast, controversies remain as to how $\mbox{EW}_{Ly\alpha}$ relates to galaxy ages. While some studies claim that stronger $\mbox{Ly}\alpha$ is typically found in older galaxies \citep[e.g.,][]{Shapley2001, Kornei2010}, others found no apparent relation between $\mbox{EW}_{Ly\alpha}$ and age \citep[e.g.,][]{Pentericci2009}, or that galaxies with stronger $\mbox{Ly}\alpha$ emission are much younger on average \citep[e.g.,][]{Pentericci2007}. In terms of the trends with UV luminosity, stellar mass, and SFR, studies have shown that stronger $\mbox{Ly}\alpha$ emission is typically associated with fainter \citep[e.g.,][]{Vanzella2009,Stark2010,Jones2012}, lower-mass galaxies \citep[e.g.,][]{Erb2006a,Pentericci2007,Pentericci2009,Jones2012} with lower SFRs \citep[e.g.,][]{Shapley2003,Kornei2010,Hathi2016}. 

In most cases, the relations described above have been investigated at individual redshifts. Now we examine whether these relations ($\mbox{EW}_{Ly\alpha}$ vs UV luminosity, stellar mass, dust extinction, SFR, and galaxy age) exhibit any evolutionary trends between $z\sim2-4$. We plot the relations between $\mbox{EW}_{Ly\alpha}$ and $E(B-V)$, $\mbox{M}_{UV}$, $\mbox{M}_{*}$, SFR, and age in Figure \ref{fig:galprop}. The composite spectra were created by binning individual spectra according to respective galaxy properties, and the median galaxy property in each bin is plotted. The choice of SED models is not the dominant factor for the relations between $\mbox{EW}_{Ly\alpha}$ and galaxy properties at $z\sim2-4$ (i.e., the results will remain qualitatively the same with properties derived from the 1.4 $Z_{\sun}$+Calzetti model for all redshift samples) except for the $\mbox{EW}_{Ly\alpha}$ vs. age relation at $z\sim3$, as discussed below.

Among all the $\mbox{Ly}\alpha$ relations with galaxy properties, the most striking one is between $\mbox{Ly}\alpha$ and $E(B-V)$. In addition to the previously observed negative correlation between $\mbox{EW}_{Ly\alpha}$ and dust extinction at individual redshifts, we find that the $\mbox{EW}_{Ly\alpha}$ vs. $E(B-V)$ relation lacks significant evolution from $z\sim4$ to $z\sim2$. This result implies that dust affects the $\mbox{Ly}\alpha$ photons in the same manner in galaxies at all redshifts studied here, and that the connection between Ly$\alpha$ emission and dust extinction is more direct than those observed between Ly$\alpha$ emission and the other galaxy properties described in this subsection. \citet{Shapley2003} proposed that the patchy neutral gas lying within the more extended ionized gas halo could also be dusty, and the $\mbox{Ly}\alpha$ photons would be absorbed if they encounter the dust grains. We revisit this physical picture in Section \ref{sec:pic_abs}.

For the relation between $\mbox{Ly}\alpha$ and UV absolute magnitude, we note that only the $z\sim4$ galaxies display noticeably stronger $\mbox{Ly}\alpha$ in fainter galaxies. The same relation at $z\sim3$ is not as strong, and at $z\sim2$ the UV absolute magnitude has a flat correlation with $\mbox{EW}_{Ly\alpha}$. Furthermore, the strength of $\mbox{Ly}\alpha$ increases with increasing redshift at fixed UV luminosity, likely as a result of lower covering fraction of both neutral gas and dust in the galaxies at fixed UV luminosity. This interpretation is consistent with the manner in which $E(B-V)$ evolves from $z\sim2$ to $z\sim4$.

Similar evolutionary trends can also be observed from the relations of $\mbox{Ly}\alpha$ with stellar mass and SFR: in general, galaxies with lower stellar mass and lower SFR have stronger $\mbox{Ly}\alpha$ emission, but these relations become weaker as redshift decreases. Higher $\mbox{EW}_{Ly\alpha}$ is also generally found in higher-redshift samples at fixed stellar mass and/or SFR. We note that the correlation among $\mbox{Ly}\alpha$ and UV luminosity, stellar mass, and SFR are not independent, since galaxies with larger stellar masses are likely to have higher UV luminosities and higher SFRs.

The relation between $\mbox{Ly}\alpha$ and galaxy age has also been explored previously. Without excluding the $K$-band photometry (where both $\mbox{H}\beta$ and $[\textrm{O}~\textsc{iii}]$ fall for the $z\sim3$ LBGs) from the SED fits, \citet{Shapley2001} and \citet{Kornei2010} found that $\mbox{Ly}\alpha$ is stronger in older galaxies. These authors interpreted the results as an evolution in time towards higher $\mbox{EW}_{Ly\alpha}$ as the neutral gas and dust are blown out by supernovae and massive stellar winds, resulting in a reduced covering fraction of both at older galaxy ages. We revisited this correlation for our three redshift samples, and found that, after excluding the photometric bands potentially contaminated by strong emission lines ($\mbox{H}\beta$, $[\textrm{O}~\textsc{iii}]$, and $\mbox{H}\alpha$), the $z\sim2$ and $z\sim3$ galaxies do not exhibit a significant positive trend between $\mbox{EW}_{Ly\alpha}$ and galaxy age. As for the $z\sim4$ galaxies, $\mbox{Ly}\alpha$ emission is more prominent in younger galaxies than in the older ones, an opposite trend from what was found by \citet{Shapley2001} and \citet{Kornei2010}. 

We further explored the connection between $\mbox{EW}_{Ly\alpha}$ and age at $z\sim3$. The positive correlation of $\mbox{Ly}\alpha$ and age can be reproduced if we do not discard the contaminated $K$-band from SED fitting at $z\sim3$ (triangles in the bottom panel of Figure \ref{fig:galprop}). It is worth noting, though, that when using the 1.4 $Z_{\sun}$+Calzetti model, which was adopted both in \citet{Shapley2001} and \citet{Kornei2010}, the derived galaxy age $z\sim3$ does reveal a positive correlation with $\mbox{EW}_{Ly\alpha}$. The exclusion of the K-band, while making the relation flatter, cannot completely remove the positive trend. On the other hand, the $\mbox{EW}_{Ly\alpha}$ vs. age relation at $z\sim2$ and $z\sim4$ is not sensitive to the choice of SED models. Although it is unclear why the adoption of the 1.4 $Z_{\sun}$+Calzetti model only affects the $\mbox{EW}_{Ly\alpha}$ relation with age specifically at $z\sim3$, galaxy age is undoubtedly the least well-determined property among those we explore here. Therefore, we conclude with caution that, with the most reasonable assumptions of our $z\sim3$ sample (0.28 $Z_{\sun}$+SMC for the majority and 1.4 $Z_{\sun}$+Calzetti for very high-mass galaxies), there is no evidence for a positive correlation between $\mbox{EW}_{Ly\alpha}$ and age at $z\sim3$. The previously reported positive $\mbox{EW}_{Ly\alpha}$ trend with age at $z\sim3$ is best explained as a combination of the choice of SED models and the contamination in the $K$ band from strong nebular $[\textrm{O}~\textsc{iii}]$ emission.

As described in Section \ref{sec:sed}, excluding the $H$- and $K$- bands for the $z\sim2$ sample and IRAC channel 1 for the $z\sim4$ sample has little impact on the age estimates. However, there are plausible factors to explain the fact that strong emission lines affect the SED fits at $z\sim3$. For the $z\sim4$ sample, only objects with $z\geqslant3.8$ (38 out of 70 objects) have SEDs that are potentially affected by emission-line contamination (i.e., $\mbox{H}\alpha$) in IRAC channel 1 while the remainder will not suffer from significant emission-line contribution to the SED. As for the $z\sim2$ sample, the best-fit SED models are less sensitive to the emission lines because (1) 75 objects ($14\%$ of the sample) already had the $H$- and $K$- bands corrected for strong nebular emission, and (2) a larger fraction of objects in the $z\sim2$ sample (474 out of 539 galaxies) than in the $z\sim3$ sample (273 out of 349 galaxies) have IRAC coverage to constrain the overall shape of the SEDs. Therefore, the relative importance of the $H$ and $K$ bands in determining the best-fit stellar population parameters is lower in the $z\sim2$ sample than at $z\sim3$. 

In summary, we do not observe an apparent correlation between $\mbox{EW}_{Ly\alpha}$ and galaxy age within the $z\sim2$ and $z\sim3$ samples, and there is no strong evidence for a physical picture in which gas and dust are systematically cleared away in galaxies as a function of time \citep[e.g., as described in][]{Kornei2010}. It is suggestive, though, that at $z\sim4$ $\mbox{Ly}\alpha$ is stronger in younger galaxies. At fixed galaxy age, the $z\sim4$ galaxies have stronger $\mbox{Ly}\alpha$ emission than galaxies at $z\sim2-3$. Furthermore, there appears to be a tendency towards smaller $\mbox{EW}_{Ly\alpha}$ (and thus increased dust content and \textrm{H}~\textsc{i} covering fraction) at older ages within the $z\sim4$ sample. In contrast to the physical picture proposed by \cite{Shapley2001} and \cite{Kornei2010}, the trend we observe at $z\sim4$ suggests that galaxies at the youngest ages in our sample ($< 200$ Myr) have a less substantial neutral ISM/CGM and are possibly experiencing the first generations of star formation. At the same time, comparably young $z\sim2-3$ galaxies may be more chemically enriched than at $z\sim4$, and the relatively higher neutral gas and dust covering fractions have reduced the observed $\mbox{EW}_{Ly\alpha}$.

In this section, we found that galaxies with stronger $\mbox{Ly}\alpha$ emission tend to be fainter, younger, lower-mass galaxies with lower SFRs. The correlations between $\mbox{Ly}\alpha$ and these galaxy properties are the strongest at $z\sim4$, and become much weaker at $z\sim2-3$. The evolutionary change of $\mbox{Ly}\alpha$ dependence on galaxy properties has also been reported in previous studies \citep[e.g.,][]{Pentericci2010}. However, before making any physical interpretation of the observed evolution, one needs to consider the larger spectroscopic incompleteness of the $z\sim4$ sample relative to that of the $z\sim2$ and $z\sim3$ samples especially at the faint, low-mass end (i.e., the $z\sim4$ galaxies have to show a stronger $\mbox{Ly}\alpha$ in order to be detected and included in the sample at faint luminosities). 

To investigate how the sample incompleteness impacts our results, we conducted two tests. First, we constructed composites from a ``complete'' subsample at each redshift. We selected galaxies with $z^{'}\leq24.75$ at $z\sim4$, where the spectroscopic completeness is nearly $100\%$, to be in the ``complete'' subsample. Galaxies in this subsample have UV absolute magnitudes extending down to M$_{UV}=-20.78$. We then constructed subsamples at $z\sim2$ and $z\sim3$ with galaxies brighter than M$_{UV}\leq-20.78$, which are also spectroscopically complete \citep{Shapley2003,Steidel2003}. The progressive evolution of $\mbox{EW}_{Ly\alpha}$ being stronger at higher redshifts still exists in the ``complete'' composites, with $\mbox{EW}_{Ly\alpha}=-4.0\mbox{\AA}$, $-2.2\mbox{\AA}$, and $6.5\mbox{\AA}$ at $z\sim2$, $z\sim3$, and $z\sim4$, respectively. In our second test we limited the $z\sim2$ and $z\sim3$ samples in each galaxy property bin such that they had the same fraction of galaxies with only $\mbox{Ly}\alpha$ redshift as the $z\sim4$ sample, and examined how the trends in Figure \ref{fig:galprop} would be affected. We note that intentionally selecting objects with ``$z_{Ly\alpha}$ only'' in the lower-redshift samples would bias the samples to objects with stronger $\mbox{Ly}\alpha$. As expected, the resulting $\mbox{EW}_{Ly\alpha}$ in the censored $z\sim2$ and $z\sim3$ composites is systemically larger, but the trends at $z\sim4$ are still steeper than those at lower redshifts across the same dynamic range in galaxy properties. Additionally, we measure larger $\mbox{EW}_{Ly\alpha}$ at $z\sim4$ in the overall composite than in the censored $z\sim2$ and $z\sim3$ stacks. Therefore, we conclude that while the difference in sample incompleteness needs to be seriously taken into account when studying galaxies at different redshifts, our key results are robust regarding the redshift evolution of $\mbox{Ly}\alpha$ emission and LIS absorption strengths, and $E(B-V)$.

The steeper $\mbox{Ly}\alpha$ trends with galaxy properties at $z\sim4$ can be attributed to a relatively larger range in \textrm{H}~\textsc{i} and dust covering fractions, which results in a larger dynamic range in $\mbox{EW}_{Ly\alpha}$ (see Figure \ref{fig:lya_spec}) especially towards the stronger-$\mbox{Ly}\alpha$ end where galaxies have very little dust. In contrast, the $z\sim2-3$ galaxies have higher $E(B-V)$ than the $z\sim4$ galaxies even in the faintest, youngest, lowest-mass, and lowest-SFR bin. The higher minimum $E(B-V)$ limits the highest $\mbox{EW}_{Ly\alpha}$ we could possibly observe at $z\sim2-3$, resulting in a much smaller range in the measured $\mbox{EW}_{Ly\alpha}$ in composites binned according to galaxy properties.

\subsection{$\mbox{Ly}\alpha$, LIS absorption features, and dust}
\label{sec:relat}

\begin{figure}
\includegraphics[width=1.0\linewidth]{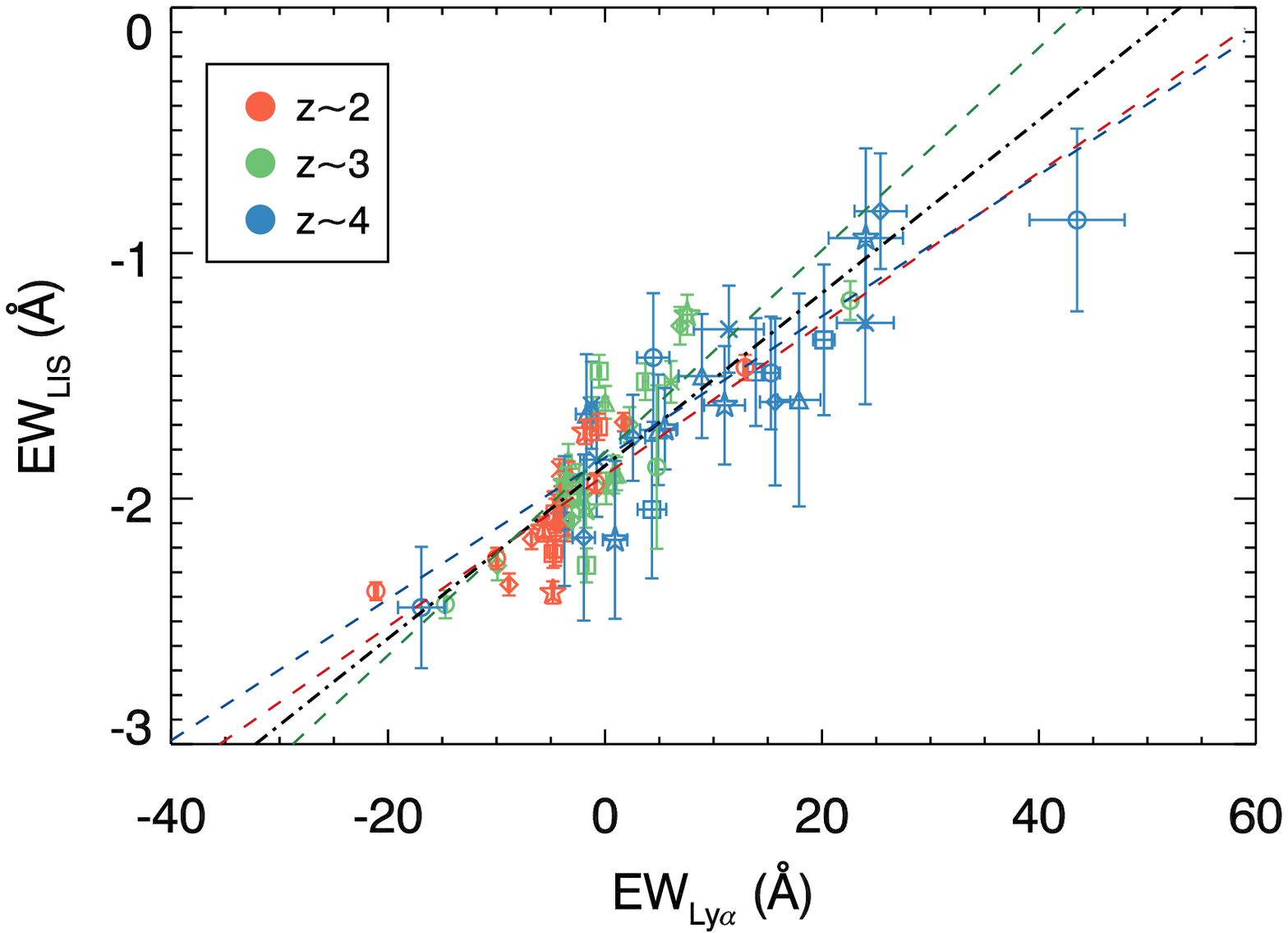}
\includegraphics[width=1.0\linewidth]{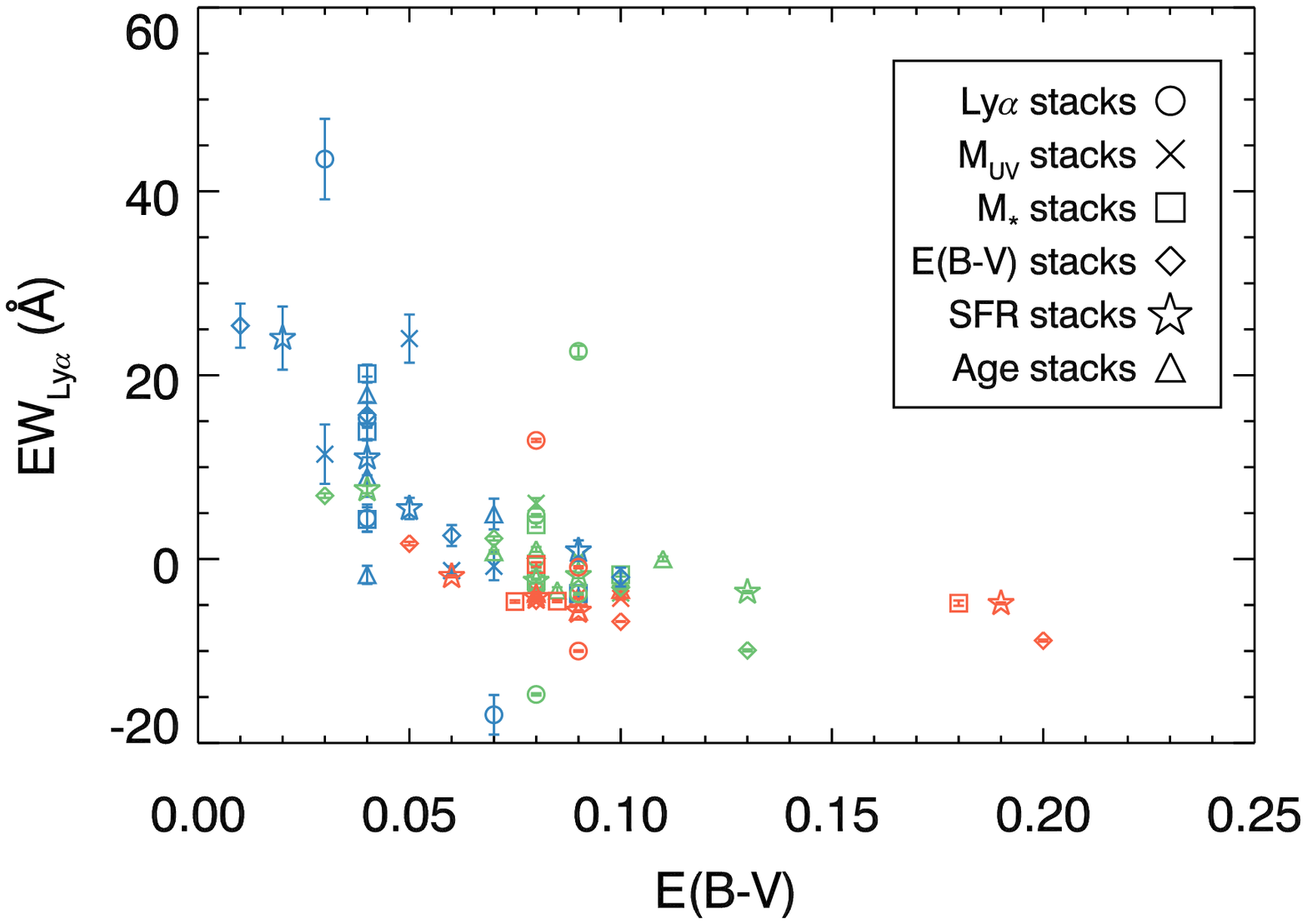}
\includegraphics[width=1.0\linewidth]{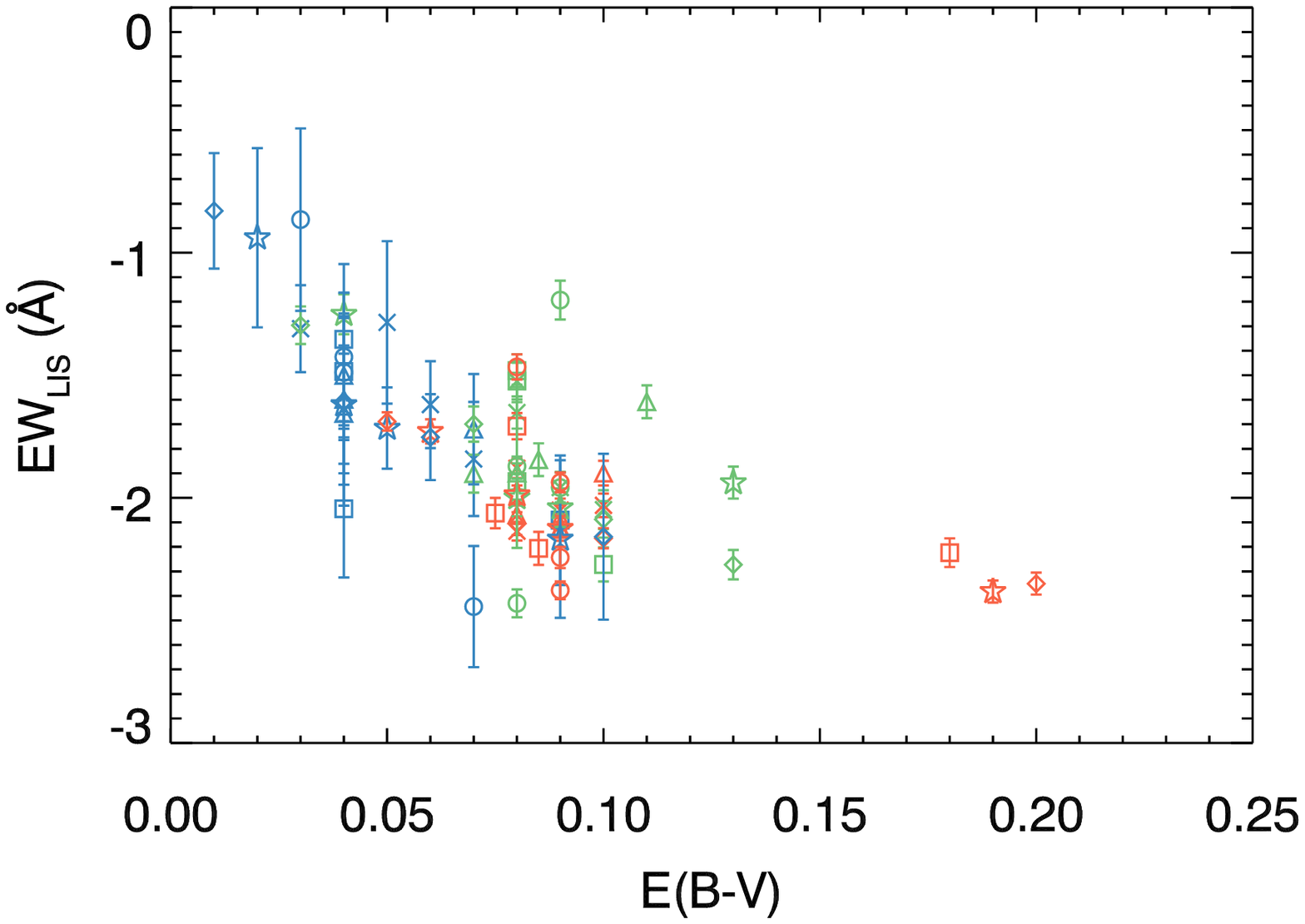}
\caption{\textbf{Top to bottom:} $\mbox{EW}_{Ly\alpha}$ vs. $\mbox{EW}_{LIS}$, $\mbox{EW}_{Ly\alpha}$ vs. $E(B-V)$, and $\mbox{EW}_{LIS}$ vs. $E(B-V)$ in composite spectra divided according to $\mbox{EW}_{Ly\alpha}$ (circles), UV absolute magnitude (crosses), stellar mass (squares), dust extinction (diamonds), star-formation rate (stars), and galaxy age (triangles). The median $E(B-V)$ in each bin is plotted. Color coding of the symbols is the same as in Figure \ref{fig:lya_spec}. In the top panel, the colored dashed lines denote the best-fit linear model for the $z\sim2$ (red), $z\sim3$ (green), and $z\sim4$ (blue) samples, and the black dash-dotted line represents the overall best-fit linear model for the $\mbox{EW}_{Ly\alpha}$ vs. $\mbox{EW}_{LIS}$ relation accounting for all three redshift samples.} 
\label{fig:lyalisdust}
\end{figure}

In Sections \ref{sec:lya_spec} and \ref{sec:lya_galprop}, we presented our result of the seemingly invariant relations between $\mbox{EW}_{Ly\alpha}$ and $\mbox{EW}_{LIS}$ (Figure \ref{fig:lya_spec}) and between $\mbox{EW}_{Ly\alpha}$ and dust extinction (Figure \ref{fig:galprop}). These non-evolving correlations indicate an interdependence among $\mbox{Ly}\alpha$ emission, LIS absorption, and dust extinction independent of redshift. If these three properties are inherently inter-related, we would expect strong correlations among any two of them regardless of galaxy properties. As a result, we utilize the composite spectra binned according to different galaxy properties (no longer restricted to the $\mbox{Ly}\alpha$ stacks as shown in Section \ref{sec:lya_spec}) and further examine the relations among $\mbox{Ly}\alpha$, the LIS lines, and the dust content. Figure \ref{fig:lyalisdust} shows the measurements of the strength of $\mbox{Ly}\alpha$ emission, LIS absorption, and $E(B-V)$ plotted against each other in composites divided according to $\mbox{M}_{UV}$, $\mbox{M}_{*}$, $E(B-V)$, SFR, age, and $\mbox{EW}_{Ly\alpha}$. While the measurements from these composite spectra are not entirely independent in the way that measurements from individual galaxies would be, the relation between the most fundamentally correlated parameters should show the least scatter in the composites binned in galaxy properties. With high $S/N$ and the inclusion of the entire samples, composite spectra provide a comprehensive view of the relative tightness and the redshift dependence, if any, of the correlations among $\mbox{EW}_{Ly\alpha}$, $\mbox{EW}_{LIS}$, and $E(B-V)$. Therefore, although we recognize that the correlation significances associated with these composite scatter plots do not hold the same meaning as for independent data points, we still use the Spearman correlation coefficient and significance as gauges of the relative tightness of these correlations. 

To quantify the $\mbox{EW}_{LIS}$ vs. $\mbox{EW}_{Ly\alpha}$ relation, we performed an inverse-squared weighted linear regression to the data points for each redshift sample. The best-fit linear models are parameterized as $\mbox{EW}_{LIS}=-1.90+0.031\times\mbox{EW}_{Ly\alpha}$, $\mbox{EW}_{LIS}=-1.81+0.041\times\mbox{EW}_{Ly\alpha}$, and $\mbox{EW}_{LIS}=-1.83+0.029\times\mbox{EW}_{Ly\alpha}$ at $z\sim2$, 3, and 4, respectively. As shown in the top panel of Figure \ref{fig:lyalisdust}, the best-fit linear model for each redshift sample predicts very similar values over the ranges of $\mbox{EW}_{Ly\alpha}$ and $\mbox{EW}_{LIS}$ probed in our study, and shows no progressive, evolutionary trend at $z\sim2-4$. The overall fit to all the $z\sim2-4$ data points yields $\mbox{EW}_{LIS}=-1.87+0.035\times\mbox{EW}_{Ly\alpha}$, with a Spearman correlation coefficient $\rho=0.87$ ($7.4\sigma$ away from the null hypothesis). 

Similar to Figure \ref{fig:galprop}, the middle panel of Figure \ref{fig:lyalisdust} shows that $\mbox{EW}_{Ly\alpha}$ decreases with increasing $E(B-V)$ in all composites binned according to galaxy properties. This correlation is less tight than the $\mbox{EW}_{Ly\alpha}$ vs. $\mbox{EW}_{LIS}$ relation, and has a Spearman correlation coefficient $\rho=-0.71$ ($5.9\sigma$ away from the null hypothesis). Given the common dependence of $\mbox{EW}_{Ly\alpha}$ on both $\mbox{EW}_{LIS}$ and dust extinction, a correlation between $\mbox{EW}_{LIS}$ and $E(B-V)$ is also expected. The tightness of the $\mbox{EW}_{LIS}$ vs. $E(B-V)$ relation is comparable to that of the $\mbox{EW}_{Ly\alpha}$ vs. $E(B-V)$ relation, with a Spearman $\rho=-0.70$ ($5.9\sigma$ away from the null hypothesis). A comparison of correlation coefficients and significances suggests that the dependence between $\mbox{Ly}\alpha$ and the LIS absorption lines is the strongest, whereas their correlation with $E(B-V)$ may be secondary. 

While the grid of discrete $E(B-V)$ values in the stellar population templates may potentially introduce larger scatter to the relations with $E(B-V)$, we argue that this discretization has a negligible impact on the resulting correlation coefficient and significance. To mimic the discrete $E(B-V)$ values and match the number of $E(B-V)$ bins, we discretized $\mbox{EW}_{Ly\alpha}$ into 21 values from $-20\mbox{\AA}$ to $40\mbox{\AA}$ with a $3\mbox{\AA}$ increment. Each $\mbox{EW}_{Ly\alpha}$ was assigned to the closest fixed value of the grid. Discretizing $\mbox{EW}_{Ly\alpha}$ results in a Spearman $\rho=0.86$ ($7.2\sigma$ away from the null hypothesis), which is almost unchanged relative to the original correlation strength. 
We therefore conclude, as in \citet{Shapley2003}, that the correlation between $\mbox{Ly}\alpha$ and the LIS lines is likely the primary one among the three. We note, however, that the larger scatter in the $\mbox{EW}_{LIS}$ and $E(B-V)$ and $\mbox{EW}_{Ly\alpha}$ vs. $E(B-V)$ relation may partially result from the uncertainty in $E(B-V)$ due to photometric errors and systematic uncertainty from the assumption of stellar population models and attenuation curves, and may not reflect the intrinsic scatter with $\mbox{EW}_{Ly\alpha}$ and $\mbox{EW}_{LIS}$. Individual measurements of $\mbox{EW}_{Ly\alpha}$, $\mbox{EW}_{LIS}$, and $E(B-V)$ from deep spectra will provide additional confirmation on which relationship is most fundamental. 

These relations, which appear to be roughly independent of redshift and other galaxy properties, have significant implications for understanding galaxy evolution. For example, the strength of both $\mbox{Ly}\alpha$ emission and LIS absorption can be predicted based on the measurement of the other, and the evolutionary trends shown in the $\mbox{EW}_{Ly\alpha}$ vs. galaxy property relations (Figure \ref{fig:galprop}) can be directly linked to the evolution of the range of LIS absorption and thus the neutral gas covering fraction at different redshifts.

\subsection{Fine-structure Emission}
\label{sec:finestruct}

\begin{figure}
\includegraphics[width=1.0\linewidth]{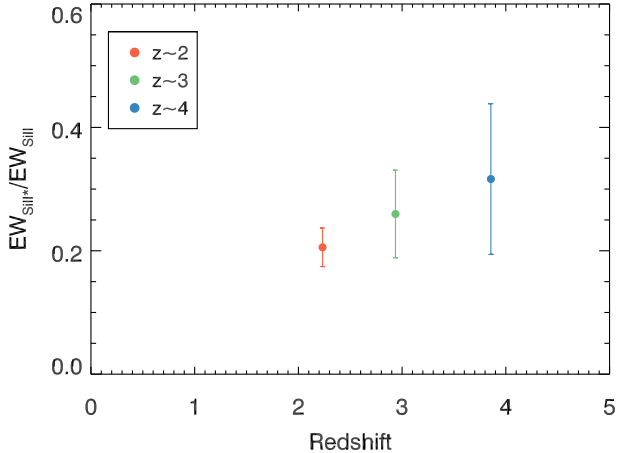}
\caption{The fine-structure-to-LIS-absorption ratio vs. redshift in the composite spectra with fixed median $E(B-V)=0.05$. The ratio is calculated by dividing the total EW of \textrm{Si}~\textsc{ii}*$\lambda$1265$+$\textrm{Si}~\textsc{ii}*$\lambda$1533 by that of \textrm{Si}~\textsc{ii}$\lambda$1260$+$\textrm{Si}~\textsc{ii}$\lambda$1527.}
\label{fig:finestruct}
\end{figure}

Aside from $\mbox{Ly}\alpha$ and LIS absorption lines, multiple fine-structure emission features, \textrm{Si}~\textsc{ii}*$\lambda$1265,1309,1533, are also detected in the LBG composite spectra at $z\sim2-4$. Fine-structure emission comes from the re-emission of photons to the excited ground state following the absorption of continuum photons, as traced by the resonant interstellar absorption lines. Accordingly, the optically thin fine-structure emission lines are considered to originate from the outflowing ISM/CGM \citep[e.g.,][]{Erb2012,Kornei2013}, and offer rich information on the spatial extent and geometry of galactic-scale outflows \citep[e.g.,][]{Prochaska2011,Finley2017}. In the absence of dust, the total strength of the fine-structure emission is expected to be equal to that of the associated interstellar absorption for optically thick gas probed by the saturated LIS lines, as the resonant photons cannot escape until they emerge as fine-structure photons. However, observations have shown that the fine-structure emission always appears to be weaker than the LIS absorption for \textrm{Si}~\textsc{ii} \citep{Shapley2003,Jones2012} and \textrm{Fe}~\textsc{ii} transitions \citep{Erb2012}. Several factors may contribute to this discrepancy: the presence of dust, the geometry of the outflows, and spectroscopic slit loss \citep{Prochaska2011,Erb2012}. By comparing the EW ratio of the \textrm{Si}~\textsc{ii}* fine-structure emission and \textrm{Si}~\textsc{ii} absorption (i.e., $R_{FS}=\mbox{EW}_{\textrm{Si}~\textsc{ii}*}$/$\mbox{EW}_{\textrm{Si}~\textsc{ii}}$) in the overall LBG composites at $z\sim3$ and $z\sim4$, \citet{Jones2012} found that this ratio is higher for the $z\sim4$ galaxies than for those at $z\sim3$. These authors interpreted such results as evidence of a smaller characteristic size for the fine-structure emitting regions in $z\sim4$ LBGs, such that a larger fraction of emission gets captured in the slit at $z\sim4$ than at $z\sim3$.

In order to investigate the effect of slit loss alone and whether the size of the fine-structure emission region evolves with redshift, we utilized the full $z\sim4$ sample and constructed composites at $z\sim2$ and $z\sim3$ with a fixed median $E(B-V)=0.05$ matched with that of the $z\sim4$ sample. We only sampled galaxies on the blue tail of the $E(B-V)$ distribution of the $z\sim2$ and $z\sim3$ samples, selecting objects from smaller to larger $E(B-V)$ until the median $E(B-V)$ is matched. This approach results in 179, 129, and 84 spectra in the $E(B-V)$-controlled subsamples at $z\sim2$, 3, and 4, respectively. We measured $R_{FS}$ using \textrm{Si}~\textsc{ii}*$\lambda$1265,1533 and \textrm{Si}~\textsc{ii}$\lambda$1260,1527. The pair of \textrm{Si}~\textsc{ii}*$\lambda$1309 and \textrm{Si}~\textsc{ii}$\lambda$1304 was excluded from the measurement because the latter is blended with \textrm{O}~\textsc{i}$\lambda$1302 in the composite spectra. We define $R_{FS}=-(\mbox{EW}_{1265}+\mbox{EW}_{1533})/(\mbox{EW}_{1260}+\mbox{EW}_{1527})$, as in \citet{Jones2012}. We measure a ratio of $0.21\pm0.03$, $0.26\pm0.07$, $0.32\pm0.12$ for the $z\sim2$, $z\sim3$, and $z\sim4$ composites, respectively. Figure \ref{fig:finestruct} shows $R_{FS}$ plotted against redshift, and does not indicate a significant evolutionary trend. Given that the composites are fixed in $E(B-V)$, our result suggests that there may not be a redshift evolution in the size of the fine-structure emission region in galaxies with similar dust extinction. In fact, if the median $E(B-V)$ is not fixed, $R_{FS}$ is measured to be $0.16\pm0.02$ and $0.18\pm0.04$ for the $z\sim2$ and $z\sim3$ samples, respectively ($R_{FS}$ at $z\sim4$ remains the same), from the overall composites. This trend is consistent with that reported by \citet{Jones2012}, that galaxies at higher redshifts on average have stronger fine-structure emission at given resonant absorption strength.
However, our results suggest that the larger $R_{FS}$ in the $z\sim4$ galaxies measured by \citet{Jones2012} reflects the lower dust content in higher-redshift galaxies on average (Table \ref{tab:galprop} and Figure \ref{fig:galprop}).

We note that while we reported $R_{FS}=0.32\pm0.12$ for the $z\sim4$ composite, \citet{Jones2012} measured a higher value, $R_{FS}=0.53\pm0.17$, using the $z\sim4$ LBG spectra from the same set of \textrm{Si}~\textsc{ii} and \textrm{Si}~\textsc{ii}* lines. Several factors may contribute to this discrepancy. First, the composites were constructed from different samples. By adding 14 objects with $z>3.4$ from the LRIS sample, removing one galaxy from the DEIMOS/FORS2 sample with $z<3.4$, and excluding 3 galaxies without an unique match in photometry, our $z\sim4$ sample includes 81 objects (91 spectra) \footnote{As we discuss below, the constraint on wavelength coverage results in 66 objects (74 spectra) contributing to the $z\sim4$ composite from which we measured $R_{FS}$. Among these individual spectra, 60 are in the DEIMOS/FORS2 sample \citet{Jones2012} used to measure $R_{FS}$, and the remaining 14 spectra are from LRIS.} while that in \citet{Jones2012} includes 70 objects (81 spectra). Second, the individual galaxy spectra were combined in different ways. In \citet{Jones2012}, all the individual spectra were normalized to $F_{\nu}=1$ over $1250-1500\mbox{\AA}$, and the mean value at each wavelength was taken with $1\sigma$ clipping. In comparison, we constructed the composites accounting for the $intrinsic$ shape of the galaxy spectra in the $L_{\lambda}$ space. We scaled all galaxy spectra in the same bin such that they have the same median $L_{\lambda}$ over $1270-1290\mbox{\AA}$, a spectral window that all individual spectra have coverage for, and performed median stacking to create the composite spectra. Additionally, we required that the same set of objects contribute to all wavelengths in a single composite. For example, to measure the features of \textrm{Si}~\textsc{ii}$\lambda$1527 and \textrm{Si}~\textsc{ii}*$\lambda$1533, we only included objects with spectral coverage at least up to $1540\mbox{\AA}$ when creating the composite. While the requirement on wavelength coverage reduced the $z\sim2$, 3, and 4 sample sizes to 173, 71, and 74, respectively, the composites represent the average of the same set of objects at all wavelengths, and the $S/N$ of the composite does not vary significantly across the spectrum due to the different numbers of objects contributing at each wavelength. 

In conclusion, we created the $z\sim4$ composite using a different method with a slightly different sample from the one in \citet{Jones2012}, and a different $R_{FS}$ value is therefore not unexpected. In fact, the $R_{FS}$ value quoted in \citet{Jones2012} can be reproduced if we use the same sample to create the composite following the same stacking approach described in \citet{Jones2012}. Since we have performed uniform stacking schemes and spectral feature measurements across the redshift samples, the results presented earlier in this subsection are free from any systematic bias and reflect the true (lack of) redshift evolution of the size of the fine-structure emitting region with fixed dust extinction. We note, however, that although the redshift rules used to determine the systemic redshift (Section \ref{sec:sample}) yield a well-established rest frame for the composite spectra, as evidenced by the lack of significant velocity offset for stellar features, the average velocity shift may not be accurate in determining the redshift of each individual galaxy. As a result, redshift errors for individual objects can potentially broaden the fine-structure emission profiles in the composites, as these features were not included in determining the redshift rules. Some fine-structure emission flux can therefore be lost during our measurements of these features due to the fixed wavelength range we used for fitting the profiles, leading to smaller measured fine-structure emission EWs than the intrinsic values. Individual spectra with high $S/N$, spectral resolution, and precise redshift measurements will provide key information on the intrinsic $R_{FS}$ of star-forming galaxies at $z\sim2-4$.

\subsection{$\mbox{Ly}\alpha$ vs. \textrm{C}~\textsc{iii}] at $z\sim2$}
\label{sec:ciii}

\begin{figure}
\includegraphics[width=1.0\linewidth]{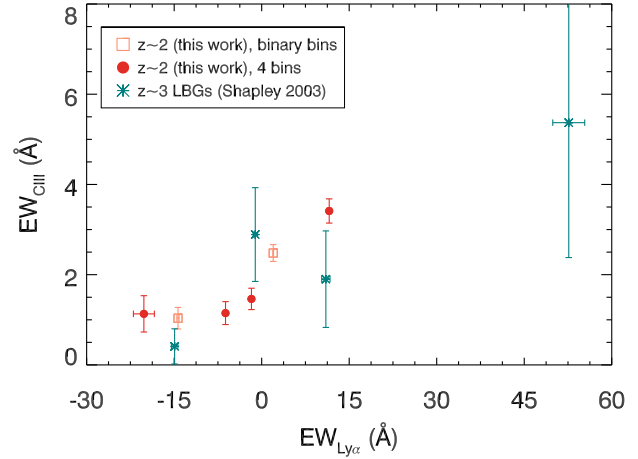}
\caption{Rest-frame EW$_{\textrm{C}~\textsc{iii}]}$ vs. EW$_{Ly\alpha}$ in the $z\sim2$ composite spectra with \textrm{C}~\textsc{iii}] coverage divided in 2 bins (light red squares) and 4 bins (dark red circles). The \textrm{C}~\textsc{iii}] emission is not significant in any of the $z\sim3$ and $z\sim4$ composites, and is therefore not shown in the figure. }
\label{fig:ciii}
\end{figure}

As typically the second strongest emission feature in the rest-frame far-UV (after $\mbox{Ly}\alpha$), the nebular \textrm{C}~\textsc{iii}]$\lambda\lambda$1907, 1909 emission doublet has received great attention as a potential redshift probe at $z>6$ \citep{Stark2014,Stark2015}, where $\mbox{Ly}\alpha$ is significantly attenuated by the neutral intergalactic medium \citep[IGM; e.g.,][]{Treu2012,Pentericci2014,Schenker2014,Konno2014}. More importantly, \textrm{C}~\textsc{iii}] is a useful probe of the physical conditions of the \textrm{H}~\textsc{ii} regions in star-forming galaxies when combined with other nebular lines from [\textrm{O}~\textsc{iii}] and \textrm{C}~\textsc{iv}. Photoionization models predict that large rest-frame \textrm{C}~\textsc{iii}] EWs, $\mbox{EW}_{\textrm{C}~\textsc{iii}]}$, result from lower gas-phase metallicity, higher ionization parameters, and harder radiation fields \citep[e.g.,][]{Erb2010,Steidel2016,Berg2016,Senchyna2017}. The common dependence of \textrm{C}~\textsc{iii}] and other emission line properties (both nebular and recombination features) on the hardness of the ionizing spectrum leads to positive predicted correlations of $\mbox{EW}_{\textrm{C}~\textsc{iii}]}$ with $\mbox{EW}_{Ly\alpha}$ \citep{Jaskot2016}, [\textrm{O}~\textsc{iii}] \citep {Stark2014,Jaskot2016,Senchyna2017,Maseda2017}, and the escape fraction of Lyman continuum (LyC) photons \citep{Vanzella2016,Jaskot2016,Debarros2016}. Previous work has suggested a connection between $\mbox{EW}_{Ly\alpha}$ and $\mbox{EW}_{\textrm{C}~\textsc{iii}]}$ \citep[e.g.,][]{Shapley2003,Stark2014,Stark2015,Rigby2015,Guaita2017}. The evidence for a positive correlation between Ly$\alpha$ and \textrm{C}~\textsc{iii}] emission EWs suggests that the observed $\mbox{EW}_{Ly\alpha}$, aside from being modulated by the covering fraction of neutral gas and dust, may also show differences in $intrinsic$ $\mbox{Ly}\alpha$ production. For example, the $z\sim4$ composite made of galaxies in the quartile with the strongest Ly$\alpha$ emission (the rightmost blue circle in the top panel of Figure \ref{fig:lyalisdust}) falls off the mean $\mbox{EW}_{LIS}$ vs. $\mbox{EW}_{Ly\alpha}$ relation, showing stronger Ly$\alpha$ emission at fixed LIS absorption strength than all other composites. This result suggests variations in the properties of \textrm{H}~\textsc{ii} regions and massive stars even among galaxies in our samples. Since \textrm{C}~\textsc{iii}] is covered in some of the individual $z\sim2-4$ LBG spectra, it is of great interest to examine how the strength of \textrm{C}~\textsc{iii}] in general relates to that of $\mbox{Ly}\alpha$, and whether such correlation evolves with time using the large datasets at hand. 

We selected a subset of objects with \textrm{C}~\textsc{iii}] coverage from each redshift sample controlled in UV luminosity and stellar mass. The selection results in 291, 37, and 34 objects in the $z\sim2$, $z\sim3$, and $z\sim4$ samples, respectively. Given the small number of objects with \textrm{C}~\textsc{iii}] coverage at $z\sim3$ and $z\sim4$, we performed binary stacks at each redshift according to $\mbox{EW}_{Ly\alpha}$, with the low and high bins containing nearly equal numbers of objects. We measured the \textrm{C}~\textsc{iii}] emission profile on the continuum-normalized spectra, in the same manner as described in Section \ref{sec:lis}.

Unfortunately, the \textrm{C}~\textsc{iii}] feature is not significantly detected in any of the $z\sim3$ and $z\sim4$ composites due to relatively low $S/N$, making it impossible to study the redshift evolution of the relation between $\mbox{EW}_{Ly\alpha}$ and $\mbox{EW}_{\textrm{C}~\textsc{iii}]}$. The $z\sim2$ composites, on the other hand, have much higher $S/N$ thanks to the significantly larger sample size, and therefore enable the first measurement of \textrm{C}~\textsc{iii}] from a statistical sample at this redshift. 

As shown in Figure \ref{fig:ciii}, the $z\sim2$ binary stacks (squares) follow a clear trend of increasing $\mbox{EW}_{\textrm{C}~\textsc{iii}]}$ at higher $\mbox{EW}_{Ly\alpha}$. In order to achieve a larger dynamic range in both \textrm{C}~\textsc{iii}] and $\mbox{EW}_{Ly\alpha}$, we further divided each $z\sim2$ $\mbox{Ly}\alpha$ bin into two (four bins in total), and investigated if any underlying trends are obscured by simple binary. From the 4-bin composites (circles in Figure \ref{fig:ciii}), we observe a fairly flat behavior between $\mbox{EW}_{\textrm{C}~\textsc{iii}]}$ and $\mbox{EW}_{Ly\alpha}$, with the exception of the strongest $\mbox{Ly}\alpha$ bin. The strongest $\mbox{Ly}\alpha$ bin has a measured $\mbox{EW}_{\textrm{C}~\textsc{iii}]}=3.41\pm0.27\mbox{\AA}$ while that in the rest of the $\mbox{Ly}\alpha$ stacks is $\sim1.25\mbox{\AA}$. This trend is consistent with what has been found in previous studies \citep[e.g.,][]{Shapley2003, Stark2014,Stark2015,Rigby2015,Jaskot2016}, suggesting $intrinsic$ difference in the $\mbox{Ly}\alpha$ photon production between galaxies showing strong and weak $\mbox{Ly}\alpha$ emission. Although we were unable to detect the \textrm{C}~\textsc{iii}] feature at $z\sim3-4$, measurements were made using the $z\sim3$ LBG composites by \citet{Shapley2003}.\footnote{The $z\sim3$ LRIS spectra presented in \citet{Shapley2003} were mainly collected before the LRIS-B upgrade, and typically have significantly redder wavelength coverage than the LRIS blue-side dichroic spectra presented here. As such the sample of $z\sim3$ spectra with \textrm{C}~\textsc{iii}] coverage is significantly larger in \citet{Shapley2003}.} These authors show a suggestive positive correlation between $\mbox{EW}_{Ly\alpha}$ and $\mbox{EW}_{\textrm{C}~\textsc{iii}]}$, which is consistent with our much cleaner trend at $z\sim2$.

The positive correlation between \textrm{C}~\textsc{iii}] and $\mbox{Ly}\alpha$ strengths suggest that environments favorable for \textrm{C}~\textsc{iii}] production are also conducive to higher $\mbox{Ly}\alpha$ production and large $\mbox{Ly}\alpha$ escape fractions. As \textrm{C}~\textsc{iii}] is a collisionally excited transition, the emission is enhanced in \textrm{H}~\textsc{ii} regions with strong radiation fields and low metallicities \citep{Jaskot2016,Senchyna2017}. In the meantime, the harder ionizing spectrum given by young, metal-poor stars leads to higher $\mbox{EW}_{Ly\alpha}$ by (1) boosting the intrinsic production rate of $\mbox{Ly}\alpha$ photons, and (2) reducing the neutral covering fraction due to the higher ionization state of the ISM \citep{Trainor2015}, allowing more $\mbox{Ly}\alpha$ photons to escape the galaxy. Therefore, the hard ionizing spectrum and the low gas-phase metallicity modulate both the \textrm{C}~\textsc{iii}] and $\mbox{Ly}\alpha$ emission in the same direction, resulting in the positive correlation we observe between these two transitions. Similarly, $\mbox{EW}_{Ly\alpha}$ has also been demonstrated to correlate with the rest-optical nebular emission properties, such as [\textrm{O}~\textsc{iii}]$\lambda5008$/$\mbox{H}\beta$ and [\textrm{N}~\textsc{ii}]$\lambda6585$/$\mbox{H}\alpha$ \citep{Erb2016,Trainor2016}, which, like \textrm{C}~\textsc{iii}], can be used to probe the physical conditions of \textrm{H}~\textsc{ii} regions (e.g., gas-phase metallicity, electron temperature and density, and ionization parameter).

\section{Kinematics}
\label{sec:res2}

The measurement of the $\mbox{Ly}\alpha$ and LIS absorption profiles in the composite spectra enables us to study not only their line strengths, but also kinematics. The line centroids measured from LIS absorption features offer at least a crude measure of the bulk velocities of outflowing ISM, and the observed $\mbox{Ly}\alpha$ centroid may further constrain the kinematics of the neutral outflows (e.g., velocity distribution). In this Section, we focus on the evolution of $\mbox{Ly}\alpha$ (Section \ref{sec:lya_kine}) and LIS (Section \ref{sec:LIS_kine}) kinematics across $z\sim2-4$.

\subsection{$\mbox{Ly}\alpha$ Kinematics}
\label{sec:lya_kine}

\begin{figure}
\includegraphics[width=1.0\linewidth]{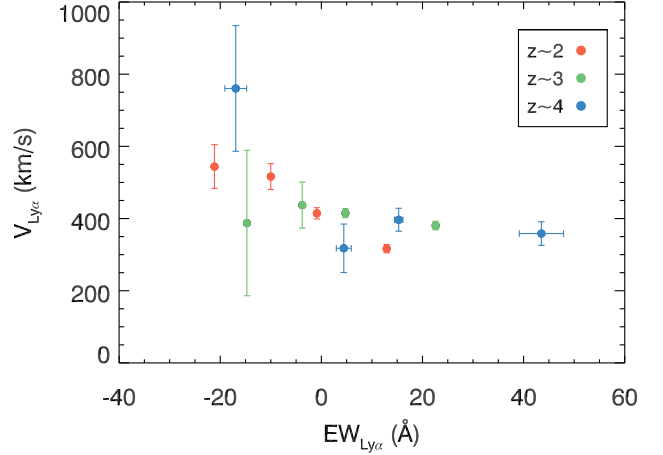}
\caption{Centroid velocity of $\mbox{Ly}\alpha$ emission and $\mbox{Ly}\alpha$ rest-frame EW in the composite spectra binned according to $\mbox{EW}_{Ly\alpha}$.}
\label{fig:lya_kine}
\end{figure}

As described in Section \ref{sec:data}, the centroids of $\mbox{Ly}\alpha$ emission (or absorption) and LIS absorption lines typically do not trace the galaxy systemic velocity because of the presence of galactic-scale outflows. Furthermore, it has been shown empirically how ISM kinematics affect the emergent $\mbox{Ly}\alpha$ profile. By studying a composite spectrum of $z\sim3$ LBGs, \citet{Shapley2003} discovered that weaker $\mbox{Ly}\alpha$ emission corresponds to a larger velocity offset between $\mbox{Ly}\alpha$ and the LIS absorption lines, $\Delta v_{em-abs}$. A similar trend has been measured in the $z\sim4$ LBGs by \citet{Jones2012}, and these authors found no redshift evolution of the $\mbox{EW}_{Ly\alpha}$ vs. $\Delta v_{em-abs}$ relation from $z\sim4$ to $z\sim3$. Motivated by these previous studies, we revisit the correlation between $\mbox{Ly}\alpha$ kinematics and strength, and investigate if the covering fraction of the neutral ISM modulates the observed $\mbox{Ly}\alpha$ profile differently at different redshifts. 

We characterized the kinematics of $\mbox{Ly}\alpha$ by measuring the centroid of the $\mbox{Ly}\alpha$ emission feature in the composite spectra binned according to $\mbox{EW}_{Ly\alpha}$. Even in the composite with the weakest $\mbox{Ly}\alpha$ where the line profile is classified as ``absorption," a small $\mbox{Ly}\alpha$ emission peak is still discernible and therefore its peak wavelength can be evaluated. We used MPFIT to fit the $\mbox{Ly}\alpha$ emission profile in a similar manner to the measurement of the LIS absorption lines, except that the fitting was performed on composites in $L_{\lambda}$ space (i.e., not continuum normalized). The initial estimate of line parameters (centroid, EW, and Gaussian FWHM) were obtained from the IRAF routine $splot$ by fitting the $\mbox{Ly}\alpha$ profile from the blue-side base to the red-side base of the line. We did not iterate the wavelength range for the $\mbox{Ly}\alpha$ emission profile in all $\mbox{Ly}\alpha$ morphological categories, given that the wavelength range bracketed by the blue- to red-bases of the $\mbox{Ly}\alpha$ profile already well defines the feature, and that iterations would not work for the ``absorption" composites as a result of the extremely weak emission line. 

We plot the centroid velocity of $\mbox{Ly}\alpha$ emission and the rest-frame $\mbox{EW}_{Ly\alpha}$ in Figure \ref{fig:lya_kine}. The three redshift samples span different dynamic ranges in $\mbox{EW}_{Ly\alpha}$, with the $z\sim2$ galaxies having on average the weakest $\mbox{Ly}\alpha$ and the $z\sim4$ showing the strongest. The $z\sim2$ and $z\sim4$ measurements show a clear trend of less redshifted $\mbox{Ly}\alpha$ centroid with stronger emission, similar to what has been found for LAEs in both individual and composite spectra at $z\sim2-4$ \citep{Erb2014,Guaita2017}, while for the $z\sim3$ sample there does not appear to be a dependence between these two parameters. Although the correlation between $\mbox{Ly}\alpha$ kinematics and strength is not conclusive at $z\sim3$ based $only$ on our measurements, given previous kinematics studies of $\mbox{Ly}\alpha$ at this redshift \citep[e.g.,][]{Shapley2003}, we believe that this relation holds at $z\sim3$ as well. Generally speaking, the data points from different redshift samples seem to follow the same curve, and we do not see any evolutionary trend of the peak velocity vs. EW relation for $\mbox{Ly}\alpha$ from $z\sim4$ to $z\sim2$. 

In explaining the trend of more redshifted $\mbox{Ly}\alpha$ with weaker $\mbox{Ly}\alpha$ emission across $z\sim2-4$, we consider two physical scenarios. The first possibility is that the larger (positive) offset of $\mbox{Ly}\alpha$ to the systemic velocity results from the faster-moving outflows on far side of the galaxies, which at the same time have a larger covering fraction, leading to weaker $\mbox{Ly}\alpha$ emission. In light of the fact that the outflow kinematics can be characterized by the centroid velocity of LIS absorption features, we can test this hypothesis by examining whether the composites with more redshifted (and therefore weaker) $\mbox{Ly}\alpha$ also show more blueshifted LIS lines. In fact, the measured LIS centroid velocities are fairly similar between composites with the strongest and weakest $\mbox{Ly}\alpha$, counter to what the ``faster outflowing gas'' hypothesis predicts. 

Alternative models are therefore needed to explain the observed correlation between $\mbox{EW}_{Ly\alpha}$ and its peak velocity. Several studies have attempted to generate the emergent $\mbox{Ly}\alpha$ profile with assumed neutral gas kinematics \citep[e.g.,][]{Verhamme2006,Verhamme2008,Steidel2010}. Specifically in \citet{Steidel2010}, where the observed profiles of $\mbox{Ly}\alpha$ and the LIS absorption lines are simultaneously recovered by a kinematic model considering the velocity distribution of the neutral gas, these authors found that the apparent velocity shift of $\mbox{Ly}\alpha$ primarily results from the absorption of $\mbox{Ly}\alpha$ photons by neutral gas near the systemic velocity. As the covering fraction of neutral gas increases around $v=0$, more $\mbox{Ly}\alpha$ photons get resonantly scattered out of the line of sight, pushing the observed $\mbox{Ly}\alpha$ centroid to a redder wavelength, resulting in a weaker $\mbox{Ly}\alpha$ emission at the same time. The model from \citet{Steidel2010} successfully explains our results, and thus the largest $\mbox{EW}_{Ly\alpha}$ seen at $z\sim4$ is likely caused by the smallest covering fraction near the systemic velocity. Furthermore, the redshift independence of the $\mbox{Ly}\alpha$ velocity vs. EW relation suggests a direct impact of the covering fraction and kinematics of the neutral ISM on the observed $\mbox{Ly}\alpha$ emission profile across $z\sim2-4$. Due to the low resolution of the spectra, we were unable to study the detailed structure of the $\mbox{Ly}\alpha$ profile (e.g., the presence and properties of multiple emission peaks). Future spectroscopic data with high spectral resolution will provide further information on the kinematics and covering fraction of the \textrm{H}~\textsc{i} gas. 

\subsection{LIS Absorption Kinematics}
\label{sec:LIS_kine}

\begin{figure}
\includegraphics[width=1.0\linewidth]{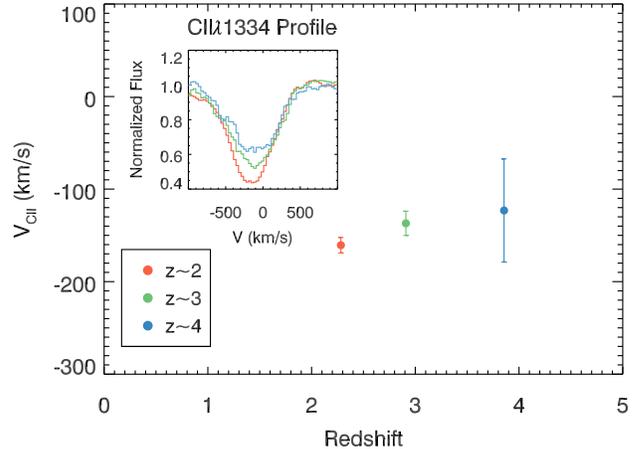}
\caption{Centroid velocity of the \textrm{C}~\textsc{ii}$\lambda$1334 absorption feature vs. redshift in the composite spectra with fixed median stellar mass, $\log(\mbox{M}_{*}/\mbox{M}_{\sun})=9.98$. The $1\sigma$ error bar on the centroid velocity was estimated based on the composite error spectra, and does not include the uncertainty on the systemic velocity. The \textrm{C}~\textsc{ii}$\lambda$1334 line profiles in the $z\sim2$, 3, and 4 fixed-mass composites are shown in the inset panel.}
\label{fig:lis_kine}
\end{figure}

While the centroid velocity of $\mbox{Ly}\alpha$ can provide valuable information on neutral gas kinematics, LIS absorption features are typically used to probe the bulk velocities of the neutral phase of ISM/CGM. Ideally, we would like to decompose the interstellar absorption profiles into a systemic and an outflow component, with the latter representing the absorption truly occurring in the outflowing gas. Unfortunately, the relatively low resolution of our data prevents us from obtaining robust measurements from a two-component fit. Therefore, we fit the absorption profiles with a single Gaussian, and used the centroid velocity of the best-fit Gaussian model as a crude estimator of the bulk kinematic properties.

In addition, it is most ideal to compare galaxies at different redshifts not only spanning the same UV luminosity and stellar mass ranges, but also having the same median values of these properties. However, due to the different distributions of the redshift samples in the $\mbox{M}_{UV}-\mbox{M}_{*}$ plane (Figure \ref{fig:cuts}), it is difficult to construct such subsamples at each redshift with the same median $\mbox{M}_{UV}$ and stellar mass at the same time. Considering that these two properties are correlated, we chose stellar mass as the ``control" and created composite spectra for the $z\sim2$, $z\sim3$ samples such that they have the same median stellar mass, $\log(\mbox{M}_{*}/\mbox{M}_{\sun})=9.98$ as the full $z\sim4$ sample. Specifically, we constructed the matching-$\mbox{M}_{*}$ subsamples by including objects starting from the lower-mass end of the $z\sim2$ and $z\sim3$ samples, until the median values of these subsamples become the closest to that of the $z\sim4$ sample. The resulting subsamples include 290, 235, and 84 spectra at $z\sim2$, 3, and 4, respectively. The higher-redshift subsamples have slightly brighter UV luminosities, with the median $\mbox{M}_{UV}$ for the $z\sim2$, $z\sim3$, and $z\sim4$ samples being -20.48, -20.96, and -21.06, respectively.

In principle, \textrm{Si}~\textsc{ii}$\lambda$1260, \textrm{C}~\textsc{ii}$\lambda$1334, and \textrm{Si}~\textsc{ii}$\lambda$1527 are all good tracers of the ISM kinematics. 
However, we chose \textrm{C}~\textsc{ii}$\lambda$1334 as the probe of the low-ion kinematics, mainly to ensure a fair comparison to the evolutionary study at $z\sim0-2$ by \citet{Sugahara} (Section \ref{sec:lowz_outflow}), in which the authors used \textrm{C}~\textsc{ii}$\lambda$1334 to trace neutral gas kinematics. Using the averaged centroid velocity of \textrm{Si}~\textsc{ii}$\lambda$1260, \textrm{C}~\textsc{ii}$\lambda$1334, and \textrm{Si}~\textsc{ii}$\lambda$1527 yields qualitatively the same answer regarding the evolution of interstellar kinematics at $z\sim2-4$ as using \textrm{C}~\textsc{ii}$\lambda$1334 alone.

Figure \ref{fig:lis_kine} shows the averaged centroid velocity plotted against redshift. The centroid velocities of \textrm{C}~\textsc{ii} are $-161\pm8$ \kms, $-137\pm13$ \kms, and $-123\pm56$ \kms, for the $z\sim2$, $z\sim3$, and $z\sim4$ samples, respectively. We note that the error bars quoted here on the velocity measurements do not include the formal uncertainty on the systemic velocity of the composites, $\Delta v_{sys}$, which is $\sim25$ \kms for $z\sim2-3$ galaxies and $\sim120$ \kms for $z\sim4$ galaxies as estimated from MPFIT.\footnote{As described in Section \ref{sec:comp}, these formal error bars were estimated by measuring the stellar feature \textrm{C}~\textsc{iii}$\lambda$1176 in the overall composites, and therefore do not reflect the systematic uncertainty associated with the redshift rules.} Taking into account $\Delta v_{sys}$ of all redshift samples, we do not observe a noticeable evolutionary trend of the neutral gas kinematics from $z\sim2$ to $z\sim4$ for galaxies at fixed stellar mass. These results, as derived from the centroid velocities, suggest that the kinematics of the neutral ISM/CGM are mostly indistinguishable in galaxies with similar UV luminosity and stellar mass in spite of being at different redshifts. Spectroscopic data with higher $S/N$ and spectral resolution will be of great help for examining the evolution of neutral gas kinematics in greater detail by enabling precise measurements of the outflow component from two-component profile fits. Additionally, high-quality imaging data will provide unique insights into the evolution of outflows through the study of the factors that modulate the strength of the outflows (e.g., size, SFR surface density).

\section{Discussion}
\label{sec:dis}

Although we have primarily investigated how the physical properties of the cool ISM/CGM evolve with redshift, the wide spectral coverage in the rest-UV and the unprecedentedly large size of our galaxy samples at $z\gtrsim2$ enable various analyses that may potentially shed light on multiple key questions in the study of galaxy evolution. Here we extend our results from Sections \ref{sec:res1} and \ref{sec:res2}, examining the redshift evolution of outflow velocities all the way from $z\sim0-4$ (Section \ref{sec:lowz_outflow}), and present a physical picture that accounts for the observed spectroscopic trends (Section \ref{sec:pic}).

\subsection{Outflow Velocity at $z\sim0-4$}
\label{sec:lowz_outflow}

\begin{figure}
\includegraphics[width=1.0\linewidth]{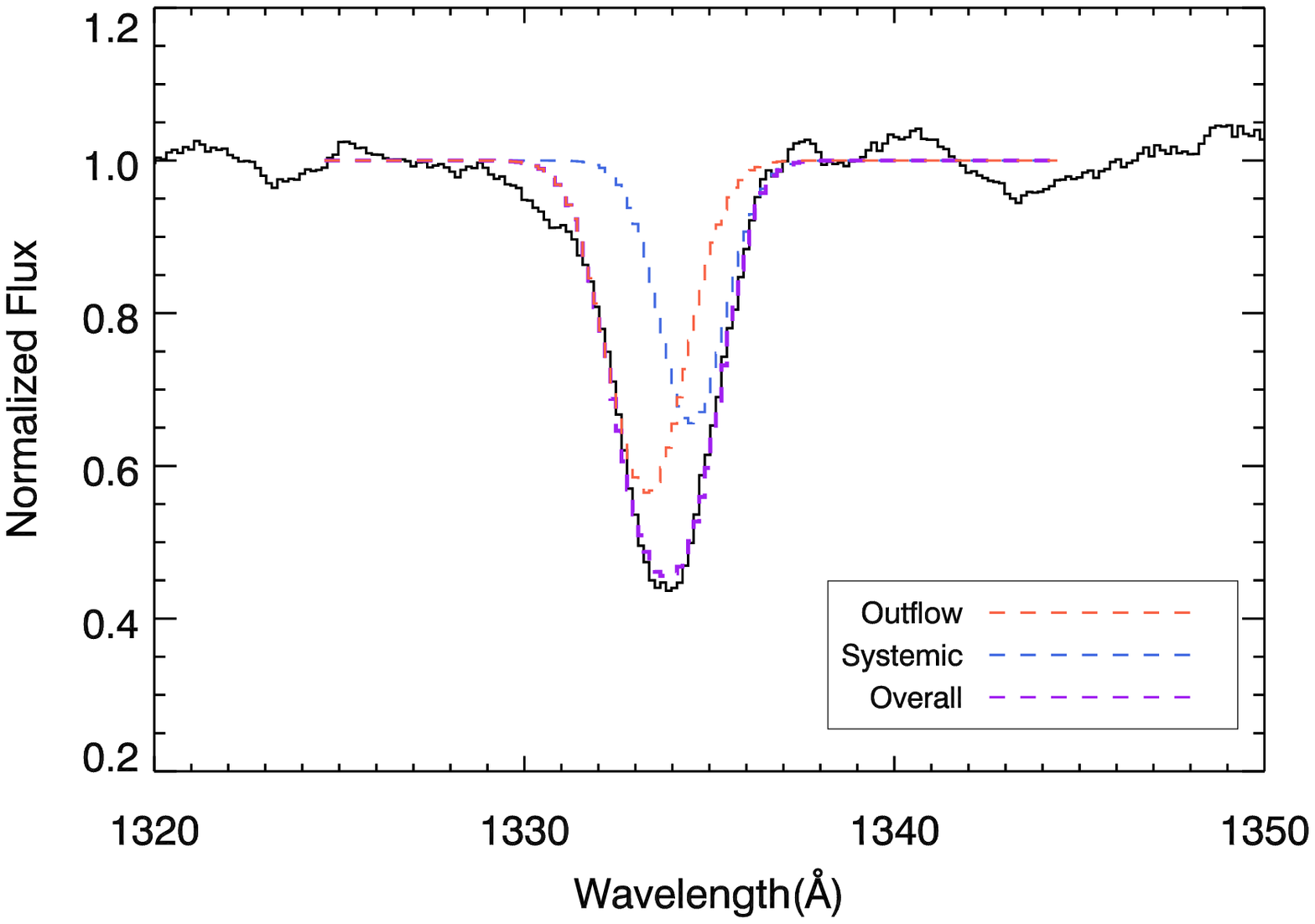}
\includegraphics[width=1.0\linewidth]{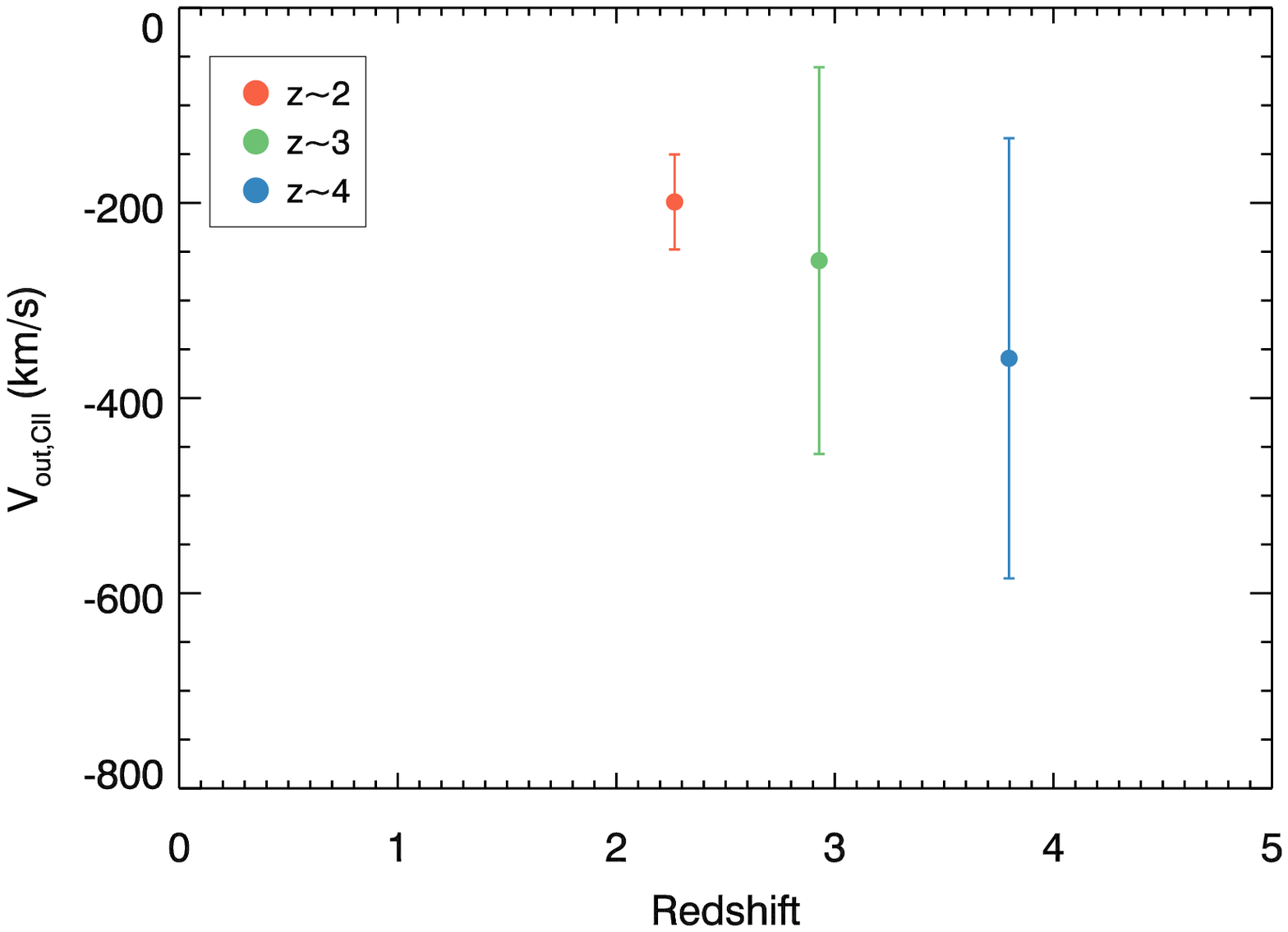}
\caption{\textbf{Top:} Best-fit outflow (orange), systemic (blue), and overall (purple) absorption profiles in the two-component fit to the $z\sim2$ overall composite. \textbf{Bottom:} Outflow velocity (determined by the centroid velocity of the outflow component) measured from the $z\sim2$, $z\sim3$, and $z\sim4$ composite spectra including all objects within the UV luminosity and stellar mass constraints.}
\label{fig:twocomp}
\end{figure}

In this paper, we have explored how the kinematics of neutral gas evolve from $z\sim2$ to $z\sim4$, as probed by the centroid velocity of LIS absorption features. While we did not find noticeable evolution of the outflow velocities of the neutral gas within the $z\sim2-4$ redshift interval, \citet{Sugahara} report evidence of increased outflow velocity from $z\sim0$ to $z\sim2$ based on the measurement of LIS absorption features. These authors studied galaxy spectra at $z\sim0$, $z\sim1$, and $z\sim2$, drawn, respectively, from the Sloan Digital Sky Survey \citep[SDSS;][]{York2000}, DEEP2, and \citet{Erb2006a,Erb2006b}, with the last comprising a subset of the $z\sim2-3$ LRIS sample presented in this paper. Considering the different rest-frame spectral coverage of the three datasets, \citet{Sugahara} used ions with similar ionization energy for the comparison at different redshifts: \textrm{Na}~\textsc{i} D$\lambda\lambda$5891, 5897 and \textrm{Mg}~\textsc{i}$\lambda$2853 from $z\sim0$ to $z\sim1$, and \textrm{Mg}~\textsc{ii}$\lambda\lambda$2796, 2803 and \textrm{C}~\textsc{ii}$\lambda$1334 from $z\sim1$ to $z\sim2$. Given that we also cover the \textrm{C}~\textsc{ii} feature in our $z\sim2-4$ spectra, it is possible to extend the $z\sim0-2$ evolutionary study of neutral gas kinematics by \citet{Sugahara} up to $z\sim4$, connecting the lower- and higher-redshift universe.

\citet{Sugahara} constructed the $z\sim0-2$ samples such that they have similar normalized stellar mass distributions. The $z\sim0$ and $z\sim1$ composites were binned according to SFRs, while the $z\sim2$ stacks included all objects in the $z\sim2$ sample. To characterize the interstellar absorption profiles in the composite spectra, \citet{Sugahara} adopted a two-component model, which includes an outflow component and a systemic component. The profile of each component in the model is described by $I(\lambda)=1-C_{f}+C_{f}e^{-\tau(\lambda)}$ and $\tau(\lambda)=\tau_{0}e^{-(\lambda-\lambda_{0})^{2}/(\lambda_{0}b_{D}/c)^2}$, where $C_{f}$ is the covering fraction, $\tau$ is the optical depth as a function of wavelength, $\tau_{0}$ is the optical depth at the rest wavelength $\lambda_{0}$, and $b_{D}$ is the Doppler parameter. Since the absorption lines were fit with a single Gaussian profile in our study, we cannot directly compare our results with those presented in \citet{Sugahara}. Instead, we conducted a crude two-component Gaussian fit to the \textrm{C}~\textsc{ii}$\lambda$1334 absorption line profile in the continuum-normalized, overall composites at $z\sim2$, 3, and 4. According to \citet{Sugahara}, the observed flux of the interstellar absorption in the continuum-normalized spectra is $I_{obs}(\lambda)=I_{out}(\lambda)I_{sys}(\lambda)$, where $I_{out}(\lambda)$ and $I_{sys}(\lambda)$ are the outflow and systemic components with the continuum normalized to unity, respectively. Accordingly, we fit the \textrm{C}~\textsc{ii} profile using MPFIT to simultaneously model the systemic component at the rest wavelength ($1334.5\mbox{\AA}$), and the blueshifted outflow component. The functional form of the model we adopted is 

\begin{equation}
\label{eq:2comp}
\begin{aligned}
I_{obs}=(1.0-P_{sys})\times(1.0-P_{out})
\end{aligned}
\end{equation}

where $P_{sys}$ and $P_{out}$ are the Gaussian profiles parameterized by line centroid, Gaussian FWHM, and enclosed area for the systemic and outflow components in the continuum-normalized composites, respectively. We list the best-fit parameters of the two-component models for the $z\sim2$, 3, and 4 overall composites in Table \ref{tab:2comp}. In contrast to \citet{Sugahara}, we did not apply additional constraints on the covering fraction, optical depth, and the impact parameter.

\begin{deluxetable*}{ccccccc}
\tablewidth{0pt}
  \tablecaption{Best-fit Parameters of 2-Component Models for $z\sim2-4$ Composites}
  \tablehead{
    \colhead{Sample} &
    \colhead{$\lambda_{sys} (\mbox{\AA})$} &
    \colhead{${\it FWHM_{sys}} (\mbox{\AA})$} &
    \colhead{${Area}_{sys} (\mbox{\AA})$} &
    \colhead{$\lambda_{out} (\mbox{\AA})$} &
    \colhead{${\it FWHM_{out}} (\mbox{\AA})$} &
    \colhead{$Area_{out} (\mbox{\AA})$} \\
    }
  \startdata
$z\sim2$ & 1334.532 & $0.79\pm0.29$ & $0.31\pm0.41$ & $1333.647\pm0.216$ & $1.25\pm0.09$ & $1.61\pm0.32$ \\ 
$z\sim3$ & 1334.532 & $0.96\pm0.26$ & $0.70\pm0.73$ & $1333.379\pm0.881$ & $1.37\pm0.36$ & $1.06\pm0.76$ \\ 
$z\sim4$ & 1334.532 & $1.04\pm0.31$ & $0.87\pm0.33$ & $1332.934\pm1.003$ & $1.03\pm0.77$ & $0.53\pm0.43$ \\ 
 \enddata
\label{tab:2comp}
\tablecomments{Column $2-4$ and $5-7$ represent the line centroid, Gaussian FWHM, and enclosed area for the systemic and outflow components in the continuum-normalized composites, respectively. The systemic component was fixed at the rest-wavelength of \textrm{C}~\textsc{ii}, $1334.532\mbox{\AA}$. }
\end{deluxetable*}

We show an example of the two-component best-fit we obtained with the $z\sim2$ composite in the top panel of Figure \ref{fig:twocomp}. The outflow velocity, $V_{out}$, can therefore be derived from the shift in centroid of the outflow component. Best-fit values of $V_{out,\textrm{C}~\textsc{ii}}$ over $z\sim2-4$ are plotted in the bottom panel of Figure \ref{fig:twocomp}. We measure an outflow velocity of $-199\pm49$ \kms, $-259\pm198$ \kms, and $-359\pm226$ \kms for the $z\sim2$, $z\sim3$, and $z\sim4$ samples, respectively. Within the error bars, \footnote{We note that the uncertainty on the outflow velocities is fairly large, which is likely due to the lack of constraints on both the outflow and systemic components, leading to numerous possible combinations near the minimum $\chi^{2}$.} no clear evolution is shown for the \textrm{C}~\textsc{ii} kinematics. Specifically, our measurement of $V_{out,\textrm{C}~\textsc{ii}}$ at $z\sim2$ is in good agreement with the value quoted in \citet{Sugahara}, $-208\pm30$ \kms, assuming ${C}_{f,sys}=0.1$ without any additional constraints. We also note that this result is consistent with what we found in Section \ref{sec:LIS_kine}, that the neutral gas kinematics do not appear to evolve significantly when being probed by the Gaussian centroid velocity of multiple LIS absorption features. While the lack of redshift evolution of the outflow velocity may not be definitive based $only$ on Figure \ref{fig:twocomp}, we can conclude that we do not see an apparent evolution of the neutral gas kinematics from $z\sim2-4$ when combining our result here with that in Section \ref{sec:LIS_kine}.

If we consider the overall evolution of the neutral gas kinematics from $z\sim0$ to $z\sim4$, combining both our results and those from \citet{Sugahara}, the outflow velocity seems to stop increasing significantly beyond $z\sim2$. A couple of factors may contribute to this ``turning point.'' First, given the positive relation between outflow velocities and the SFR surface density observed in star-forming galaxies \citep{Heckman2000,Chen2010,Kornei2012}, the flattening of the outflow velocity at higher redshifts may result from a smaller change in SFR surface density at fixed mass by a factor of $\sim2.5$ at $z\sim2$ to $z\sim4$ relative to that at $z\sim0$ to $z\sim2$ \citep[e.g.,][]{Shibuya2015}. Secondly, galaxies at $z\leqslant1.5$ tend to show a bipolar outflow geometry \citep[e.g.,][]{Rupke2005,Martin2012,Rubin2014}, as inferred by the outflow detection rate among galaxies with similar star-forming properties as well as the detection of extraplanar gas perpendicular to the plane \citep[e.g.,][]{Bordoloi2011}. Galactic-scale outflows at $z\gtrsim2$, on the other hand, are much more ubiquitous \citep[e.g.,][]{Pettini2002,Shapley2003,Steidel2010}, suggesting a more spherical geometry. Therefore, the smaller detection fraction of blueshifted interstellar absorption lines may make those line profiles less blueshifted in the composites at $z\sim0-2$ than at $z\sim2-4$, resulting in an apparent plateau in outflow speed at $z\sim2$ and higher. Since both factors can collectively affect the observed interstellar absorption profiles in the composites, individual spectra with significant outflow detections or composite spectra constructed by $only$ stacking those with individual detections of outflows would help determine if the SFR surface density or the outflow geometry plays a more important role in this ``turning point'' at $z\sim2$.

\subsection{A physical picture}
\label{sec:pic}

We now interpret our results in a physical picture for the evolution of Ly$\alpha$ emission, and the properties of the absorbing ISM. We also consider the implications of our results for estimates of the Lyman-continuum escape fraction during the epoch of reionization.

\subsubsection{Ly$\alpha$ Emission}

Ly$\alpha$ emission is commonly the strongest feature in the UV spectra of distant star-forming galaxies and has been used to confirm the spectroscopic redshifts of galaxies out to $z>8$ \citep{Zitrin2015}. Furthermore, large samples of LAEs have been assembled at $z\sim 6-7$ in order to probe the neutral fraction of the IGM \citep[e.g.,][]{Konno2018}. Given the prevalence of Ly$\alpha$ measurements, it is important to describe how $\mbox{EW}_{Ly\alpha}$ relates to other galaxy properties, and, more fundamentally, to understand the factors controlling the observed strength of Ly$\alpha$ emission. This knowledge will enable us to relate LAEs to the star-forming galaxy population in general over a wide range of redshifts, and to use the measurement of Ly$\alpha$ as a probe of other key galaxy properties that cannot be directly measured (e.g., escaping Lyman-continuum radiation during the epoch of reionization).

There are three key factors to highlight when attempting to explain the observed Ly$\alpha$ emission properties of high-redshift star-forming galaxies: (1) the intrinsic production of Ly$\alpha$ photons through nebular recombination emission; (2) the radiative transfer of Ly$\alpha$ photons through the ISM and CGM; (3) the radiative transfer of Ly$\alpha$ photons through the increasingly neutral IGM at higher redshifts. Here we consider the importance of these factors, as suggested by our results.

We have established the constancy of the relations among Ly$\alpha$ emission, LIS absorption, and dust extinction (as parametrized by $E(B-V)$), extending over the widest redshift baseline ($z\sim2-4$), and using the most systematically controlled samples to date. In fact, a similar relationship between EW$_{Ly\alpha}$, and EW$_{LIS}$ has even been observed in $z\sim0$ star-forming galaxies \citep[e.g.,][]{Henry2015,Chisholm2017}. The direct and non-evolving connection between EW$_{Ly\alpha}$, EW$_{LIS}$, and $E(B-V)$ suggests that the evolving relations between Ly$\alpha$ and other galaxy properties (e.g., $M_{UV}$, age, $M_*$, SFR) arise as the LIS-absorption gas and dust reddening (at fixed galaxy properties) evolve with redshift. Furthermore, according to one simple interpretation of the trends among EW$_{Ly\alpha}$, EW$_{LIS}$, and $E(B-V)$, galaxies in our samples all have similar $intrinsic$ Ly$\alpha$ EWs, as determined by the ionizing photon production efficiency for a given mass of stars formed, and a relatively constant (and small) escape fraction of Lyman-continuum photons. Together these two factors determine the ratio of Ly$\alpha$ recombination luminosity to UV continuum luminosity density \citep[i.e., the rest-frame Ly$\alpha$ EW][]{Reddy2016}. Accordingly, in this simple picture, the observed range of EW$_{Ly\alpha}$ at $z\sim2-4$ is entirely determined by the range of LIS-absorbing gas covering factors and dust reddenings -- i.e., the radiative transfer of Ly$\alpha$ through the ISM and CGM.

However, our results on the correlation between Ly$\alpha$ and \textrm{C}~\textsc{iii}] emission suggest that, at the highest Ly$\alpha$ EWs, the \textrm{H}~\textsc{ii} region properties and the massive stars that ionize them, are intrinsically different. Specifically, in our $z\sim 2$ sample, galaxies in the quartile with the strongest Ly$\alpha$ emission (median EW$_{Ly\alpha}\sim10$\AA) also have significantly stronger \textrm{C}~\textsc{iii}] emission than in the remainder of the sample. As shown in theoretical and observational work \citep[e.g.,][]{Rigby2015,Jaskot2016,Senchyna2017} stronger \textrm{C}~\textsc{iii}] EWs are typically produced in lower metallicity (i.e., significantly sub-solar) \textrm{H}~\textsc{ii} regions where the electron temperature is higher and the stellar ionizing radiation is stronger. Accordingly, this result indicates a connection between Ly$\alpha$ emission strength and nebular metallicity. This result is consistent with the work of \citet{Trainor2016}, who show that EW$_{Ly\alpha}$ is strongly correlated with nebular metallicity at $z\sim2$, as probed by rest-optical emission line ratios indicating the degree of excitation and ionization. Therefore, the strength of Ly$\alpha$ emission, at least at the highest values of EW$_{Ly\alpha}$, reflects not only the transfer Ly$\alpha$ photons through the ISM but also the ionizing spectra, ionization parameters, and nebular metallicities of the \textrm{H}~\textsc{ii} regions where Ly$\alpha$ photons are originally produced.

The increasing neutral hydrogen opacity in the IGM is evident in the $z\sim2$, 3, and 4 composite spectra shown in Figure \ref{fig:spec}. Specifically, the ratio of continuum flux density bluewards of Ly$\alpha$ relative to that redwards of Ly$\alpha$ decreases significantly from $z\sim2$ to $z\sim4$. Simulations \citep[e.g.,][]{Laursen2011} suggest that increasing IGM 
opacity will lead to a measurably reduced Ly$\alpha$ EW, by $\sim$25\% at $z=3.5$, and by $\sim 75$\% at $z=5.8$. \citet{Jones2012} in fact find evidence for this effect when comparing the Ly$\alpha$ and LIS properties of composite spectra constructed from LBG subsamples at $\langle z \rangle =3.8$ and $\langle z \rangle =4.7$. For roughly the same EW$_{Ly\alpha}$, the $\langle z \rangle =4.7$ sample has a significantly weaker EW$_{LIS}$ ($\sim1\mbox{\AA}$ at $\langle z \rangle =4.7$ vs. $\sim1.7\mbox{\AA}$ at $\langle z \rangle =3.8$). This difference can be explained if the $\langle z \rangle =3.8$ and $\langle z \rangle =4.7$ galaxies follow the same intrinsic relationship between EW$_{Ly\alpha}$ and EW$_{LIS}$, but the average EW$_{Ly\alpha}$ in the subset of LBGs at $\langle z \rangle =4.7$ is suppressed by roughly a factor of two by IGM absorption -- consistent with the predictions by \citet{Laursen2011}. In our work, we find no significant difference among the EW$_{LIS}$ vs. EW$_{Ly\alpha}$ relations from $z\sim2$ to $z\sim4$. This result suggests that, even in our $z\sim4$ sample, the IGM absorption of Ly$\alpha$ photons produced in LBGs is not significant, or that we are not measuring the part of the Ly$\alpha$ profile that is affected by IGM absorption. Alternatively, $z\sim4$ galaxies may follow a different intrinsic relation between EW$_{Ly\alpha}$ and EW$_{LIS}$ relative to the $z\sim2-3$ samples, characterized by higher EW$_{Ly\alpha}$ for a given EW$_{LIS}$. The 
increased IGM absorption at $z\sim 4$ then attenuates the EW$_{Ly\alpha}$ to the point that the $z\sim 4$ sample follows the same patterns as the lower redshift galaxies, mimicking a non-evolving EW$_{Ly\alpha}$ vs. EW$_{LIS}$ relation. For now, we favor the simpler explanation that the EW$_{LIS}$ vs. EW$_{Ly\alpha}$ relation does not evolve between $z\sim2$ and $z\sim4$ and that the effects of IGM absorption are not significant within our $z\sim2-4$ samples. We also note that the lack of evolution between Ly$\alpha$ kinematics ($v_{Ly\alpha}$) and EW$_{Ly\alpha}$ (Section \ref{sec:lya_kine}) are consistent with the subdominance of evolving IGM absorption in our $z\sim2-4$ measurements. However, it will be extremely valuable to investigate the combined possible effects of IGM and ISM evolution out to $z\sim5-6$ using existing rest-UV spectroscopic samples \citep[e.g.,][]{Faisst2016}.

\subsubsection{The Absorbing ISM}
\label{sec:pic_abs}

Although $z\sim2-4$ galaxies follow the same relations among EW$_{Ly\alpha}$, EW$_{LIS}$, and $E(B-V)$, the $average$ EW$_{Ly\alpha}$ increases from $z\sim2-4$, while the average EW$_{LIS}$ and $E(B-V)$ decrease. This evolution in average properties suggests that (at fixed UV luminosity), the covering fraction of LIS-absorbing gas and dust decreases from $z\sim2-4$. As demonstrated by \citet{Reddy2016}, the covering fractions of various low ions is correlated with, but systematically lower than, the covering fraction of \textrm{H}~\textsc{i} ($f_{cov}(\textrm{H}~\textsc{i})$) by a factor of 2--3. Furthermore, the ratio between the covering fractions of low ions and \textrm{H}~\textsc{i} increases as $f_{cov}(\textrm{H}~\textsc{i})$ increases. These trends, as described in \citet{Reddy2016}, are consistent with a scenario in which discrete metal-enriched regions of interstellar gas are contained within the outflowing interstellar HI, whose dust-to-gas (i.e., metal-to-gas) ratio increases with increasing covering fraction of HI. Accordingly, the decrease in average EW$_{LIS}$ from $z\sim2$ to $z\sim4$ corresponds to a decrease in the average covering fraction of interstellar HI.

Two factors that may contribute to the decrease in the covering fraction of interstellar \textrm{H}~\textsc{i} (at fixed galaxy properties such as UV luminosity) are (1) an evolving average stellar radiation field corresponding to a harder ionizing spectrum at $z\sim4$ than at $z\sim 2$, and (2) a decrease in the overall covering fraction of both neutral and ionized interstellar gas. We can test for the presence of both of these effects in our data.

To investigate the nature of the ionizing radiation field, we used the relative column densities of \textrm{Si}~\textsc{ii} and \textrm{Si}~\textsc{iv} as a proxy of the ISM ionization state. As discussed in Section \ref{sec:lya_spec}, we found that the column density of \textrm{Si}~\textsc{iv} appears to increase from $z\sim2$ to $z\sim3-4$, At the same time, the variation in the EW ratio of \textrm{Si}~\textsc{ii}$\lambda$1260 and \textrm{Si}~\textsc{ii}$\lambda$1527 can be used to infer a change in the column density of \textrm{Si}~\textsc{ii}. In the optically thin regime, the EW of \textrm{Si}~\textsc{ii}$\lambda$1260 should be $\sim 5.3$ times larger than that of SiII$\lambda$1527. We observe EW$_{\textrm{Si}~\textsc{ii}\lambda1260}$/EW$_{\textrm{Si}~\textsc{ii}\lambda1527}$ ratios of $0.93\pm0.03\mbox{\AA}$, $0.84\pm0.07\mbox{\AA}$, and $1.30\pm0.25\mbox{\AA}$ at $z \sim 2$, 3, and 4, respectively. This progression suggests a slight decrease in \textrm{Si}~\textsc{ii} column density from $z\sim2$ to $z\sim4$. The combined increase and decrease in the respective column densities of \textrm{Si}~\textsc{iv} and \textrm{Si}~\textsc{ii} suggest a more highly ionized ISM at $z\sim4$ than at $z\sim2$. We also note that, in simple, empirical terms, the ratio of the EWs of \textrm{Si}~\textsc{iv} and \textrm{Si}~\textsc{ii} lines increases from $z\sim 2-4$, as seen in the composite spectra of Figure \ref{fig:spec}: compare the relative depths of \textrm{Si}~\textsc{iv} and \textrm{Si}~\textsc{ii} features in the $z\sim2$ (red) and the $z\sim4$ (blue) composites.

While the balance between high and low ions in the ISM of star-forming galaxies at $z\sim2-4$ may evolve, it also appears that the covering fraction of both neutral and ionized phases decreases. This decrease is traced by the decreasing EW of saturated LIS lines, and by the roughly constant EW of \textrm{Si}~\textsc{iv} absorption as the column density increases (Section \ref{sec:lya_spec}). The evolution in the covering fraction of interstellar gas must reflect an evolving balance among gas accretion, star formation, and feedback. For example, star-forming galaxies at fixed UV luminosity are smaller at $z\sim4$ than at $z\sim2-3$, with approximately doubled SFR surface densities \citep{Shibuya2015}. Given the connection between outflow properties and SFR surface density \citep{Heckman2000,Chen2010,Kornei2012}, the higher typical SFR surface densities at $z\sim4$ may correspond to outflows that more efficiently clear out channels through the ISM, reducing the covering fraction of gas, and allowing for the escape of Ly$\alpha$ photons. Alternatively, if outflows have similar properties at $z\sim2-4$ (at fixed UV luminosity) in terms of energy and momentum input, and ability to evacuate holes in the ISM, we may be observing the accumulation of gas in the ISM and CGM as cosmic time advances. At least in our results, we do not detect significant evolution in the bulk outflow kinematics from 
$z\sim2-4$ (Section \ref{sec:LIS_kine}, Section \ref{sec:lowz_outflow}), so this second alternative is also plausible. A comparison with gas in galaxies modeled in cosmological zoom-in simulations \citep[e.g.,][]{Feldmann2017}, as well as observations at both lower and higher redshifts, will be crucial for understanding the evolution of the covering fraction of metals and \textrm{H}~\textsc{i} in the neutral ISM of star-forming galaxies.

We now consider the connection between the covering fraction of \textrm{H}~\textsc{i} and dust reddening. As shown in this work and previously \citep[e.g.,][]{Shapley2003,Jones2012,Reddy2016}, there is a strong correlation between EW$_{LIS}$ and $E(B-V)$, such that stronger LIS absorption is associated with increased dust reddening. At the same time, there is no significant connection between the high-ionization absorption strength, EW$_{HIS}$ and $E(B-V)$. These results suggest that dust responsible for reddening the stellar continuum in the outflowing ISM/CGM is confined to the neutral regions probed by the LIS features. We would also like to understand what fraction of the reddening and attenuation of the UV continuum takes place in the outflowing ISM/CGM as opposed to gas that is much more local to \textrm{H}~\textsc{ii} 
regions and not participating in large-scale outflows. For example, does the evolution at $z\sim2-4$ towards lower $E(B-V)$ at fixed galaxy properties such as $M_{UV}$ and SFR reflect more than simply the evolution towards lower \textrm{H}~\textsc{i} gas covering fraction, but also the overall lower degree of \textrm{H}~\textsc{ii} region chemical enrichment (and dust content) at higher redshift (at fixed galaxy properties)? Spatially resolved high-$S/N$ measurements of EW$_{LIS}$ and $E(B-V)$ indicating the degree of correlation between small-scale EW$_{LIS}$ and $E(B-V)$ variations may help address this important question of where significant dust attenuation occurs.

\subsubsection{Lyman-continuum Emission}

One of the frontiers for galaxy formation studies is to quantify the contribution of star-forming galaxies to reionization \citep[for a review see][]{Stark2016}. Measurements of the non-ionizing rest-frame UV luminosity function are extending to extremely faint magnitudes based on Hubble Frontier Fields observations \citep[e.g.,][]{Livermore2017,Bouwens2017}. However, since neither Lyman-continuum radiation nor any of the Lyman series absorption features bluewards of Ly$\alpha$ can be measured directly during the epoch of reionization \citep{Vanzella2012,Shapley2016}, we require another proxy for the Lyman-continuum radiation properties of $z>6$ galaxies. Recent work at both low-redshift and $z\sim3$ has highlighted the connection between the strength of Ly$\alpha$ emission, the \textrm{H}~\textsc{i} covering fraction, and the escape of Lyman-continuum radiation \citep[][Steidel et al., in prep]{Shapley2003,Jones2013,Trainor2015,Reddy2016,Marchi2017,Chisholm2017}. Galaxies with higher ratios of Lyman-continuum to non-ionizing UV radiation tend to show stronger Ly$\alpha$ emission and weaker LIS absorption, and models of their UV spectra suggest lower \textrm{H}~\textsc{i} covering fractions.  

Here we suggest that measuring Ly$\alpha$ and LIS absorption properties of $z>6$ star-forming galaxies holds great promise for 
estimating how Lyman-continuum escape varies with UV luminosity during the Epoch of Reionization. In particular, one promising path forward consists of using the Ly$\alpha$ escape fraction as a proxy for the \textrm{H}~\textsc{i} covering fraction \citep[and, accordingly the Lyman-continuum escape fraction which is $1-f_{cov}(\textrm{H}~\textsc{i})$;][]{Jones2013,Lee2016} in bins of UV luminosity. For this method to be effective, there are key uncertainties that will need to be addressed: (1) we will need to estimate and correct for the typical IGM absorption of the Ly$\alpha$ emission profile at $z\sim6$; (2) we will require constraints on the intrinsic production rate of Ly$\alpha$ photons in $z\sim6$ star-forming galaxies in order to convert the observed EW$_{Ly\alpha}$ (corrected for IGM absorption) to a Ly$\alpha$ escape fraction. It is also tempting to use measurements of LIS absorption at $z\sim6$ to predict $f_{cov}(\textrm{H}~\textsc{i})$. However, given the hints of significant metallicity evolution out to $z\sim6$ in star-forming galaxies \citep[e.g.,][]{Stark2016}, typical LIS absorption features at $z\sim6$ may not be saturated, and therefore not offer a direct probe of the \textrm{H}~\textsc{i} covering fraction. More generally, the \textrm{H}~\textsc{i} gas may be optically thinner at $z\sim6$ than at $z\sim3$, which indicates that the absolute escape fraction of LyC photons would no longer be $1-f_{cov}(\textrm{H}~\textsc{i})$ as stated earlier. In order to understand the utility of EW$_{LIS}$ for probing the \textrm{H}~\textsc{i} covering fraction at $z\sim6$, we should first trace the EW$_{Ly\alpha}$ vs. EW$_{LIS}$ relation out to $z\sim6$, using high-quality $z\sim4-6$ ground-based galaxy spectra. At even higher redshifts, rest-UV spectroscopy with the $James$ $Webb$ $Space$ $Telescope$ ({\it JWST}) will prove a powerful tool for constraining the Lyman-continuum emission properties of star-forming galaxies.

\section{Summary}
\label{sec:sum}

Rest-UV spectroscopic features provide rich insights into the key properties (structure, kinematics, covering fraction) of the ISM/CGM in star-forming galaxies. By constructing LBG samples controlled in UV absolute magnitude and stellar mass at $z\sim2-4$, and performing systematic measurements of the spectral and galaxy properties in a uniform manner, we primarily studied the evolution in line strengths and kinematics of both emission and absorption features in the rest-UV spectra. We have found the following key results:

1. No noticeable redshift evolution is found for the relations among $\mbox{EW}_{Ly\alpha}$, $\mbox{EW}_{LIS}$, and $E(B-V)$. In other words, $\mbox{Ly}\alpha$, LIS lines, and dust extinction are fundamentally inter-correlated. Specifically, the redshift independence of the $\mbox{EW}_{Ly\alpha}$ vs. $\mbox{EW}_{LIS}$ relation suggests either that the covering fraction of neutral gas is the dominant factor for modulating both $\mbox{EW}_{Ly\alpha}$ and $\mbox{EW}_{LIS}$ in the same manner at all redshifts, or that the intrinsically higher $\mbox{EW}_{Ly\alpha}$ due to lower metallicity and stronger radiation fields at higher redshifts at fixed galaxy property is compensated by increased absorption from the IGM. In contrast, the strength of $\mbox{Ly}\alpha$ exhibits no correlation with the HIS absorption strength, suggesting that $\mbox{Ly}\alpha$ and the HIS absorption features trace different phases of gas. The approximately constant $\mbox{EW}_{HIS}$ over $z\sim2-4$ suggests a lower covering fraction of the ionized gas at higher redshifts, given the apparent increasing \textrm{Si}~\textsc{iv} column density.

2. We observe a redshift-invariant correlation between $\mbox{EW}_{LIS}$ and $E(B-V)$, which is consistent with a physical picture in which dust and metal ions are embedded in the outflowing \textrm{H}~\textsc{i} gas. Both the covering fraction of neutral gas, as probed by $\mbox{EW}_{LIS}$, and dust extinction seem to decrease with increasing redshift. We interpret this redshift evolution with two possible scenarios: either the $z\sim4$ galaxies are more effective in reducing the \textrm{H}~\textsc{i} covering fraction and dust content by increasing the ionization state of the ISM and driving powerful outflows, or the $z\sim2-3$ galaxies are more efficient in accumulating gas and dust as they evolve. Individual spectra with high $S/N$, spectral and spatial resolution will shed light on the physical location of the dust with respect to the neutral gas.

3. The strength of $\mbox{Ly}\alpha$ emission decreases with decreasing redshift at fixed stellar mass, UV luminosity, and SFR, which likely results from a larger covering fraction of the neutral gas and higher $E(B-V)$ at lower redshifts. In the $z\sim2$ and $z\sim3$ samples, $\mbox{EW}_{Ly\alpha}$ shows little to no correlation with stellar mass, UV luminosity, and SFR. However, stronger $\mbox{EW}_{Ly\alpha}$ is observed at fainter galaxies with lower stellar mass and SFR at $z\sim4$. The weakening of these trends from $z\sim4$ to $z\sim2$ is likely caused by a larger dynamic range of the neutral gas covering fraction at $z\sim4$ than at $z\sim2$ and $z\sim3$. Although the sample incompleteness enhances the observed $\mbox{Ly}\alpha$ trends, the evolution of these relations (i.e., stronger $\mbox{Ly}\alpha$ emission at higher redshifts at fixed galaxy properties, steeper $\mbox{Ly}\alpha$ trends at $z\sim4$ than at $z\sim2-3$) still exists after we have accounted for differential sample incompleteness. We thus conclude that the trends are real and robust, and not simply an effect of increasing sample incompleteness to galaxies lacking $\mbox{Ly}\alpha$ emission at higher redshift and fainter magnitude.

4. Younger galaxies at $z\sim4$ show stronger $\mbox{Ly}\alpha$, which can be attributed to their lower gas-phase metallicities and harder ionizing spectra. On the other hand, no visible trends are seen at $z\sim2$ and $z\sim3$ between $\mbox{EW}_{Ly\alpha}$ and galaxy age. The different behavior of the EW$_{Ly\alpha}$ vs. age relation can be explained by the $z\sim2-3$ galaxies being more chemically enriched at the youngest ages than the $z\sim4$ galaxies, which are possibly experiencing their first generation of star formation. Our results here do not support the positive correlation between $\mbox{Ly}\alpha$ and age, previously reported in some studies of $z\sim3$ LBGs \citep{Shapley2001,Kornei2010}. In this earlier work, the authors modeled galaxy SEDs assuming 1.4 solar metallicity and the Calzetti extinction curve, and did not account for the contamination of strong nebular emission lines in the photometric bands. We found that with the most reasonable description of the stellar populations in the $z\sim3$ galaxies in our sample (0.28 solar metallicity and the SMC extinction curve except for the highest-mass objects), the positive correlation between $\mbox{EW}_{Ly\alpha}$ and age disappears after removing the contaminated $K_{s}$-band from SED modeling.

5. The ratio of fine-structure emission to corresponding resonant absorption does not evolve significantly with redshift at fixed dust extinction. While on average larger $\mbox{EW}_{\textrm{Si}~\textsc{ii}*}/\mbox{EW}_{\textrm{Si}~\textsc{ii}}$ is observed at $z\sim4$, this trend is a result of the lower dust extinction in higher-redshift galaxies, rather than an evolution with cosmic time towards larger size of the fine-structure emission region.

6. We find a flat trend between $\mbox{EW}_{\textrm{C}~\textsc{iii}]}$ and $\mbox{EW}_{Ly\alpha}$ at $z\sim2$ except in the quartile with the strongest $\mbox{Ly}\alpha$ emission, where the \textrm{C}~\textsc{iii}] emission is significantly higher than in the remaining $\mbox{Ly}\alpha$ stacks. Given that $\mbox{EW}_{\textrm{C}~\textsc{iii}]}$ depends on the physical properties of the \textrm{H}~\textsc{ii} regions (e.g., gas-phase metallicity, ionization parameter), this result suggests variation in the $intrinsic$ $\mbox{Ly}\alpha$ production among galaxies in our samples. Galaxies with larger observed $\mbox{EW}_{Ly\alpha}$ may not only have lower \textrm{H}~\textsc{i} gas covering fractions, but also intrinsically produce more ionizing (and $\mbox{Ly}\alpha$) photons per unit mass of stars formed.

7. We measure no strong evolution in the key rest-UV spectroscopic trends tracing interstellar kinematics. The trend that strong $\mbox{EW}_{Ly\alpha}$ corresponds to more redshifted $\mbox{Ly}\alpha$ emission profile seems to be universal across $z\sim2-4$, which suggests that the covering fraction of the neutral gas near $v=0$ modifies the $\mbox{Ly}\alpha$ profile in the same manner across $z\sim2-4$. On the other hand, the outflow velocities traced by the centroids of the LIS absorption features do not exhibit any evolutionary trend with redshift. We therefore speculate that the kinematics of the neutral gas are similar in LBGs in fixed UV luminosity and stellar mass ranges at $z\sim2-4$.

Answering the key outstanding questions in the study of the evolving ISM/CGM at high redshift will require spectroscopic data with high $S/N$ and spectral resolution, along with Integral Field Unit (IFU) spectroscopic maps. For example, rest-UV and optical spectral maps of lensed, spatially-resolved galaxies will provide key information dissecting the relative physical distribution of neutral gas and dust. Furthermore, the near-IR capabilities of the {\it JWST} will enable rest-UV and rest-optical spectroscopic studies of star-forming galaxies out to $z>6$. With these high-quality individual, deep spectra, detailed, quantitative conclusions can be drawn on topics including: (1) the nature of the relation between dust and neutral gas; (2) the intrinsic production and escape of $\mbox{Ly}\alpha$ and Lyman-continuum photons; and (3) the detailed kinematics and spatial extent of the neutral and ionized phases of outflows. Making progress on these questions from an empirical standpoint is essential for our understanding of where the ISM/CGM absorption arises in galaxies, and a complete model of feedback in galaxy formation.

\acknowledgements We acknowledge support from the David $\&$ Lucile Packard Foundation (A.E.S). 
We are grateful to the 3D-HST team for providing ancillary data on galaxy properties. This paper also includes data gathered with the 6.5 meter Magellan Telescopes located at Las Campanas Observatory, Chile.
We wish to extend special thanks to those of Hawaiian ancestry on whose sacred mountain we are privileged to be guests. Without their generous hospitality, most of the observations presented herein would not have been possible.

\bibliographystyle{apj}
\bibliography{DU17_new}

\begin{thebibliography}{}
\expandafter\ifx\csname natexlab\endcsname\relax\def\natexlab#1{#1}\fi

\bibitem[{{Berg} {et~al.}(2016){Berg}, {Skillman}, {Henry}, {Erb}, \&
  {Carigi}}]{Berg2016}
{Berg}, D.~A., {Skillman}, E.~D., {Henry}, R.~B.~C., {Erb}, D.~K., \& {Carigi},
  L. 2016, \apj, 827, 126

\bibitem[{{Berry} {et~al.}(2012){Berry}, {Gawiser}, {Guaita}, {Padilla},
  {Treister}, {Blanc}, {Ciardullo}, {Francke}, \& {Gronwall}}]{Berry2012}
{Berry}, M., {Gawiser}, E., {Guaita}, L., {et~al.} 2012, \apj, 749, 4

\bibitem[{{Bordoloi} {et~al.}(2011){Bordoloi}, {Lilly}, {Knobel}, {Bolzonella},
  {Kampczyk}, {Carollo}, {Iovino}, {Zucca}, {Contini}, {Kneib}, {Le Fevre},
  {Mainieri}, {Renzini}, {Scodeggio}, {Zamorani}, {Balestra}, {Bardelli},
  {Bongiorno}, {Caputi}, {Cucciati}, {de la Torre}, {de Ravel}, {Garilli},
  {Kova{\v c}}, {Lamareille}, {Le Borgne}, {Le Brun}, {Maier}, {Mignoli},
  {Pello}, {Peng}, {Perez Montero}, {Presotto}, {Scarlata}, {Silverman},
  {Tanaka}, {Tasca}, {Tresse}, {Vergani}, {Barnes}, {Cappi}, {Cimatti},
  {Coppa}, {Diener}, {Franzetti}, {Koekemoer}, {L{\'o}pez-Sanjuan},
  {McCracken}, {Moresco}, {Nair}, {Oesch}, {Pozzetti}, \&
  {Welikala}}]{Bordoloi2011}
{Bordoloi}, R., {Lilly}, S.~J., {Knobel}, C., {et~al.} 2011, \apj, 743, 10

\bibitem[{{Bouwens} {et~al.}(2017){Bouwens}, {van Dokkum}, {Illingworth},
  {Oesch}, {Maseda}, {Ribeiro}, {Stefanon}, \& {Lam}}]{Bouwens2017}
{Bouwens}, R.~J., {van Dokkum}, P.~G., {Illingworth}, G.~D., {et~al.} 2017,
  ArXiv e-prints, arXiv:1711.02090

\bibitem[{{Bouwens} {et~al.}(2016){Bouwens}, {Aravena}, {Decarli}, {Walter},
  {da Cunha}, {Labb{\'e}}, {Bauer}, {Bertoldi}, {Carilli}, {Chapman}, {Daddi},
  {Hodge}, {Ivison}, {Karim}, {Le Fevre}, {Magnelli}, {Ota}, {Riechers},
  {Smail}, {van der Werf}, {Weiss}, {Cox}, {Elbaz}, {Gonzalez-Lopez},
  {Infante}, {Oesch}, {Wagg}, \& {Wilkins}}]{Bouwens2016}
{Bouwens}, R.~J., {Aravena}, M., {Decarli}, R., {et~al.} 2016, \apj, 833, 72

\bibitem[{{Brammer} {et~al.}(2012){Brammer}, {van Dokkum}, {Franx},
  {Fumagalli}, {Patel}, {Rix}, {Skelton}, {Kriek}, {Nelson}, {Schmidt},
  {Bezanson}, {da Cunha}, {Erb}, {Fan}, {F{\"o}rster Schreiber}, {Illingworth},
  {Labb{\'e}}, {Leja}, {Lundgren}, {Magee}, {Marchesini}, {McCarthy},
  {Momcheva}, {Muzzin}, {Quadri}, {Steidel}, {Tal}, {Wake}, {Whitaker}, \&
  {Williams}}]{Brammer2012}
{Brammer}, G.~B., {van Dokkum}, P.~G., {Franx}, M., {et~al.} 2012, \apjs, 200,
  13

\bibitem[{{Bruzual} \& {Charlot}(2003)}]{BC03}
{Bruzual}, G., \& {Charlot}, S. 2003, \mnras, 344, 1000

\bibitem[{{Calzetti} {et~al.}(2000){Calzetti}, {Armus}, {Bohlin}, {Kinney},
  {Koornneef}, \& {Storchi-Bergmann}}]{Calzetti2000}
{Calzetti}, D., {Armus}, L., {Bohlin}, R.~C., {et~al.} 2000, \apj, 533, 682

\bibitem[{{Chabrier}(2003)}]{Chabrier2003}
{Chabrier}, G. 2003, \pasp, 115, 763

\bibitem[{{Chen} {et~al.}(2010){Chen}, {Tremonti}, {Heckman}, {Kauffmann},
  {Weiner}, {Brinchmann}, \& {Wang}}]{Chen2010}
{Chen}, Y.-M., {Tremonti}, C.~A., {Heckman}, T.~M., {et~al.} 2010, \aj, 140,
  445

\bibitem[{{Chisholm} {et~al.}(2017){Chisholm}, {Orlitov{\'a}}, {Schaerer},
  {Verhamme}, {Worseck}, {Izotov}, {Thuan}, \& {Guseva}}]{Chisholm2017}
{Chisholm}, J., {Orlitov{\'a}}, I., {Schaerer}, D., {et~al.} 2017, \aap, 605,
  A67

\bibitem[{{de Barros} {et~al.}(2016){de Barros}, {Vanzella}, {Amor{\'{\i}}n},
  {Castellano}, {Siana}, {Grazian}, {Suh}, {Balestra}, {Vignali}, {Verhamme},
  {Zamorani}, {Mignoli}, {Hasinger}, {Comastri}, {Pentericci},
  {P{\'e}rez-Montero}, {Fontana}, {Giavalisco}, \& {Gilli}}]{Debarros2016}
{de Barros}, S., {Vanzella}, E., {Amor{\'{\i}}n}, R., {et~al.} 2016, \aap, 585,
  A51

\bibitem[{{Donley} {et~al.}(2012){Donley}, {Koekemoer}, {Brusa}, {Capak},
  {Cardamone}, {Civano}, {Ilbert}, {Impey}, {Kartaltepe}, {Miyaji}, {Salvato},
  {Sanders}, {Trump}, \& {Zamorani}}]{Donley2012}
{Donley}, J.~L., {Koekemoer}, A.~M., {Brusa}, M., {et~al.} 2012, \apj, 748, 142

\bibitem[{{Du} {et~al.}(2016){Du}, {Shapley}, {Martin}, \& {Coil}}]{Du2016}
{Du}, X., {Shapley}, A.~E., {Martin}, C.~L., \& {Coil}, A.~L. 2016, \apj, 829,
  64

\bibitem[{{Du} {et~al.}(2017){Du}, {Shapley}, {Martin}, \& {Coil}}]{Du2017}
---. 2017, \apj, 838, 63

\bibitem[{{Erb} {et~al.}(2010){Erb}, {Pettini}, {Shapley}, {Steidel}, {Law}, \&
  {Reddy}}]{Erb2010}
{Erb}, D.~K., {Pettini}, M., {Shapley}, A.~E., {et~al.} 2010, \apj, 719, 1168

\bibitem[{{Erb} {et~al.}(2016){Erb}, {Pettini}, {Steidel}, {Strom}, {Rudie},
  {Trainor}, {Shapley}, \& {Reddy}}]{Erb2016}
{Erb}, D.~K., {Pettini}, M., {Steidel}, C.~C., {et~al.} 2016, \apj, 830, 52

\bibitem[{{Erb} {et~al.}(2012){Erb}, {Quider}, {Henry}, \& {Martin}}]{Erb2012}
{Erb}, D.~K., {Quider}, A.~M., {Henry}, A.~L., \& {Martin}, C.~L. 2012, \apj,
  759, 26

\bibitem[{{Erb} {et~al.}(2006{\natexlab{a}}){Erb}, {Steidel}, {Shapley},
  {Pettini}, {Reddy}, \& {Adelberger}}]{Erb2006b}
{Erb}, D.~K., {Steidel}, C.~C., {Shapley}, A.~E., {et~al.} 2006{\natexlab{a}},
  \apj, 647, 128

\bibitem[{{Erb} {et~al.}(2006{\natexlab{b}}){Erb}, {Steidel}, {Shapley},
  {Pettini}, {Reddy}, \& {Adelberger}}]{Erb2006a}
---. 2006{\natexlab{b}}, \apj, 646, 107

\bibitem[{{Erb} {et~al.}(2014){Erb}, {Steidel}, {Trainor}, {Bogosavljevi{\'c}},
  {Shapley}, {Nestor}, {Kulas}, {Law}, {Strom}, {Rudie}, {Reddy}, {Pettini},
  {Konidaris}, {Mace}, {Matthews}, \& {McLean}}]{Erb2014}
{Erb}, D.~K., {Steidel}, C.~C., {Trainor}, R.~F., {et~al.} 2014, \apj, 795, 33

\bibitem[{{Faber} {et~al.}(2003){Faber}, {Phillips}, {Kibrick}, {Alcott},
  {Allen}, {Burrous}, {Cantrall}, {Clarke}, {Coil}, {Cowley}, {Davis}, {Deich},
  {Dietsch}, {Gilmore}, {Harper}, {Hilyard}, {Lewis}, {McVeigh}, {Newman},
  {Osborne}, {Schiavon}, {Stover}, {Tucker}, {Wallace}, {Wei}, {Wirth}, \&
  {Wright}}]{Faber2003}
{Faber}, S.~M., {Phillips}, A.~C., {Kibrick}, R.~I., {et~al.} 2003, in
  \procspie, Vol. 4841, Instrument Design and Performance for Optical/Infrared
  Ground-based Telescopes, ed. M.~{Iye} \& A.~F.~M. {Moorwood}, 1657--1669

\bibitem[{{Faisst} {et~al.}(2016){Faisst}, {Capak}, {Davidzon}, {Salvato},
  {Laigle}, {Ilbert}, {Onodera}, {Hasinger}, {Kakazu}, {Masters}, {McCracken},
  {Mobasher}, {Sanders}, {Silverman}, {Yan}, \& {Scoville}}]{Faisst2016}
{Faisst}, A.~L., {Capak}, P.~L., {Davidzon}, I., {et~al.} 2016, \apj, 822, 29

\bibitem[{{Feldmann} {et~al.}(2017){Feldmann}, {Quataert}, {Hopkins},
  {Faucher-Gigu{\`e}re}, \& {Kere{\v s}}}]{Feldmann2017}
{Feldmann}, R., {Quataert}, E., {Hopkins}, P.~F., {Faucher-Gigu{\`e}re}, C.-A.,
  \& {Kere{\v s}}, D. 2017, \mnras, 470, 1050

\bibitem[{{Finley} {et~al.}(2017){Finley}, {Bouch{\'e}}, {Contini}, {Epinat},
  {Bacon}, {Brinchmann}, {Cantalupo}, {Erroz-Ferrer}, {Marino}, {Maseda},
  {Richard}, {Schroetter}, {Verhamme}, {Weilbacher}, {Wendt}, \&
  {Wisotzki}}]{Finley2017}
{Finley}, H., {Bouch{\'e}}, N., {Contini}, T., {et~al.} 2017, \aap, 605, A118

\bibitem[{{Giavalisco} {et~al.}(2004){Giavalisco}, {Ferguson}, {Koekemoer},
  {Dickinson}, {Alexander}, {Bauer}, {Bergeron}, {Biagetti}, {Brandt},
  {Casertano}, {Cesarsky}, {Chatzichristou}, {Conselice}, {Cristiani}, {Da
  Costa}, {Dahlen}, {de Mello}, {Eisenhardt}, {Erben}, {Fall}, {Fassnacht},
  {Fosbury}, {Fruchter}, {Gardner}, {Grogin}, {Hook}, {Hornschemeier}, {Idzi},
  {Jogee}, {Kretchmer}, {Laidler}, {Lee}, {Livio}, {Lucas}, {Madau},
  {Mobasher}, {Moustakas}, {Nonino}, {Padovani}, {Papovich}, {Park},
  {Ravindranath}, {Renzini}, {Richardson}, {Riess}, {Rosati}, {Schirmer},
  {Schreier}, {Somerville}, {Spinrad}, {Stern}, {Stiavelli}, {Strolger},
  {Urry}, {Vandame}, {Williams}, \& {Wolf}}]{Gia2004}
{Giavalisco}, M., {Ferguson}, H.~C., {Koekemoer}, A.~M., {et~al.} 2004, \apjl,
  600, L93

\bibitem[{{Guaita} {et~al.}(2017){Guaita}, {Talia}, {Pentericci}, {Verhamme},
  {Cassata}, {Lemaux}, {Orlitova}, {Ribeiro}, {Schaerer}, {Zamorani},
  {Garilli}, {Le Brun}, {Le F{\`e}vre}, {Maccagni}, {Tasca}, {Thomas},
  {Vanzella}, {Zucca}, {Amorin}, {Bardelli}, {Castellano}, {Grazian}, {Hathi},
  {Koekemoer}, \& {Marchi}}]{Guaita2017}
{Guaita}, L., {Talia}, M., {Pentericci}, L., {et~al.} 2017, \aap, 606, A19

\bibitem[{{Hainline} {et~al.}(2016){Hainline}, {Reines}, {Greene}, \&
  {Stern}}]{Hainline2016}
{Hainline}, K.~N., {Reines}, A.~E., {Greene}, J.~E., \& {Stern}, D. 2016, \apj,
  832, 119

\bibitem[{{Hainline} {et~al.}(2012){Hainline}, {Shapley}, {Greene}, {Steidel},
  {Reddy}, \& {Erb}}]{Hainline2012}
{Hainline}, K.~N., {Shapley}, A.~E., {Greene}, J.~E., {et~al.} 2012, \apj, 760,
  74

\bibitem[{{Hathi} {et~al.}(2016){Hathi}, {Le F{\`e}vre}, {Ilbert}, {Cassata},
  {Tasca}, {Lemaux}, {Garilli}, {Le Brun}, {Maccagni}, {Pentericci}, {Thomas},
  {Vanzella}, {Zamorani}, {Zucca}, {Amor{\'{\i}}n}, {Bardelli}, {Cassar{\`a}},
  {Castellano}, {Cimatti}, {Cucciati}, {Durkalec}, {Fontana}, {Giavalisco},
  {Grazian}, {Guaita}, {Koekemoer}, {Paltani}, {Pforr}, {Ribeiro}, {Schaerer},
  {Scodeggio}, {Sommariva}, {Talia}, {Tresse}, {Vergani}, {Capak}, {Charlot},
  {Contini}, {Cuby}, {de la Torre}, {Dunlop}, {Fotopoulou},
  {L{\'o}pez-Sanjuan}, {Mellier}, {Salvato}, {Scoville}, {Taniguchi}, \&
  {Wang}}]{Hathi2016}
{Hathi}, N.~P., {Le F{\`e}vre}, O., {Ilbert}, O., {et~al.} 2016, \aap, 588, A26

\bibitem[{{Heckman} {et~al.}(2000){Heckman}, {Lehnert}, {Strickland}, \&
  {Armus}}]{Heckman2000}
{Heckman}, T.~M., {Lehnert}, M.~D., {Strickland}, D.~K., \& {Armus}, L. 2000,
  \apjs, 129, 493

\bibitem[{{Henry} {et~al.}(2015){Henry}, {Scarlata}, {Martin}, \&
  {Erb}}]{Henry2015}
{Henry}, A., {Scarlata}, C., {Martin}, C.~L., \& {Erb}, D. 2015, \apj, 809, 19

\bibitem[{{Jaskot} \& {Ravindranath}(2016)}]{Jaskot2016}
{Jaskot}, A.~E., \& {Ravindranath}, S. 2016, \apj, 833, 136

\bibitem[{{Jones} {et~al.}(2012){Jones}, {Stark}, \& {Ellis}}]{Jones2012}
{Jones}, T., {Stark}, D.~P., \& {Ellis}, R.~S. 2012, \apj, 751, 51

\bibitem[{{Jones} {et~al.}(2013){Jones}, {Ellis}, {Schenker}, \&
  {Stark}}]{Jones2013}
{Jones}, T.~A., {Ellis}, R.~S., {Schenker}, M.~A., \& {Stark}, D.~P. 2013,
  \apj, 779, 52

\bibitem[{{Konno} {et~al.}(2014){Konno}, {Ouchi}, {Ono}, {Shimasaku},
  {Shibuya}, {Furusawa}, {Nakajima}, {Naito}, {Momose}, {Yuma}, \&
  {Iye}}]{Konno2014}
{Konno}, A., {Ouchi}, M., {Ono}, Y., {et~al.} 2014, \apj, 797, 16

\bibitem[{{Konno} {et~al.}(2018){Konno}, {Ouchi}, {Shibuya}, {Ono},
  {Shimasaku}, {Taniguchi}, {Nagao}, {Kobayashi}, {Kajisawa}, {Kashikawa},
  {Inoue}, {Oguri}, {Furusawa}, {Goto}, {Harikane}, {Higuchi}, {Komiyama},
  {Kusakabe}, {Miyazaki}, {Nakajima}, \& {Wang}}]{Konno2018}
{Konno}, A., {Ouchi}, M., {Shibuya}, T., {et~al.} 2018, \pasj, 70, S16

\bibitem[{{Kornei} {et~al.}(2010){Kornei}, {Shapley}, {Erb}, {Steidel},
  {Reddy}, {Pettini}, \& {Bogosavljevi{\'c}}}]{Kornei2010}
{Kornei}, K.~A., {Shapley}, A.~E., {Erb}, D.~K., {et~al.} 2010, \apj, 711, 693

\bibitem[{{Kornei} {et~al.}(2012){Kornei}, {Shapley}, {Martin}, {Coil}, {Lotz},
  {Schiminovich}, {Bundy}, \& {Noeske}}]{Kornei2012}
{Kornei}, K.~A., {Shapley}, A.~E., {Martin}, C.~L., {et~al.} 2012, \apj, 758,
  135

\bibitem[{{Kornei} {et~al.}(2013){Kornei}, {Shapley}, {Martin}, {Coil}, {Lotz},
  \& {Weiner}}]{Kornei2013}
---. 2013, \apj, 774, 50

\bibitem[{{Kulas} {et~al.}(2012){Kulas}, {Shapley}, {Kollmeier}, {Zheng},
  {Steidel}, \& {Hainline}}]{Kulas2012}
{Kulas}, K.~R., {Shapley}, A.~E., {Kollmeier}, J.~A., {et~al.} 2012, \apj, 745,
  33

\bibitem[{{Laursen} {et~al.}(2011){Laursen}, {Sommer-Larsen}, \&
  {Razoumov}}]{Laursen2011}
{Laursen}, P., {Sommer-Larsen}, J., \& {Razoumov}, A.~O. 2011, \apj, 728, 52

\bibitem[{{Leethochawalit} {et~al.}(2016){Leethochawalit}, {Jones}, {Ellis},
  {Stark}, \& {Zitrin}}]{Lee2016}
{Leethochawalit}, N., {Jones}, T.~A., {Ellis}, R.~S., {Stark}, D.~P., \&
  {Zitrin}, A. 2016, \apj, 831, 152

\bibitem[{{Livermore} {et~al.}(2017){Livermore}, {Finkelstein}, \&
  {Lotz}}]{Livermore2017}
{Livermore}, R.~C., {Finkelstein}, S.~L., \& {Lotz}, J.~M. 2017, \apj, 835, 113

\bibitem[{{Madau} \& {Dickinson}(2014)}]{Madau2014}
{Madau}, P., \& {Dickinson}, M. 2014, \araa, 52, 415

\bibitem[{{Marchi} {et~al.}(2017){Marchi}, {Pentericci}, {Guaita}, {Ribeiro},
  {Castellano}, {Schaerer}, {Hathi}, {Lemaux}, {Grazian}, {Le F{\`e}vre},
  {Garilli}, {Maccagni}, {Amorin}, {Bardelli}, {Cassata}, {Fontana},
  {Koekemoer}, {Le Brun}, {Tasca}, {Thomas}, {Vanzella}, {Zamorani}, \&
  {Zucca}}]{Marchi2017}
{Marchi}, F., {Pentericci}, L., {Guaita}, L., {et~al.} 2017, \aap, 601, A73

\bibitem[{{Markwardt}(2009)}]{Mark2009}
{Markwardt}, C.~B. 2009, in Astronomical Society of the Pacific Conference
  Series, Vol. 411, Astronomical Data Analysis Software and Systems XVIII, ed.
  D.~A. {Bohlender}, D.~{Durand}, \& P.~{Dowler}, 251

\bibitem[{{Martin} {et~al.}(2012){Martin}, {Shapley}, {Coil}, {Kornei},
  {Bundy}, {Weiner}, {Noeske}, \& {Schiminovich}}]{Martin2012}
{Martin}, C.~L., {Shapley}, A.~E., {Coil}, A.~L., {et~al.} 2012, \apj, 760, 127

\bibitem[{{Maseda} {et~al.}(2017){Maseda}, {Brinchmann}, {Franx}, {Bacon},
  {Bouwens}, {Schmidt}, {Boogaard}, {Contini}, {Feltre}, {Inami},
  {Kollatschny}, {Marino}, {Richard}, {Verhamme}, \& {Wisotzki}}]{Maseda2017}
{Maseda}, M.~V., {Brinchmann}, J., {Franx}, M., {et~al.} 2017, \aap, 608, A4

\bibitem[{{Newman} {et~al.}(2013){Newman}, {Cooper}, {Davis}, {Faber}, {Coil},
  {Guhathakurta}, {Koo}, {Phillips}, {Conroy}, {Dutton}, {Finkbeiner}, {Gerke},
  {Rosario}, {Weiner}, {Willmer}, {Yan}, {Harker}, {Kassin}, {Konidaris},
  {Lai}, {Madgwick}, {Noeske}, {Wirth}, {Connolly}, {Kaiser}, {Kirby},
  {Lemaux}, {Lin}, {Lotz}, {Luppino}, {Marinoni}, {Matthews}, {Metevier}, \&
  {Schiavon}}]{Newman2013}
{Newman}, J.~A., {Cooper}, M.~C., {Davis}, M., {et~al.} 2013, \apjs, 208, 5

\bibitem[{{Oesch} {et~al.}(2013){Oesch}, {Labb{\'e}}, {Bouwens}, {Illingworth},
  {Gonzalez}, {Franx}, {Trenti}, {Holden}, {van Dokkum}, \&
  {Magee}}]{Oesch2013}
{Oesch}, P.~A., {Labb{\'e}}, I., {Bouwens}, R.~J., {et~al.} 2013, \apj, 772,
  136

\bibitem[{{Oke} {et~al.}(1995){Oke}, {Cohen}, {Carr}, {Cromer}, {Dingizian},
  {Harris}, {Labrecque}, {Lucinio}, {Schaal}, {Epps}, \& {Miller}}]{Oke1995}
{Oke}, J.~B., {Cohen}, J.~G., {Carr}, M., {et~al.} 1995, \pasp, 107, 375

\bibitem[{{Onodera} {et~al.}(2016){Onodera}, {Carollo}, {Lilly}, {Renzini},
  {Arimoto}, {Capak}, {Daddi}, {Scoville}, {Tacchella}, {Tatehora}, \&
  {Zamorani}}]{Onodera2016}
{Onodera}, M., {Carollo}, C.~M., {Lilly}, S., {et~al.} 2016, \apj, 822, 42

\bibitem[{{Pentericci} {et~al.}(2009){Pentericci}, {Grazian}, {Fontana},
  {Castellano}, {Giallongo}, {Salimbeni}, \& {Santini}}]{Pentericci2009}
{Pentericci}, L., {Grazian}, A., {Fontana}, A., {et~al.} 2009, \aap, 494, 553

\bibitem[{{Pentericci} {et~al.}(2007){Pentericci}, {Grazian}, {Fontana},
  {Salimbeni}, {Santini}, {de Santis}, {Gallozzi}, \&
  {Giallongo}}]{Pentericci2007}
---. 2007, \aap, 471, 433

\bibitem[{{Pentericci} {et~al.}(2010){Pentericci}, {Grazian}, {Scarlata},
  {Fontana}, {Castellano}, {Giallongo}, \& {Vanzella}}]{Pentericci2010}
{Pentericci}, L., {Grazian}, A., {Scarlata}, C., {et~al.} 2010, \aap, 514, A64

\bibitem[{{Pentericci} {et~al.}(2014){Pentericci}, {Vanzella}, {Fontana},
  {Castellano}, {Treu}, {Mesinger}, {Dijkstra}, {Grazian}, {Brada{\v c}},
  {Conselice}, {Cristiani}, {Dunlop}, {Galametz}, {Giavalisco}, {Giallongo},
  {Koekemoer}, {McLure}, {Maiolino}, {Paris}, \& {Santini}}]{Pentericci2014}
{Pentericci}, L., {Vanzella}, E., {Fontana}, A., {et~al.} 2014, \apj, 793, 113

\bibitem[{{Pettini} {et~al.}(2002){Pettini}, {Rix}, {Steidel}, {Adelberger},
  {Hunt}, \& {Shapley}}]{Pettini2002}
{Pettini}, M., {Rix}, S.~A., {Steidel}, C.~C., {et~al.} 2002, \apj, 569, 742

\bibitem[{{Prochaska} {et~al.}(2011){Prochaska}, {Kasen}, \&
  {Rubin}}]{Prochaska2011}
{Prochaska}, J.~X., {Kasen}, D., \& {Rubin}, K. 2011, \apj, 734, 24

\bibitem[{{Quider} {et~al.}(2009){Quider}, {Pettini}, {Shapley}, \&
  {Steidel}}]{Quider2009}
{Quider}, A.~M., {Pettini}, M., {Shapley}, A.~E., \& {Steidel}, C.~C. 2009,
  \mnras, 398, 1263

\bibitem[{{Reddy} {et~al.}(2012){Reddy}, {Pettini}, {Steidel}, {Shapley},
  {Erb}, \& {Law}}]{Reddy2012}
{Reddy}, N.~A., {Pettini}, M., {Steidel}, C.~C., {et~al.} 2012, \apj, 754, 25

\bibitem[{{Reddy} {et~al.}(2006){Reddy}, {Steidel}, {Erb}, {Shapley}, \&
  {Pettini}}]{Reddy2006}
{Reddy}, N.~A., {Steidel}, C.~C., {Erb}, D.~K., {Shapley}, A.~E., \& {Pettini},
  M. 2006, \apj, 653, 1004

\bibitem[{{Reddy} {et~al.}(2008){Reddy}, {Steidel}, {Pettini}, {Adelberger},
  {Shapley}, {Erb}, \& {Dickinson}}]{Reddy2008}
{Reddy}, N.~A., {Steidel}, C.~C., {Pettini}, M., {et~al.} 2008, \apjs, 175, 48

\bibitem[{{Reddy} {et~al.}(2016){Reddy}, {Steidel}, {Pettini},
  {Bogosavljevi{\'c}}, \& {Shapley}}]{Reddy2016}
{Reddy}, N.~A., {Steidel}, C.~C., {Pettini}, M., {Bogosavljevi{\'c}}, M., \&
  {Shapley}, A.~E. 2016, \apj, 828, 108

\bibitem[{{Reddy} {et~al.}(2017){Reddy}, {Oesch}, {Bouwens}, {Montes},
  {Illingworth}, {Steidel}, {van Dokkum}, {Atek}, {Carollo}, {Cibinel},
  {Holden}, {Labbe}, {Magee}, {Morselli}, {Nelson}, \& {Wilkins}}]{Reddy2017}
{Reddy}, N.~A., {Oesch}, P.~A., {Bouwens}, R.~J., {et~al.} 2017, ArXiv
  e-prints, arXiv:1705.09302

\bibitem[{{Rigby} {et~al.}(2015){Rigby}, {Bayliss}, {Gladders}, {Sharon},
  {Wuyts}, {Dahle}, {Johnson}, \& {Pe{\~n}a-Guerrero}}]{Rigby2015}
{Rigby}, J.~R., {Bayliss}, M.~B., {Gladders}, M.~D., {et~al.} 2015, \apjl, 814,
  L6

\bibitem[{{Rix} {et~al.}(2004){Rix}, {Pettini}, {Leitherer}, {Bresolin},
  {Kudritzki}, \& {Steidel}}]{Rix2004}
{Rix}, S.~A., {Pettini}, M., {Leitherer}, C., {et~al.} 2004, \apj, 615, 98

\bibitem[{{Rubin} {et~al.}(2014){Rubin}, {Prochaska}, {Koo}, {Phillips},
  {Martin}, \& {Winstrom}}]{Rubin2014}
{Rubin}, K.~H.~R., {Prochaska}, J.~X., {Koo}, D.~C., {et~al.} 2014, \apj, 794,
  156

\bibitem[{{Rudie} {et~al.}(2012){Rudie}, {Steidel}, {Trainor}, {Rakic},
  {Bogosavljevi{\'c}}, {Pettini}, {Reddy}, {Shapley}, {Erb}, \&
  {Law}}]{Rudie2012}
{Rudie}, G.~C., {Steidel}, C.~C., {Trainor}, R.~F., {et~al.} 2012, \apj, 750,
  67

\bibitem[{{Rupke} {et~al.}(2005){Rupke}, {Veilleux}, \& {Sanders}}]{Rupke2005}
{Rupke}, D.~S., {Veilleux}, S., \& {Sanders}, D.~B. 2005, \apjs, 160, 115

\bibitem[{{Sanders} {et~al.}(2015){Sanders}, {Shapley}, {Kriek}, {Reddy},
  {Freeman}, {Coil}, {Siana}, {Mobasher}, {Shivaei}, {Price}, \& {de
  Groot}}]{Sanders2015}
{Sanders}, R.~L., {Shapley}, A.~E., {Kriek}, M., {et~al.} 2015, \apj, 799, 138

\bibitem[{{Schenker} {et~al.}(2014){Schenker}, {Ellis}, {Konidaris}, \&
  {Stark}}]{Schenker2014}
{Schenker}, M.~A., {Ellis}, R.~S., {Konidaris}, N.~P., \& {Stark}, D.~P. 2014,
  \apj, 795, 20

\bibitem[{{Senchyna} {et~al.}(2017){Senchyna}, {Stark}, {Vidal-Garc{\'{\i}}a},
  {Chevallard}, {Charlot}, {Mainali}, {Jones}, {Wofford}, {Feltre}, \&
  {Gutkin}}]{Senchyna2017}
{Senchyna}, P., {Stark}, D.~P., {Vidal-Garc{\'{\i}}a}, A., {et~al.} 2017,
  \mnras, 472, 2608

\bibitem[{{Shapley} {et~al.}(2001){Shapley}, {Steidel}, {Adelberger},
  {Dickinson}, {Giavalisco}, \& {Pettini}}]{Shapley2001}
{Shapley}, A.~E., {Steidel}, C.~C., {Adelberger}, K.~L., {et~al.} 2001, \apj,
  562, 95

\bibitem[{{Shapley} {et~al.}(2003){Shapley}, {Steidel}, {Pettini}, \&
  {Adelberger}}]{Shapley2003}
{Shapley}, A.~E., {Steidel}, C.~C., {Pettini}, M., \& {Adelberger}, K.~L. 2003,
  \apj, 588, 65

\bibitem[{{Shapley} {et~al.}(2016){Shapley}, {Steidel}, {Strom},
  {Bogosavljevi{\'c}}, {Reddy}, {Siana}, {Mostardi}, \& {Rudie}}]{Shapley2016}
{Shapley}, A.~E., {Steidel}, C.~C., {Strom}, A.~L., {et~al.} 2016, \apjl, 826,
  L24

\bibitem[{{Shibuya} {et~al.}(2015){Shibuya}, {Ouchi}, \&
  {Harikane}}]{Shibuya2015}
{Shibuya}, T., {Ouchi}, M., \& {Harikane}, Y. 2015, \apjs, 219, 15

\bibitem[{{Skelton} {et~al.}(2014){Skelton}, {Whitaker}, {Momcheva}, {Brammer},
  {van Dokkum}, {Labb{\'e}}, {Franx}, {van der Wel}, {Bezanson}, {Da Cunha},
  {Fumagalli}, {F{\"o}rster Schreiber}, {Kriek}, {Leja}, {Lundgren}, {Magee},
  {Marchesini}, {Maseda}, {Nelson}, {Oesch}, {Pacifici}, {Patel}, {Price},
  {Rix}, {Tal}, {Wake}, \& {Wuyts}}]{Skelton2014}
{Skelton}, R.~E., {Whitaker}, K.~E., {Momcheva}, I.~G., {et~al.} 2014, \apjs,
  214, 24

\bibitem[{{Stark} {et~al.}(2009){Stark}, {Ellis}, {Bunker}, {Bundy}, {Targett},
  {Benson}, \& {Lacy}}]{Stark2009}
{Stark}, D.~P., {Ellis}, R.~S., {Bunker}, A., {et~al.} 2009, \apj, 697, 1493

\bibitem[{{Stark} {et~al.}(2010){Stark}, {Ellis}, {Chiu}, {Ouchi}, \&
  {Bunker}}]{Stark2010}
{Stark}, D.~P., {Ellis}, R.~S., {Chiu}, K., {Ouchi}, M., \& {Bunker}, A. 2010,
  \mnras, 408, 1628

\bibitem[{{Stark} {et~al.}(2014){Stark}, {Richard}, {Siana}, {Charlot},
  {Freeman}, {Gutkin}, {Wofford}, {Robertson}, {Amanullah}, {Watson}, \&
  {Milvang-Jensen}}]{Stark2014}
{Stark}, D.~P., {Richard}, J., {Siana}, B., {et~al.} 2014, \mnras, 445, 3200

\bibitem[{{Stark} {et~al.}(2015){Stark}, {Richard}, {Charlot}, {Cl{\'e}ment},
  {Ellis}, {Siana}, {Robertson}, {Schenker}, {Gutkin}, \&
  {Wofford}}]{Stark2015}
{Stark}, D.~P., {Richard}, J., {Charlot}, S., {et~al.} 2015, \mnras, 450, 1846

\bibitem[{{Stark} {et~al.}(2017){Stark}, {Ellis}, {Charlot}, {Chevallard},
  {Tang}, {Belli}, {Zitrin}, {Mainali}, {Gutkin}, {Vidal-Garc{\'{\i}}a},
  {Bouwens}, \& {Oesch}}]{Stark2016}
{Stark}, D.~P., {Ellis}, R.~S., {Charlot}, S., {et~al.} 2017, \mnras, 464, 469

\bibitem[{{Steidel} {et~al.}(2003){Steidel}, {Adelberger}, {Shapley},
  {Pettini}, {Dickinson}, \& {Giavalisco}}]{Steidel2003}
{Steidel}, C.~C., {Adelberger}, K.~L., {Shapley}, A.~E., {et~al.} 2003, \apj,
  592, 728

\bibitem[{{Steidel} {et~al.}(2010){Steidel}, {Erb}, {Shapley}, {Pettini},
  {Reddy}, {Bogosavljevi{\'c}}, {Rudie}, \& {Rakic}}]{Steidel2010}
{Steidel}, C.~C., {Erb}, D.~K., {Shapley}, A.~E., {et~al.} 2010, \apj, 717, 289

\bibitem[{{Steidel} {et~al.}(2004){Steidel}, {Shapley}, {Pettini},
  {Adelberger}, {Erb}, {Reddy}, \& {Hunt}}]{Steidel2004}
{Steidel}, C.~C., {Shapley}, A.~E., {Pettini}, M., {et~al.} 2004, \apj, 604,
  534

\bibitem[{{Steidel} {et~al.}(2016){Steidel}, {Strom}, {Pettini}, {Rudie},
  {Reddy}, \& {Trainor}}]{Steidel2016}
{Steidel}, C.~C., {Strom}, A.~L., {Pettini}, M., {et~al.} 2016, \apj, 826, 159

\bibitem[{{Steidel} {et~al.}(2014){Steidel}, {Rudie}, {Strom}, {Pettini},
  {Reddy}, {Shapley}, {Trainor}, {Erb}, {Turner}, {Konidaris}, {Kulas}, {Mace},
  {Matthews}, \& {McLean}}]{Steidel2014}
{Steidel}, C.~C., {Rudie}, G.~C., {Strom}, A.~L., {et~al.} 2014, \apj, 795, 165

\bibitem[{{Strom} {et~al.}(2017){Strom}, {Steidel}, {Rudie}, {Trainor},
  {Pettini}, \& {Reddy}}]{Strom2017}
{Strom}, A.~L., {Steidel}, C.~C., {Rudie}, G.~C., {et~al.} 2017, \apj, 836, 164

\bibitem[{{Sugahara} {et~al.}(2017){Sugahara}, {Ouchi}, {Lin}, {Martin}, {Ono},
  {Harikane}, {Shibuya}, \& {Yan}}]{Sugahara}
{Sugahara}, Y., {Ouchi}, M., {Lin}, L., {et~al.} 2017, \apj, 850, 51

\bibitem[{{Trainor} {et~al.}(2015){Trainor}, {Steidel}, {Strom}, \&
  {Rudie}}]{Trainor2015}
{Trainor}, R.~F., {Steidel}, C.~C., {Strom}, A.~L., \& {Rudie}, G.~C. 2015,
  \apj, 809, 89

\bibitem[{{Trainor} {et~al.}(2016){Trainor}, {Strom}, {Steidel}, \&
  {Rudie}}]{Trainor2016}
{Trainor}, R.~F., {Strom}, A.~L., {Steidel}, C.~C., \& {Rudie}, G.~C. 2016,
  \apj, 832, 171

\bibitem[{{Treu} {et~al.}(2012){Treu}, {Trenti}, {Stiavelli}, {Auger}, \&
  {Bradley}}]{Treu2012}
{Treu}, T., {Trenti}, M., {Stiavelli}, M., {Auger}, M.~W., \& {Bradley}, L.~D.
  2012, \apj, 747, 27

\bibitem[{{Tumlinson} {et~al.}(2017){Tumlinson}, {Peeples}, \&
  {Werk}}]{Tumlinson2017}
{Tumlinson}, J., {Peeples}, M.~S., \& {Werk}, J.~K. 2017, \araa, 55, 389

\bibitem[{{Vanzella} {et~al.}(2005){Vanzella}, {Cristiani}, {Dickinson},
  {Kuntschner}, {Moustakas}, {Nonino}, {Rosati}, {Stern}, {Cesarsky}, {Ettori},
  {Ferguson}, {Fosbury}, {Giavalisco}, {Haase}, {Renzini}, {Rettura}, {Serra},
  \& {GOODS Team}}]{Vanzella2005}
{Vanzella}, E., {Cristiani}, S., {Dickinson}, M., {et~al.} 2005, \aap, 434, 53

\bibitem[{{Vanzella} {et~al.}(2006){Vanzella}, {Cristiani}, {Dickinson},
  {Kuntschner}, {Nonino}, {Rettura}, {Rosati}, {Vernet}, {Cesarsky},
  {Ferguson}, {Fosbury}, {Giavalisco}, {Grazian}, {Haase}, {Moustakas},
  {Popesso}, {Renzini}, {Stern}, \& {GOODS Team}}]{Vanzella2006}
---. 2006, \aap, 454, 423

\bibitem[{{Vanzella} {et~al.}(2008){Vanzella}, {Cristiani}, {Dickinson},
  {Giavalisco}, {Kuntschner}, {Haase}, {Nonino}, {Rosati}, {Cesarsky},
  {Ferguson}, {Fosbury}, {Grazian}, {Moustakas}, {Rettura}, {Popesso},
  {Renzini}, {Stern}, \& {GOODS Team}}]{Vanzella2008}
---. 2008, \aap, 478, 83

\bibitem[{{Vanzella} {et~al.}(2009){Vanzella}, {Giavalisco}, {Dickinson},
  {Cristiani}, {Nonino}, {Kuntschner}, {Popesso}, {Rosati}, {Renzini}, {Stern},
  {Cesarsky}, {Ferguson}, \& {Fosbury}}]{Vanzella2009}
{Vanzella}, E., {Giavalisco}, M., {Dickinson}, M., {et~al.} 2009, \apj, 695,
  1163

\bibitem[{{Vanzella} {et~al.}(2012){Vanzella}, {Guo}, {Giavalisco}, {Grazian},
  {Castellano}, {Cristiani}, {Dickinson}, {Fontana}, {Nonino}, {Giallongo},
  {Pentericci}, {Galametz}, {Faber}, {Ferguson}, {Grogin}, {Koekemoer},
  {Newman}, \& {Siana}}]{Vanzella2012}
{Vanzella}, E., {Guo}, Y., {Giavalisco}, M., {et~al.} 2012, \apj, 751, 70

\bibitem[{{Vanzella} {et~al.}(2016){Vanzella}, {De Barros}, {Cupani}, {Karman},
  {Gronke}, {Balestra}, {Coe}, {Mignoli}, {Brusa}, {Calura}, {Caminha},
  {Caputi}, {Castellano}, {Christensen}, {Comastri}, {Cristiani}, {Dijkstra},
  {Fontana}, {Giallongo}, {Giavalisco}, {Gilli}, {Grazian}, {Grillo},
  {Koekemoer}, {Meneghetti}, {Nonino}, {Pentericci}, {Rosati}, {Schaerer},
  {Verhamme}, {Vignali}, \& {Zamorani}}]{Vanzella2016}
{Vanzella}, E., {De Barros}, S., {Cupani}, G., {et~al.} 2016, \apjl, 821, L27

\bibitem[{{Verhamme} {et~al.}(2008){Verhamme}, {Schaerer}, {Atek}, \&
  {Tapken}}]{Verhamme2008}
{Verhamme}, A., {Schaerer}, D., {Atek}, H., \& {Tapken}, C. 2008, \aap, 491, 89

\bibitem[{{Verhamme} {et~al.}(2006){Verhamme}, {Schaerer}, \&
  {Maselli}}]{Verhamme2006}
{Verhamme}, A., {Schaerer}, D., \& {Maselli}, A. 2006, \aap, 460, 397

\bibitem[{{York} {et~al.}(2000){York}, {Adelman}, {Anderson}, {Anderson},
  {Annis}, {Bahcall}, {Bakken}, {Barkhouser}, {Bastian}, {Berman}, {Boroski},
  {Bracker}, {Briegel}, {Briggs}, {Brinkmann}, {Brunner}, {Burles}, {Carey},
  {Carr}, {Castander}, {Chen}, {Colestock}, {Connolly}, {Crocker}, {Csabai},
  {Czarapata}, {Davis}, {Doi}, {Dombeck}, {Eisenstein}, {Ellman}, {Elms},
  {Evans}, {Fan}, {Federwitz}, {Fiscelli}, {Friedman}, {Frieman}, {Fukugita},
  {Gillespie}, {Gunn}, {Gurbani}, {de Haas}, {Haldeman}, {Harris}, {Hayes},
  {Heckman}, {Hennessy}, {Hindsley}, {Holm}, {Holmgren}, {Huang}, {Hull},
  {Husby}, {Ichikawa}, {Ichikawa}, {Ivezi{\'c}}, {Kent}, {Kim}, {Kinney},
  {Klaene}, {Kleinman}, {Kleinman}, {Knapp}, {Korienek}, {Kron}, {Kunszt},
  {Lamb}, {Lee}, {Leger}, {Limmongkol}, {Lindenmeyer}, {Long}, {Loomis},
  {Loveday}, {Lucinio}, {Lupton}, {MacKinnon}, {Mannery}, {Mantsch}, {Margon},
  {McGehee}, {McKay}, {Meiksin}, {Merelli}, {Monet}, {Munn}, {Narayanan},
  {Nash}, {Neilsen}, {Neswold}, {Newberg}, {Nichol}, {Nicinski}, {Nonino},
  {Okada}, {Okamura}, {Ostriker}, {Owen}, {Pauls}, {Peoples}, {Peterson},
  {Petravick}, {Pier}, {Pope}, {Pordes}, {Prosapio}, {Rechenmacher}, {Quinn},
  {Richards}, {Richmond}, {Rivetta}, {Rockosi}, {Ruthmansdorfer}, {Sandford},
  {Schlegel}, {Schneider}, {Sekiguchi}, {Sergey}, {Shimasaku}, {Siegmund},
  {Smee}, {Smith}, {Snedden}, {Stone}, {Stoughton}, {Strauss}, {Stubbs},
  {SubbaRao}, {Szalay}, {Szapudi}, {Szokoly}, {Thakar}, {Tremonti}, {Tucker},
  {Uomoto}, {Vanden Berk}, {Vogeley}, {Waddell}, {Wang}, {Watanabe},
  {Weinberg}, {Yanny}, {Yasuda}, \& {SDSS Collaboration}}]{York2000}
{York}, D.~G., {Adelman}, J., {Anderson}, Jr., J.~E., {et~al.} 2000, \aj, 120,
  1579

\bibitem[{{Zitrin} {et~al.}(2015){Zitrin}, {Ellis}, {Belli}, \&
  {Stark}}]{Zitrin2015}
{Zitrin}, A., {Ellis}, R.~S., {Belli}, S., \& {Stark}, D.~P. 2015, \apjl, 805,
  L7

\end{thebibliography}
\end{CJK}
\appendix

\section{Appendix A: Individual and Composite Error Spectra}
\label{sec:comperr}

Individual error spectra are essential for creating the composite error spectra, which ideally account for both sample variance and measurement uncertainty. However, the individual error spectra were not available for all the $z\sim2$ and $z\sim3$ galaxies, and for the FORS2 objects in the $z\sim4$ sample. In this subsection, we describe how we constructed the composite error spectra for the $z\sim2-3$ and the $z\sim4$ samples, respectively, despite the lack of some individual error spectra.

To create the composite error spectra for the $z\sim2$ and $z\sim3$ samples, we first reconstructed individual error spectra for all $z\sim2-3$ galaxies by utilizing the vast library of the individual LRIS error spectra attained in the $z\sim1$ outflow kinematic study presented in \citet{Martin2012}. In this study, a sample of 208 objects were drawn from the Deep Extragalatic Evolutionary Probe 2 \citep[DEEP2;][]{Newman2013} galaxy redshift survey spanning the range $0.4 \leqslant z \leqslant 1.4$ and apparent $B$-band magnitude $B<24.0$, and were observed with LRIS. The 400 lines $\mbox{mm}^{-1}$ and 600 lines $\mbox{mm}^{-1}$ grisms were used to obtain the blue side of the spectra. 145 objects were observed with the 400-lines $\mbox{mm}^{-1}$ grism and 63 objects were observed with the 600-lines $\mbox{mm}^{-1}$ grism.
The individual one-dimensional error spectra for the $z\sim1$ galaxies were determined by the standard deviation at each wavelength of the corresponding science spectra of the same object from multiple exposures.

To reconstruct the individual error spectra of the $z\sim2-3$ LRIS spectra, the main goal is to recover the shape of the error spectra, which is set by the sensitivity of the instrument in the observed frame. We first made the 400-line and 600-line error spectrum templates by stacking individual error spectra with respective spectral resolution. When being combined, these individual error spectra were scaled to a common median over $4000-5000\mbox{\AA}$ in the observed frame. We then used the IRAF $continuum$ routine to fit the general shape of the two error spectrum templates with an order of 5. After obtaining the smoothed error spectrum templates, for each object in the $z\sim2-3$ LBG sample that was observed with the 400-line (600-line) grism, we shifted the 400-line (600-line) template to the rest-frame, and scaled it such that the pixel-to-pixel noise over $1250-1400\mbox{\AA}$ in the rest-frame science spectrum matched the average flux level of the rest-frame error template across the same wavelength range. 

Unfortunately, we were unable to precisely simulate the overall shape of the 300-line grating and grism error spectra, given that the $z\sim1$ DEEP2/LRIS data were not obtained with these configurations on LRIS. For this small fraction of our sample ($\sim7\%$ combined at $z\sim2-3$), we adopted the shape of the 400-line grism error template for the 300-line grating and grism objects as a crude estimate. For individual 300-line galaxies, the 400-line error template was also transformed into their respective rest frames, and scaled to match the rms of corresponding science spectra over $1250-1400\mbox{\AA}$. We note that approximating the individual 300-line error spectra with the 400-line error template does not affect the line measurements in the composite spectra significantly. Removing the 300-line grating and grism objects from the composites (and therefore having ``clean'' $z\sim2$ and $z\sim3$ samples of only 400- and 600-line objects with well-reconstructed individual error spectra) results in a change of only $\lesssim8\%$ or within the $1\sigma$ uncertainty for the $\mbox{EW}_{Ly\alpha}$ measurements, and $\lesssim5\%$ of change in the $\mbox{EW}_{LIS}$.

With estimates of the individual error spectra of the $z\sim2$ and $z\sim3$ galaxies in hand, we bootstrap-resampled the objects in each bin and perturbed each spectrum (in the $L_{\lambda}$ space) in the bootstrap sample according to its own error spectrum. The perturbed spectra in the bootstrap sample were then scaled and combined (following the same procedure of constructing the science composites) to create a new composite spectrum. The process was repeated 100 times and the standard deviation of these 100 fake composites at each wavelength was adopted as the composite error spectrum for each bin.

The construction of the composite error spectra for the $z\sim4$ sample is a bit different. While the DEIMOS objects have available individual error spectra, the FORS2 objects do not. Considering that the individual error spectra were only needed for creating the composite error spectra, we evaluated the relative contribution of bootstrap resampling and individual error spectra to the overall uncertainty for both the DEIMOS and FORS2 objects, based on their proportion in the $z\sim4$ sample and typical $S/N$.\footnote{As the $z\sim4$ sample mainly consists of DEIMOS and FORS2 objects (80 out of 91 spectra), we did not take into account the $z>3.4$ LRIS galaxies and their reconstructed individual error spectra when estimating the overall uncertainty.} We first calculated the ratio of the uncertainty estimated from `bootstrap only' to that from both bootstrap and individual error spectra for the DEIMOS data. The median value of the ratio is 0.404, which was estimated over $1250-1400\mbox{\AA}$, a spectral region within which DEIMOS has a decent sensitivity. Given that the DEIMOS spectra on average have lower $S/N$ at these wavelengths than the FORS2 spectra (median $S/N=2.76$ and 4.07, respectively), we scaled the individual DEIMOS error spectra down by a factor of $4.07/2.76=1.47$ to match the $S/N$ of the FORS2 spectra, assuming that the FORS2 error spectra have the same shape as the DEIMOS error spectra. We repeated the same process, estimating the ratio between the `bootstrap only' and `bootstrap $+$ individual error spectra' for the DEIMOS spectra scaled to the $S/N$ of the FORS2 spectra and the resulting median ratio is 0.564. These two ratios were combined with weights that correspond to the proportion of the DEIMOS and FORS spectra in the entire $z\sim4$ sample, yielding a overall ratio of 0.460. As a result, the overall error composite spectra of the $z\sim4$ sample was determined by dividing the error spectra evaluated from `bootstrap only' by a factor 0.460 to account for the contribution of individual error spectra with different $S/N$.

\section{Appendix B: IR-excess galaxies}
\label{sec:AGN}

\begin{figure}
\includegraphics[width=1.0\linewidth]{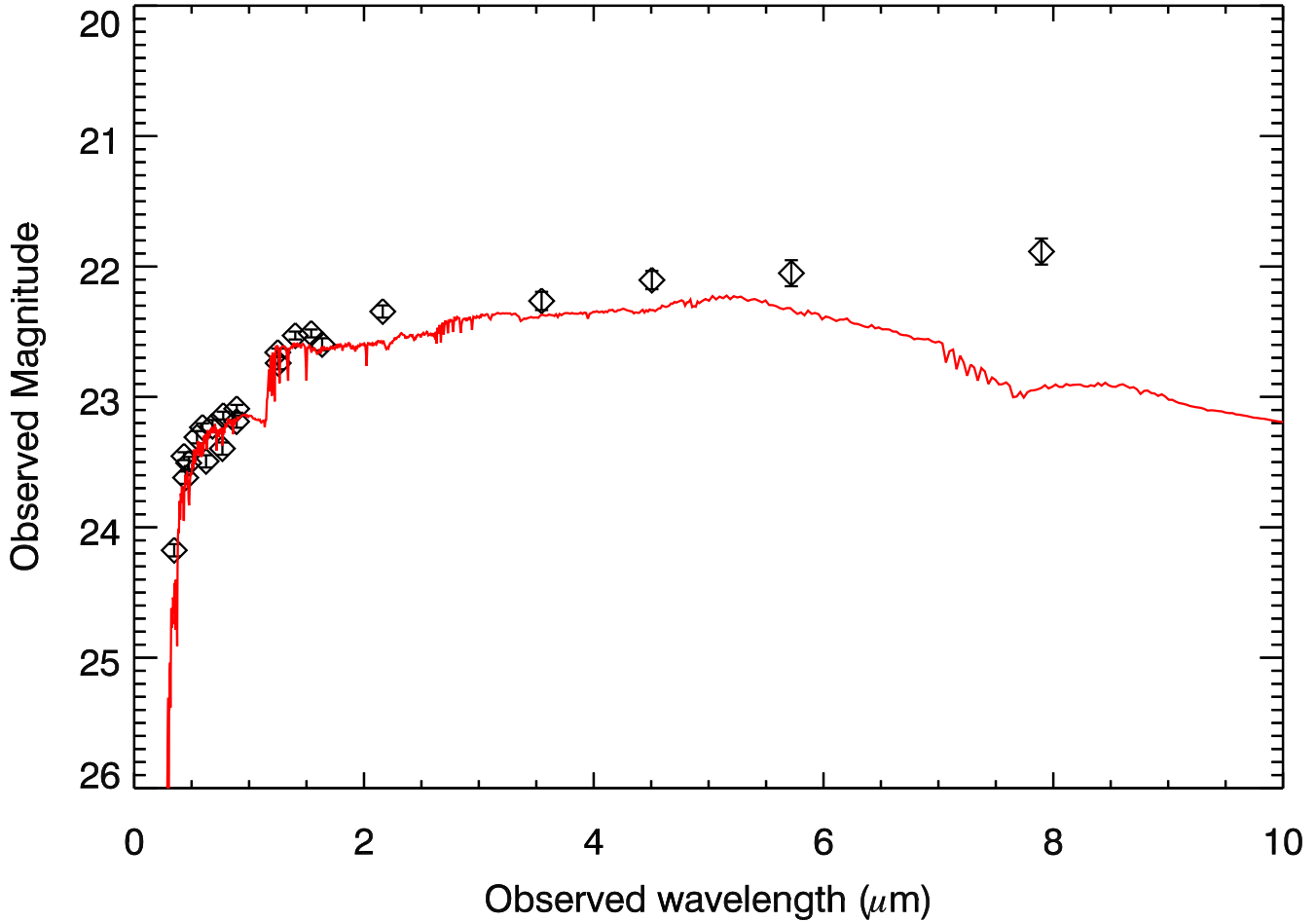}
\caption{Example SED of one of the IR-excess galaxies, GOODS-BX1100 at $z=2.08$, selected based on the $\geqslant2\sigma$ flux excess criteria in the IRAC 3 and 4 channels, as described in Section \ref{sec:AGN}. The diamonds represent the observed magnitudes in respective photometric bands used for SED modeling (Table \ref{tab:phot}), and the red curve shows the best-fit galaxy SED excluding photometric data from IRAC 3 and 4.}
\label{fig:sedex}
\end{figure}

\begin{figure*}
\includegraphics[width=1.0\linewidth]{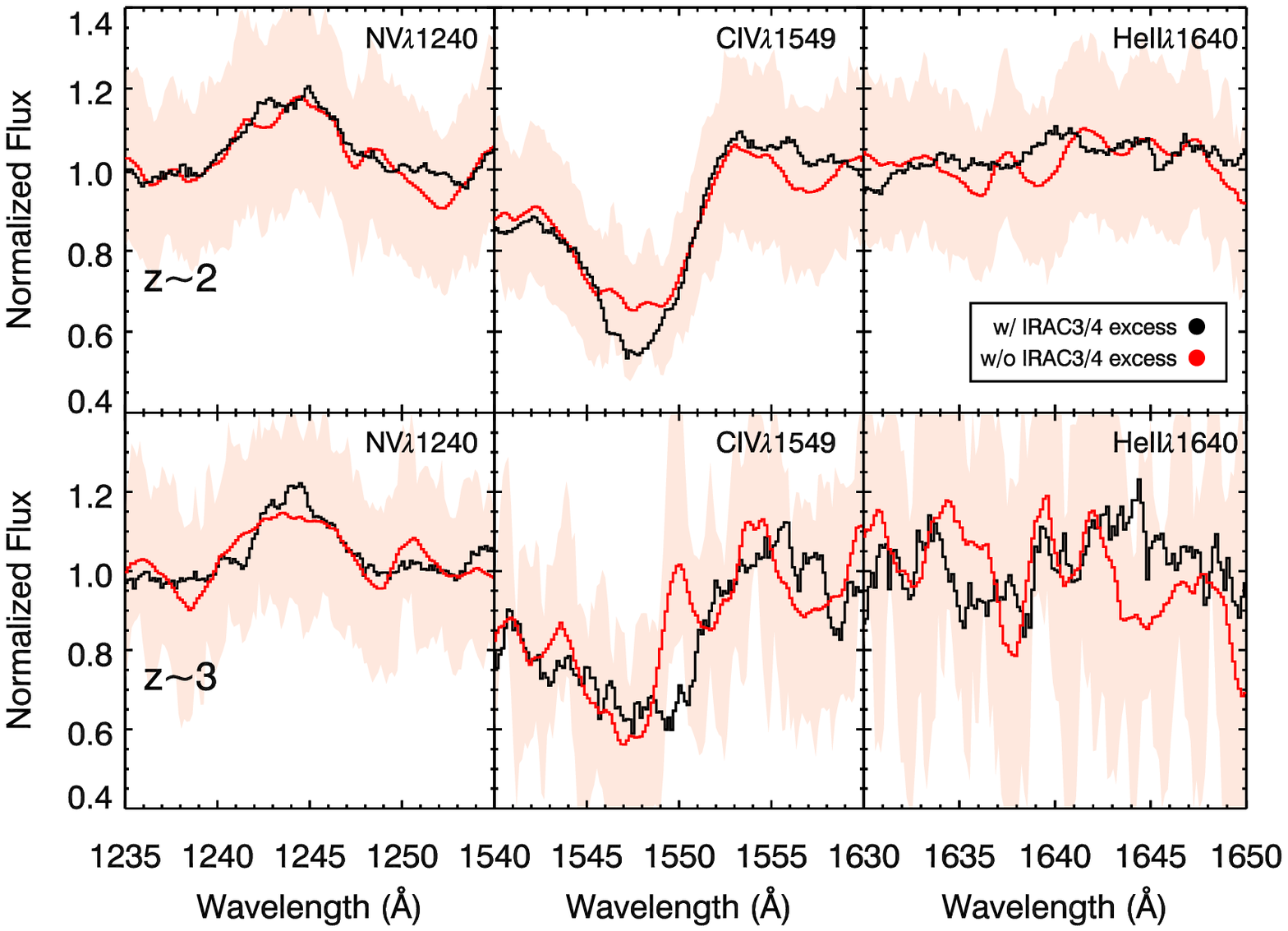}
\caption{Comparison of the AGN signature profiles (\textrm{N}~\textsc{v}, \textrm{C}~\textsc{iv}, and \textrm{He}~\textsc{ii}) between the stacks of objects showing a $\geqslant2\sigma$ excess in either IRAC channel 3 or 4 (red) and those without (black) for the $z\sim2$ (top) and $z\sim3$ (bottom) samples. The light red shaded area represents the $1\sigma$ uncertainty level of the ``excess'' composites. The $z\sim4$ composites are not plotted because of the small number of objects meeting the flux excess threshold. Given the much smaller size of the ``excess'' sample compared to that of the ``non-excess'' sample, the ``excess'' composites have been smoothed to match the continuum $S/N$ of the counterpart ``non-excess'' composites. This smoothing minimizes the visual difference caused by the pixel-to-pixel rms.}
\label{fig:agn}
\end{figure*}

The process of SED modeling reveals a small fraction of objects that stand out because of their flux excess in the IRAC channels relative to the best-fit galaxy SED template. Studies have shown that monotonically rising IRAC SEDs can be an effective tool for selecting AGNs \citep{Reddy2006,Donley2012,Hainline2012}, complementary to other AGN identification methods based on X-ray properties or rest-frame optical and UV spectra. By studying a sample of 33 AGNs identified on the basis of narrow, high-ionization emission lines in the rest-frame UV, \citet{Hainline2012} found that 11 out of 16 objects with IRAC photometry show a monotonically increasing flux in the IRAC channels towards longer wavelength. These authors discovered that the ``excess" in the IRAC bands could be fit by an additional power-law on top of the best-fit galaxy SED to represent the emission from hot dust associated with the AGN. Motivated by the idea that the flux excess in the IRAC bands can possibly be used as a indicator of AGN emission, we select galaxies with IR-excess in our samples, and compare their rest-UV spectra with those showing no flux excess in the mid-IR. 

Fortunately, the $z\sim2-4$ LBG spectra cover multiple high-ionization emission lines characteristic of AGN spectra, including \textrm{N}~\textsc{v}$\lambda$1240, \textrm{C}~\textsc{iv}$\lambda$1549, and \textrm{He}~\textsc{ii}$\lambda$1640. These emission features are clear indicators of photoionization by a non-stellar source, and are therefore typically weak and undetected individually in our redshift samples of star-forming galaxies. However, the use of composite spectra makes it possible to study these lines in greater detail and examine potential AGN activity in the LBGs in our samples.

Although objects with high-ionization UV emission features detected on an individual basis were classified as AGNs and therefore already removed from the samples presented in this work prior to the construction of composite spectra, there may exist an underlying, low-level AGN contribution in the remaining sample that can only be identified within composite spectra. In order to select objects with potential low-level AGN activity, we searched for objects in our redshift samples based on a $\geqslant2\sigma$ flux excess threshold relative to the best-fit SED in 1) either IRAC channel 3 or channel 4 if only one band is detected, or 2) channel 3 $and$ channel 4 if both are detected. In the latter case, we further required that channel 4 has a higher flux density (lower value in magnitude) than channel 3 to ensure the shape of a power-law. The flux density excess was calculated by comparing the observed magnitude in corresponding bands with the theoretical magnitude derived from the best-fit galaxy SED fit excluding IRAC channels 3 and 4. Figure \ref{fig:sedex} shows the SED of an IR-excess galaxy (GOODS-BX1100 in the $z\sim2$ sample) selected based on these criteria.

A total of 55, 28, and 4 objects were selected based on the flux density excess for the $z\sim2$, $z\sim3$, and $z\sim4$ samples, respectively. We did not stack the $z\sim4$ ``excess'' spectra given that this sample is too small to be considered a statistical one. Figure \ref{fig:agn} shows the comparison of the composite spectra with IRAC excess (red) and without (black) for the $z\sim2$ (top) and $z\sim3$ samples (bottom) near the wavelengths of \textrm{N}~\textsc{v}, \textrm{C}~\textsc{iv}, and \textrm{He}~\textsc{ii}. For a better visual comparison, we have smoothed the ``excess" spectrum to match the pixel-to-pixel rms of its ``non-excess" counterpart in order to account for the difference in continuum $S/N$. We found that among the 3 high-ionization emission features, \textrm{N}~\textsc{v} is the strongest in all composites. The \textrm{C}~\textsc{iv} emission is weak, and \textrm{He}~\textsc{ii} is almost not detected at all. It is worth noting that the ``excess" and ``non-excess" composites look fairly similar in that the former do not show specifically stronger \textrm{N}~\textsc{v}, \textrm{C}~\textsc{iv}, or \textrm{He}~\textsc{ii} emission. 
This result suggests that the IR-excess galaxies in our star-forming galaxy samples - which are already cleaned of IR-excess galaxies identified individually as AGNs based on strong rest-UV high-ionization lines - may not be AGNs. The flux density excess in the mid-IR may suggest evidence for hot dust due to active star formation, as observed in some local dwarf galaxies \citep{Hainline2016}. Alternatively, the minor AGN activity in these galaxies, if any, is highly obscured and does not result in distinct features in the rest-frame UV.

\end{document}